\documentclass[12pt]{article}
\usepackage[margin=2.5cm]{geometry}  % add also "showframe" options  
\usepackage{epsfig, graphics}

\usepackage{subcaption}

\usepackage{caption}

\usepackage{amsfonts}
\usepackage{amssymb}
\usepackage{amsmath}
\usepackage[latin1]{inputenc}

 \usepackage[table,xcdraw]{xcolor}
\usepackage{xcolor}
\usepackage{multirow}
\usepackage{hyperref}

\usepackage{float}
\newcommand{\nin}{\noindent}

\def\bkR{{\rm I\kern-.17em R}}

\def \1n{1\hskip -3pt \mbox{N}}

\def \Frac {\displaystyle \frac }

\def \Sum {\displaystyle \sum }

\newfont{\bbf}{cmbx12 scaled 1435}

\begin{document}
\setlength{\baselineskip}{.26in}
\thispagestyle{empty}
\renewcommand{\thefootnote}{\fnsymbol{footnote}}
\vspace*{0cm}
\begin{center}

\setlength{\baselineskip}{.32in}
{\bbf GCov-Based Portmanteau Test
}\\

\vspace{0.5in}

\large{Joann Jasiak}\footnote{York University, 
e-mail:{\it jasiakj@yorku.ca}},
\large{ and Aryan Manafi Neyazi }\footnote{York University, e-mail: {\it aryanmn@yorku.ca }\\

}

\setlength{\baselineskip}{.26in}

 \vspace{0.4in}

This version: \today\\

\medskip

\vspace{0.3in}
\begin{minipage}[t]{12cm}
\small
\begin{center}
Abstract \\
\end{center}

We study nonlinear serial dependence tests based on portmanteau statistics with nonlinear autocovariances for non-Gaussian time series and residuals of dynamic models. A new test with an asymptotic $\chi^2$ distribution is introduced for testing nonlinear serial dependence (NLSD) in time series. It stems from the Generalized Covariance (GCov) residual-based specification test with an asymptotic $\chi^2$ distribution for semi-parametric dynamic models with i.i.d. non-Gaussian errors.
%when the model is correctly specified and estimated by the GCov estimator.  
We derive new asymptotic results under local alternative hypotheses on the parameters of a dynamic model, and extend the GCov test to an infinite set of nonlinear autocovariance conditions. A GCov bootstrap test is introduced
for size adjustments in finite samples, or residual diagnostics 
in models estimated parametrically by a maximum likelihood method.
%by a different method, such as the maximum likelihood 
%under a distributional assumption on the errors.
A simulation study shows that the tests perform well in applications to mixed causal-noncausal autoregressive models. The GCov specification test is used to assess the fit of a mixed causal-noncausal model of aluminum prices with locally explosive patterns, such as bubbles and spikes, between 2005 and 2024.

\bigskip

\end{minipage}

\end{center}
\renewcommand{\thefootnote}{\arabic{footnote}}

\newpage
\section{Introduction}

Diagnostic checking is an important component of the traditional Box-Jenkins procedure for the identification and estimation of stationary linear ARMA models with serially uncorrelated Gaussian errors. It has been routinely conducted by applying the Box-Pierce or Ljung-Box tests to the residuals of an ARMA model to detect serial correlation. Analogously, the diagnostic checking of nonlinear ARCH-type models is often conducted by applying the Box-Pierce or Ljung-Box test twice to the residuals and their squares.
Recently, there has been a growing interest in dynamic models with independent, identically distributed (i.i.d.) non-Gaussian errors, such as the structural vector autoregressive (SVAR) models for macro-economic data, and the mixed causal-noncausal autoregressive (MAR), vector autoregressive (VAR) and double autoregressive (DAR) models for processes with local explosive features, such as spikes and bubbles, displayed by the commodity and cryptocurrency price processes, for example. Diagnostic checking in these models is commonly based on the Ljung Box tests applied to the residuals, their squares, and potentially higher powers, which is a cumbersome multi-step procedure. 

This paper reviews 
convenient one-step diagnostic procedures for jointly testing various forms of nonlinear serial dependence in non-Gaussian time series before estimation, as well as in the residuals ex-post, as a specification test for dynamic models with i.i.d. non-Gaussian errors.

We introduce a new (non)linear serial dependence (NLSD) test for strictly stationary time series with non-Gaussian distributions. We show that in strictly stationary processes, the hypothesis of the absence of linear and nonlinear serial dependence can be tested by a multivariate portmanteau test involving nonlinear autocovariances, i.e., the autocovariance (matrices) of nonlinear transformations of a time series. This approach is available for univariate or multivariate processes, and the test statistic follows asymptotically a chi-square distribution under the null hypothesis \textcolor{black}{of serial independence}.

The NLSD test is inspired by the Generalized Covariance (GCov) specification test introduced by Gourieroux and Jasiak (2023). The GCov specification test allows for detecting nonlinear serial correlation in the residuals. It is a convenient one-step procedure that can replace multiple Box-Pierce and Ljung-Box tests for diagnostic checking of strictly stationary semi-parametric models with i.i.d. non-Gaussian errors. The advantage of the
semi-parametric approach is that it does not require any distributional assumptions on the errors other than being i.i.d. and non-Gaussian. The GCov specification test statistic has an asymptotic chi-square distribution under the null when the dynamic model is estimated by the semi-parametric Generalized Covariance (GCov)  \textcolor{black}{estimator}, which is consistent, asymptotically normally distributed, and semi-parametrically efficient\footnote{It achieves the parametric efficiency
 for a well-chosen set of nonlinear transformations.} [Gourieroux and Jasiak (2023)]. We study analytically and through simulations the properties of \textcolor{black}{the GCov specification test} under local alternative hypotheses to provide convincing arguments and empirical evidence of its potential as a widely applicable diagnostic tool.  \textcolor{black}{We also extend the GCov specification test by considering a set of nonlinear autocovariance conditions increasing to infinity.}
 For size adjustments and power computation in finite samples, we introduce a new GCov bootstrap test. It enlarges the domain of applications to residual diagnostics in parametric models estimated by a Maximum Likelihood (ML) method under a distributional assumption on the errors.

\color{black}
%The limitation of the GCov-specification test is that its asymptotic distribution is known only when the dynamic model is estimated by the Generalized Covariance (GCov) estimator.  It is only then that the multivariate portmanteau test statistic computed from the nonlinear autocovariance (matrices) of residuals follows asymptotically a $\chi^2$ distribution under the null hypothesis.
%This motivates us to introduce a new bootstrap-based GCov test
%that allows for using the GCov specification test in models with i.i.d. errors estimated consistently by a different method, such as the generalized Method of Moments (GMM) or the maximum likelihood (ML), approximate (AML) or quasi maximum likelihood (QML or PML) methods under the parametric assumptions on the error distribution. 

\color{black}
This paper contributes to the literature on univariate and multivariate portmanteau tests. The GCov specification test can be compared to the test of the martingale difference hypothesis of De Gooijer (2023), which relies on a portmanteau test statistic computed from the residuals and squared residuals. The GCov spefication test is more general in the sense that it can include the autocovariances of various nonlinear functions of residuals rather than the residuals and their squares only. An alternative approach for specification testing is based on the auto-distance covariance \textcolor{black}{[Fokianos and Pitsillou (2018) and references therein], or its auto-distance correlation (ADCF) counterpart [Davis et al. (2018)].} Wan and Davis (2022) consider 
the auto-distance covariance (ADCV) function and propose a specification test of the null hypothesis of serial independence of errors [see also Chu  (2023)].  The asymptotic distribution of
the ADVC test statistic is generally intractable, and the asymptotic critical values have to be approximated by bootstrap. In comparison, the GCov specification test based on the GCov estimator has the advantage of a known asymptotic distribution. \textcolor{black}{Another difference between the ADCV and GCov methods is in the weights, which in the GCov procedure are optimally selected and estimated rather than being predetermined}. For data analysis in the frequency domain, the inference methods of Velasco and Lobato (2018) are available, but are not discussed here as the focus of our paper is on the time domain only.

This paper also makes a contribution to the practice of time series analysis. There exists a growing literature on the SVAR models, motivated by the fact that shock identification in these models is feasible by assuming that the structural shocks are independent
and non-Gaussian. 
%The methods used for the estimation of the SVAR models with i.i.d non-Gaussian errors are
%the semi-parametric GMM and the parametric Pseudo Maximum Likelihood PML estimators. 
In applied research, residual diagnostics are often disregarded, likely because applying multiple Ljung-Box tests to the data and then to the model residuals and their powers, for example, is too cumbersome.
Then, the model selection is based only on the goodness-of-fit criteria. For example, Gourieroux, Monfort, and Renne (2017), Lanne and Luoto (2019), Guay (2021), and Keweloh (2020) assume in their application that the errors are independent and non-Gaussian without reporting the results of tests for the absence of (non)linear serial dependence. Gourieroux, Monfort, and Renne (2018) write only that the Box-Pierce (1970) and Ljung-Box (1978) tests of the null hypothesis of absence of autocorrelation were applied to the residuals.  Lanne, Meitz, and Saikkonen (2017) provide the most detailed description of applying the Ljung-Box test to residuals and report p-values. Thus, applied research would benefit from a convenient one-step test designed to detect nonlinear dependence in residuals.

%For example,  Guay (2021) and Keweloh (2020) assume in their application that the errors are independent and non-Gaussian without reporting the results of tests for the absence of (non)linear serial dependence. Gourieroux, Monfort and Renne (2017)  use portmanteau tests to justify the fit of their model without providing the results of residual diagnostic checks. Gourieroux, Monfort and Renne (2018) write only that the Box and Pierce (1970) and Ljung and Box (1978) tests of the null hypothesis of absence of autocorrelation were applied to the residuals. Lanne and Luoto (2019) comment on the absence of serial correlation of errors without specifying how it was determined. Lanne, Meitz, and Saikkonen (2017) provide the most detailed description of applying the Ljung-Box test to the residuals and provide the p-values. This applied research would benefit from a convenient one-step test designed for detecting nonlinear dependence in the residuals.

 The GCov-based specification test, in addition to being applicable to a variety of models with i.i.d. non-Gaussian errors, is particularly useful for testing the fit of causal-noncausal dynamic models with non-Gaussian errors. The reason is that 
nonlinear autocovariances identify both causal and noncausal dynamics [Chan et al. (2006)]. There is a growing interest in univariate and multivariate mixed causal-noncausal models for processes with locally explosive patterns, such as short-lasting trends, bubbles and spikes, and time-varying volatility [see e.g. Hecq et al. (2016), (2020),  Gourieroux and Jasiak (2017), (2023), Gourieroux and Zakoian (2017),  Fries and Zakoian (2019), Davis and Song (2020), Hecq and Voisin (2021), Swensen (2022)]. In practice, locally explosive patterns characterize the time series of commodity prices, including oil, soybean, nickel, and aluminum prices, as well as cryptocurrency prices of native cryptocurrencies, tokens, and some stablecoins. The estimators available for this class of models are the aforementioned semi-parametric GCov estimator [Gourieroux and Jasiak (2017), (2023)] and the parametric (Approximate) Maximum Likelihood (AML), or ML estimators [Lanne and Saikkonen (2011), Davis and Song (2020)]. When the error distribution of a mixed model is misspecified, the AML estimator can be unreliable [Hecq, Lieb, and Telg (2016)],
adversely affecting the outcomes of any ML-based fit criteria, such as the AIC, \textcolor{black}{Schwarz}, and Hannan-Quinn criteria. Hence, in these models, the proposed tests arise as convenient tools for both detecting nonlinear serial dependence in the data and testing the goodness of fit.
 
%We examine the finite sample performance of the GCov specification test analytically under a sequence of local alternatives to study the hypotheses about the parameters of a semi-parametric model. Our simulation study shows that the NLSD, GCov specification, and GCov bootstrap tests perform well in detecting nonlinear serial dependence in mixed causal-noncausal processes and the residuals of univariate (MAR) and multivariate autoregressive VAR models, respectively. The semi-parametric GCov specification test successfully detects the correct fit in various models and is a valuable diagnostic tool.  We also evidence the good performance of the GCov bootstrap test applied to the residuals of a model estimated by the AML estimator. We illustrate empirically the GCov specification test applied to a mixed causal-noncausal autoregressive (MAR) model of commodity prices estimated by the AML method.

The following notation is used. For any $m \times K$ matrix $A$ with the $k$th column  $a_k, k= 1, . . ., K$ $vec(A)$ will denote the column vector of dimension $mK$ defined as: 
$$vec(A) = (a_1'...,a_k',...,a_K')',$$
 \nin where the prime denotes a transpose. For any two matrices $A \equiv (a_{ik})$ and $B$, the Kronecker product $(A \otimes B)$ is the block matrix with the $(i,k)$th block denoted by $a_{ik}B$.

The paper is organized as follows. Section 2 introduces the NLSD test for time series inspired by the classical multivariate portmanteau test, which is briefly reviewed. Section 3 reviews the GCov estimator and describes the GCov specification test.
%based on the portmanteau test statistic with nonlinear autocovariances of residuals of a model estimated by the GCov estimator. 
It provides new results on the asymptotic properties of this test under a sequence of local alternatives. Section 4 introduces the GCov bootstrap test. Section 5 examines how the asymptotic performance of the GCov specification test can be improved by increasing the set of nonlinear autocovariances. Section 6 presents the simulation results. Section 7 presents an empirical application of the GCov specification test to a mixed causal-noncausal model fitted to monthly aluminum prices recorded between 2005 and 2024. Section 8 concludes. The Online Appendix contains the technical results in Appendices A, B, and additional simulations and empirical results in Appendix C.

\setcounter{equation}{0}\def\theequation{2.\arabic{equation}}
\section{NLSD Test for Non-Gaussian Processes}

In this section, we first recall the existing results on the weak white noise hypothesis test. Next, we introduce a new test of the absence of linear and nonlinear serial dependence (NLSD) for non-Gaussian processes based on a portmanteau statistic involving nonlinear transformations of strictly stationary univariate or multivariate time series with non-Gaussian marginal distributions and nonlinear dynamics. 

\subsection{Linear Serial Dependence in Time Series}

This section summarizes the results 
%existing in the literature 
on the test of the weak white noise hypothesis, i.e., the absence of linear serial dependence in time series. 

\subsubsection{Univariate Time Series}

Let us consider a univariate stationary time series $(y_t)$ with finite fourth-order moments\footnote{The existence of moments up to order 4 is needed to derive the asymptotic variance of $\hat{\gamma}(h)$ under the asymptotic normality.}. The test of the weak  white noise hypothesis $H_0= \{ \gamma(h) = 0, h=1,....,H \}$, with $\gamma(h) = Cov(y_t, y_{t-h})$ is commonly based on the test statistic:
\begin{equation}
\label{1.1}
\hat{\xi}_T(H) = T \sum_{h=1}^H \hat{\rho}(h)^2 = T \sum_{h=1}^H \frac{\hat{\gamma}(h)^2}{\hat{\gamma}(0)^2},
\end{equation}
\nin where $\hat{\gamma}(h)$ and $\hat{\rho}(h)$ are the sample autocovariance and autocorrelation of order $h$, respectively, computed from a sample of $T$ observations\footnote{In this Section the index $T$ of the estimators, e.g., $\hat{\gamma}_T(h)$, is omitted to simplify the notation.}.

Under the null hypothesis of serial independence and standard regularity conditions, this statistic follows asymptotically a chi-square distribution $\chi^2(H)$ with $H$ degrees of freedom [see Box and Pierce ( 1970)]. 
%The following two subsections introduce an analogous test statistic for testing for the absence of
%(non)linear serial dependence in strictly stationary univariate or multivariate processes.

\subsubsection{Multivariate Time Series}

Let us now consider a strictly stationary time series $(Y_t)$ of dimension $m$ with finite fourth-order moments. The null hypothesis is  $H_0 = \{ \Gamma(h) = 0, h=1,....,H \}$, where $\Gamma(h)= Cov(Y_t, Y_{t-h})$ is the autocovariance (matrix) of order $h$. The multivariate equivalent of the test statistic (2.1) is:
\begin{equation}
\hat{\xi}_T(H) = T  \sum_{h=1}^H Tr [ \hat{R}^2(h)],
\end{equation} 
\nin where $\hat{R}^2(h)$ is the sample analogue of the multivariate R-square defined by:
\begin{equation}
R^2(h) =  \Gamma(h) \Gamma(0)^{-1} \Gamma(h)' \Gamma(0)^{-1}. 
\end{equation} 
\nin Since 
\begin{equation}
\hat{R}^2(h) = \hat{\Gamma}(0)^{1/2} [ \hat{\Gamma}(0)^{-1/2} \hat{\Gamma}(h) \hat{\Gamma}(0)^{-1} \hat{\Gamma}(h)'\hat{\Gamma}(0)^{-1/2}] \hat{\Gamma}(0)^{-1/2},
\end{equation}
\nin this matrix is equivalent up to a change of basis to the matrix within the brackets, which is symmetric and positive-definite.  Therefore, it is diagonalizable, with a trace equal to the sum of its eigenvalues, which are the squares of the empirical canonical correlations between $Y_t$ and $Y_{t-h}$, denoted by $\hat{\rho}_i^2(h), i=1,....,m$ [Hotelling (1936)]:
\begin{equation}
\hat{\xi}_T(H)  =  T \sum_{h=1}^H  Tr [ \hat{\Gamma}(h) \hat{\Gamma}(0)^{-1} \hat{\Gamma}(h)'\hat{\Gamma}(0)^{-1}] \nonumber = T \sum_{h=1}^H [ \sum_{i=1}^m \hat{\rho}_i(h)^2].
\end{equation}
\nin Under the null hypothesis of strong white noise, this statistic follows asymptotically a chi-square distribution $\chi^2(mH)$ [see, e.g., Robinson (1973), Anderson (1999), Section 7, Anderson (2002), Section 5].

 %There exist alternative test statistics that have been introduced in the literature and are asymptotically equivalent to the test statistic $\hat{\xi}_T(H)$ under the null hypothesis. 
% For example, we can consider the Seemingly Unrelated Regression (SUR) model:

%\begin{equation}
%Y_t = \alpha +  B_1 Y_{t-1}+ \cdots + B_H Y_{t-H} + u_t,
%\end{equation}

%\nin and introduce the statistic by using Frisch-Waugh-Lovell theorem:

%\begin{equation}
%\tilde{\xi}_{1T}(H) = T\; Tr [ \hat{\Gamma}^*(1) \hat{\Gamma}^*(0)^{-1} %\hat{\Gamma}^*(1)'\hat{\Gamma}^*(0)^{-1} ],
%\end{equation}

%\nin where $\Gamma^*(1) = Cov(Y_t, \underline{Y_{t-1}}), \Gamma^*(0) = V(\underline{Y_{t-1}})$ and
%$\underline{Y_{t-1}} = (Y_{t-1}',....,Y'_{t-H})'$.

%Under the null hypothesis of strong white noise, the explanatory variables in (2.6) are (asymptotically) uncorrelated, which explains the possibility of  replacing the canonical correlation analysis of dimension $nH$ by $H$ canonical correlations of dimension $n$ only. 

%\nin The statistics $\hat{\xi}_T(H)_{1T}$ and $\hat{\xi}_T(H)$ are asymptotically equivalent, that is:

%$$ \hat{\xi}_{1,T}(H) - \hat{\xi}_T(H) = o_p(1).$$

%\nin Then, these test statistics have  the same asymptotic distribution under the serial independence condition. This result is demonstrated in Appendix A.

\subsection{Nonlinear Serial Dependence in Time Series}

The NLSD test based on nonlinear functions of a strictly stationary process with a non-Gaussian distribution extends the approach presented in the previous section. For processes with Gaussian distributions, zero-valued autocovariances are equivalent to serial independence, which becomes the null hypothesis of interest. Then, the asymptotic distribution of the test statistics 
%for testing the serial independence hypothesis 
is determined under that null hypothesis. In the case of non-Gaussian processes, zero-valued linear autocovariances do not imply serial independence.
Chan et al. (2006) and Gourieroux and Jasiak (2023) show that the nonlinear autocovariances, i.e., the autocovariances of nonlinear functions of non-Gaussian processes reveal and identify the nonlinear and noncausal serial dependencies. This
%\footnote{The test is also applicable to stationary Gaussian time series.}.
suggests that the null hypothesis of the absence of (non)linear serial dependence in univariate or multivariate time series $Y_t$ can be tested by applying the test statistic $\hat{\xi}_T(H)$ to a set of nonlinear transformations of $Y_t$. Thus, the NLSD test concerns the null hypothesis of zero autocovariances of the transformed series.  
 
%The NLSD test statistic is denoted by $\hat{\xi}_T(H)_a$ and computed from 
We consider nonlinear transformations $Y_t^a$ of a univariate or multivariate non-Gaussian process $Y_t$, where $a=\{ a_1, a_2,... \}$ is a set of nonlinear scalar functions satisfying the regularity conditions given in Gourieroux and Jasiak (2023).
%Let us consider a vector of such nonlinear functions $a$ of a strictly stationary process $y_t$. That vector increases the dimension of $y_t$ by appending it with the nonlinear functions of $y_t$, such as the squares or logarithms. In particular if
%$(y_t)$ has no finite fourth-order moment, then it can be replaced by a transformed multivariate process $Y_t^a$ with a finite fourth-order moment to ensure the validity of the asymptotic distributional results under the null hypothesis.
\textcolor{black}{The transformations include $Y_t$ to accommodate linear serial dependence and the powers of $Y_t$, for example, to account for nonlinear serial dependence. If the process has fat tails, the existence of the fourth-moment condition may not be satisfied by $Y_t$, but instead by a well-chosen transformation $a_k(Y_t)$. Hence, transformations such as $\sqrt{Y_t}$, $\log Y_t$, or $1/(1+\exp(-Y_t))$ of univariate positive processes can be used depending on the tails of their distributions.}

For ease of exposition, let us introduce $K$ nonlinear scalar functions $a_1,...,a_K$ of the process ($Y_t$), transforming it into a multivariate process of dimension $K$ with components $a_k(Y_t)$:
$$Y_t^a = \left( \begin{array}{c} a_1(Y_{t}) \\
                          \vdots \\
                          a_K (Y_t)
                          \end{array}\right),$$

%\nin \textcolor{black}{where for a univariate time series $y_t$, the transformation} $a_1(y_t) = y_t$ is the time series itself, allowing the test to capture the linear dependence. 

\nin We compute the sample autocovariances of the transformations $a_k(Y_t), k=1,...,K$:
$$\hat{\Gamma}^a(h) = \frac{1}{T} \sum_{t=h}^T  Y_t^a Y_{t-h}^{a'} - \frac{1}{T} \sum_{t=h}^{T-1} Y_t^{a'} \frac{1}{T} \sum_{t=h+1}^T  Y_{t-h}^a.$$
%\nin  assume that $a_k(y_t), \; k=1,...,K$ have finite variances. 
\nin Once a set of transformations is determined, the null hypothesis:
$$H_{0,a} = (\Gamma^a (h) = 0, \; h=1,...,H),$$
\nin is used to test for the absence of (non)linear serial dependence. Because one cannot consider all the lags and nonlinear transformations, the null hypothesis $H_{0,a}$ is not equivalent to the serial independence condition, but is arbitrarily close to it, depending on the lag $H$ and the set of nonlinear transformations considered [see Section 5]. 
Then, the NLSD test statistic:
\begin{equation}
\label{NLSD}
\hat{\xi}_T(H)_a = T \sum_{h=1}^H Tr \textcolor{black}{[}\hat{R}_a^2(h)\textcolor{black}{]},
\end{equation}
\nin where:
$$\hat{R}_a^2(h) =  \hat{\Gamma}_a(h) \hat{\Gamma}_a(0)^{-1} \hat{\Gamma}_a(h)'\hat{\Gamma}_a(0)^{-1}],$$
\nin is computed from the nonlinear sample autocovariances, i.e., sample autocovariance matrices of a transformed univariate or multivariate process. In each case, the dimension of the process can be increased, i.e., become higher than that of the initial time series of interest. By analogy to the traditional literature, we use the independence hypothesis to obtain the asymptotic distribution of the test statistic.  If the dimension of the transformed process is $K$, then under serial independence, the NLSD test statistic 
(\ref{NLSD}) 
follows asymptotically a $\chi^2(K^2 H)$ distribution [see, e.g., Robinson (1973), Chitturi (1976), Anderson (1999), Section 7, Anderson (2002), Section 5]. The test of the null hypothesis $H_{0,a}$ at level $\alpha$ is performed as follows: the null hypothesis $H_{0,a}$ is rejected when $\hat{\xi}_T (H)_a > \chi^2 _{1-\alpha}(K^2H)$ and $H_{0,a}$ is not rejected, otherwise.

%For some practical implementations, it is easy to show that for univariate processes and their quadratic transformations, the NLSD test statistic is invariant with respect to the scale effect and change of sign of $y^a_t$. For example, let us consider a univariate $(y_t)$ and a diagonal matrix $\Lambda$:
%$$\Lambda= \left[ \begin{array}{cc} \lambda & 0 \\ 0 & \lambda^2 \end{array} \right], $$
%\nin where $\lambda$ represents the scale effect or the change of sign effect for $\lambda = -1$. Then the multivariate R-square of process $y_t^a = [y_t, y_t^2]'$:
%$$
%R_a^2(1)  =  \Gamma_a(1) \Gamma_a(0)^{-1} \Gamma_a(1)' \Gamma_a(0)^{-1}
%$$
%\nin computed for the rescaled process $\Lambda y_t^a$ is:
%\begin{eqnarray*}
%\tilde{R}_a^2(1) & = & \Lambda \, \Gamma_a(1) \, \Lambda [\Lambda \, \Gamma_a(0) \, \Lambda]^{-1} \, \Lambda' \Gamma_a(1)'\, \Lambda' \, [\Lambda \, \Gamma_a(0) \, \Lambda]^{-1}  \nonumber \\
%& = & \Lambda \Gamma_a(1) \Lambda \Lambda^{-1}  \Gamma_a(0)^{-1} \Lambda^{-1} \Lambda \Gamma_a(1)' \Lambda  \Lambda^{-1} \Gamma_a(0)^{-1} \Lambda^{-1}  \nonumber \\
%& = & \Lambda \Gamma_a(1) \Gamma_a(0)^{-1} \Gamma_a(1)' \Gamma_a(0)^{-1} \Lambda^{-1}.
%\end{eqnarray*}
%\nin Hence, the multivariate R-square of the transformed process is $\Lambda R_a^2(1) \Lambda^{-1}$. We see that its trace is:
%$$Tr(\tilde{R}_a^2(1)) = Tr(\Lambda R_a^2(1) \Lambda^{-1}) = Tr (R_a^2(1) \Lambda \Lambda^{-1}) = Tr R_a^2(1), $$
%\nin which implies that the test statistic $T Tr \tilde{R}_a^2(1) = T Tr(\Lambda R_a^2(1) \Lambda^{-1}) $ remains unchanged and is equal for $y_t^a$ and $\Lambda y_t^a$ .

\setcounter{equation}{0}\def\theequation{3.\arabic{equation}}

\section{GCov Specification Test for Semi-Parametric Models}

The NLSD test introduced in Section 2.2 is a special case of the Generalized Covariance (GCov) specification test introduced by Gourieroux and Jasiak (2023) for semi-parametric nonlinear models of strictly stationary time series with i.i.d. errors and parameter vector $\theta$ describing their dynamics. In this context, $\hat{\xi}_T(H)$ is computed from a multivariate time series of residuals and their nonlinear transforms instead of an observed time series, which \textcolor{black}{reduces the degrees of freedom of its limiting  $\chi^2$ distribution} [Gourieroux and Jasiak (2023)]. The model and GCov test are reviewed below.

\subsection{The Semi-Parametric Model}
%{\bf rewrite to change a bit the wording}

Let us consider a strictly stationary process $(Y_t)$ satisfying a semi-parametric model studied in Gourieroux and Jasiak (2023):
\begin{equation}
\label{3.1}
g(\tilde{Y}_t; \theta) = u_t,
\end{equation}

\nin where $g$ is a known function with $dim(g) = dim(u_t)$, $\tilde{Y}_t= (Y_t, Y_{t-1},\ldots, Y_{t-p})$, $p$ is a non-negative integer, $(u_t)$ is an i.i.d. sequence of errors (not necessarily with mean zero) with a common marginal density function $f$ and $\theta$ is an unknown parameter vector of dimension $dim(\theta)$. We assume that the model is well-specified, the true value of parameter $\theta$ is $\theta_0$, and the true error density is $f_0$. 
%Model (\ref{3.1}) does not imply a nonlinear causal autoregressive specification of order $p$ for process $(Y_t)$ because the dimension of $Y_t$ can be strictly larger (resp. smaller) than the dimension of $u_t$. Hence, model (\ref{3.1}) is not directly invertible with respect to $Y_t$. 
Moreover, $u_t$ is not assumed to be independent of $\tilde{Y}_{t-1}$.
%Therefore, the information generated by the current and lagged values of $Y_t$  does not necessarily coincide with the information generated by the current and lagged values of $u_t$. 
\textcolor{black}{
Hence, errors $u_t$ are not necessarily interpretable as 
%either causal or non-causal (nonlinear) 
innovations, assuming that $g(Y_t, Y_{t-1},..; \theta)$ is an invertible function of $Y_t$.}

Similarly, as was done in Section 2.2, a process following model (\ref{3.1}) can be transformed into a system of dimension $K$ by considering (linear and) nonlinear scalar transformations of $u_t$. Let us introduce $K$ nonlinear transformations $a_1,...,a_K$. Then we have:
\begin{eqnarray}
\label{3.2}
a_k[g(\tilde{y}_t; \theta)] & = & a_k (u_t) , \; k=1,...,K, \nonumber \\
  \mbox{or, equivalently} \;\;     a[g(\tilde{y}_t; \theta)]  &=& a (u_t) \equiv v_t,
\end{eqnarray}
\nin where the transformed process $(v_t)$ is also an i.i.d. process. Henceforth, the subscripts of transformations $a$ are omitted for clarity.

\subsection{The GCov Specification Test}

We consider the null hypothesis of the absence of (non)linear serial dependence in i.i.d. errors whose distribution is left unspecified. The portmanteau test statistic for testing the model specification is:
    \begin{equation}
    \label{2.17}
      \hat{\xi}_T(H) =T L_T( \hat{\theta}_T), \;\; \mbox{where} \;\;
L_T( \hat{\theta}_T) = \sum\limits_{h=1}^{H} Tr[\hat{R}_T^2(h, \hat{\theta}_T)],    
        \end{equation}
\nin and
    \begin{equation}
        \hat{R}_T^2(h,\theta)=
        \hat{\Gamma}_T(h; \hat{\theta}_T) \hat{\Gamma}_T(0; \hat{\theta}_T)^{-1}
        \hat{\Gamma}_T(h; \hat{\theta}_T)'
        \hat{\Gamma}_T(0; \hat{\theta}_T)^{-1},
    \end{equation}
\nin and the estimated autocovariances $\hat{\Gamma}_T(h;\hat{\theta}_T)$ are the sample autocovariances\footnote{They have to be divided by $T$ instead of 
$(T-H-p)$ to ensure that the sequence of multivariate sample autocovariances remains positive semi-definite.} of the residuals $\hat{u}_{t,T} = u_t (\hat{\theta}_T) = g(\tilde{Y}_t, \hat{\theta}_T)$ evaluated at the GCov estimator $\hat{\theta}_T$ of $\theta$ defined by:
    \begin{equation}
        \hat{\theta}_T(H)= Argmin_{\theta} \sum\limits_{h=1}^{H} Tr[\hat{R}_T^2(h,\theta)].
    \end{equation}
\nin When model (3.1) is well-specified, \textcolor{black}{and the true value $\theta_0$ is identifiable from the zero nonlinear autocovariance conditions}, the one-step GCov estimator is consistent, asymptotically normally distributed, and attains a semi-parametric efficiency bound, under standard regularity conditions [Gourieroux and Jasiak (2023), assumptions A.1, A.2 and Proposition 3]. Then, the GCov portmanteau test statistic
$\hat{\xi}_T(H)$ follows asymptotically the chi-square distribution with degrees of freedom equal to  $K^2H-dim(\theta)$ [see Gourieroux and Jasiak (2023), Proposition 4]. This result holds only for the GCov estimator $\hat{\theta}_T$.
The GCov specification test can be applied as a diagnostic tool to a variety of dynamic models, with i.i.d. non-Gaussian errors, including the following ones:

\medskip
\nin {\bf Example 1: Double Autoregressive DAR(1) Model}
$$y_ t = \phi y_{t-1} + u_t \sqrt{w+ \alpha y_{t-1}^2},$$
\nin where $w >0, \phi \geq0, \alpha \geq 0$, $\theta= (w, \phi, \alpha)'$ and the $u_t$'s are i.i.d. with a Gaussian or non-Gaussian distribution [Ling (2007)]. We assume that $E(log|\phi +\sqrt{\alpha} u|)<0$ and the regularity conditions on functions $\theta, f$ are satisfied, ensuring the existence of a strictly stationary solution. 
%Moreover, the initial value $y_0$, is assumed drawn in the stationary distribution. 
Then, the semi-parametric representation (\ref{3.1}) of this model is
$$ g (\tilde{y}_t ; \theta) = [(y_t - \phi y_{t-1})/\sqrt{w+ \alpha y_{t-1}^2} = u_t,$$
\nin In particular, this process is strictly stationary for $\phi=1$ (i.e., in the unit root case) due to the volatility induced \textcolor{black}{mean reversion}. 

\medskip
\nin {\bf Example 2:  Mixed (Causal-Noncausal) Autoregressive MAR(r,s) Model}
\begin{equation}
\label{mar(r,s)}
    (1-\phi_1 L - \phi_2 L^2 - ... -\phi_r L^r)(1- \psi_1 L^{-1} - \psi_2 L^{-2} - ... - \psi_s L^{-s})y_t = u_t,
\end{equation}
\nin where the errors are i.i.d., non-Gaussian and such that $E(|u_t|^{\delta})<\infty$ for $\delta>0$. The polynomials $\Phi(L)$ and $\Psi(L)$ in $L$ are of degrees $r$ and $s$, respectively, with roots strictly outside the unit circle and such that $\Phi(0)=\Psi(0)=1$. For $r=s=1$, the MAR(1,1) model is obtained:
\begin{equation}
\label{mar(1,1)}
(1- \phi L)(1-\psi L^{-1}) y_t = u_t, 
\end{equation}
\nin where the autoregressive coefficients satisfy the condition $|\phi|<1$ and $|\psi|<1$ for strict stationarity. Coefficient $\phi$ represents the standard causal persistence, while coefficient $\psi$ determines noncausal persistence, along with
locally explosive patterns and conditional heteroscedasticity. For the MAR(1,1) we have $\theta=(\phi, \psi)'$, and the semi-parametric representation (\ref{3.1}) of this model is:
$$
g (\tilde{y}_t ; \theta) = \Phi(L)\Psi(L^{-1})y_t = u_t.
$$

\medskip
\nin {\bf Example 3: Causal \textcolor{black}{SVAR(p)} Model} 
$$Y_t = \Phi_1 Y_{t-1}+ \cdots + \Phi_p Y_{t-p} + D u_t,$$
\nin where $\theta = [vec \Phi_1', ..., vec \Phi_p']'$, $u_t$ is a multivariate, non-Gaussian, serially and cross-sectionally i.i.d. error \textcolor{black}{ and matrix $D$ is square and invertible.}
The roots of the characteristic equation $det(Id - \Phi_1 \lambda - \cdots  \Phi_p \lambda^p) = 0$ are of modulus strictly greater than one. We have
$$g(\tilde{Y}_t, \theta)  = D^{-1} [Y_t - \Phi_1 Y_{t-1} - \cdots  - \Phi_p Y_{t-p}] = u_t.$$
\nin This model is commonly used in macroeconomic applications.

\medskip
\nin {\bf Example 4: Causal-Noncausal VAR(p) Model} 

\nin The model is specified as above except that the roots of the characteristic equation $det(Id - \Phi_1 \lambda - \cdots  \Phi_p \lambda^p) = 0$
are of modulus either strictly greater, or smaller than one
and the errors are not necessarily assumed to be cross-sectionally independent. The roots located inside the unit circle create locally explosive patterns and conditional heteroscedasticity, like in the MAR(r,s) model. There exists a unique (strictly) stationary solution $(Y_t)$ with a two-sided $MA(\infty)$ representation, which satisfies model (\ref{3.1}) with:
$$g(\tilde{Y}_t, \theta)  =  Y_t - \Phi_1 Y_{t-1} - \cdots  - \Phi_p Y_{t-p} = u_t.$$
\nin The causal-noncausal VAR(p) model has been studied in Gourieroux and Jasiak (2017),(2023), Davis and Song (2020), and Swenson (2020).  
%The error $u_t$ of this model cannot be interpreted as an innovation. 

%Even though the function $g$ is linear in the current and lagged values of $Y_t$, the assumption of strict stationarity of $Y_t$ implies a nonlinear causal dynamics of $Y_t$ with predictions $E(Y_t|\underline{Y_{t-1}})$ nonlinear in $\underline{Y_{t-1}} = (Y_t, Y_{t-1},...)$ and  past-conditional heteroscedasticity $V(Y_t| \underline{Y_{t-1}})$.

\subsection{Asymptotic Properties of the GCov Specification Test}

This section presents new results on the asymptotic properties of the GCov specification test under local alternatives. 
%to show its validity as a diagnostic test for semi-parametric models with i.i.d. errors. 
By focusing on the local alternatives, we can study hypotheses on the parameters of a semi-parametric model, given the marginal error distribution and extend the results on the asymptotic distribution of the GCov test statistic derived in Gourieroux and Jasiak (2023).

\subsubsection{Null Hypothesis and Asymptotic Size}

Let us discuss the null hypothesis in the semi-parametric framework. As mentioned earlier, there are two types of parameters: vector $\theta$ defining the dynamics, and functional parameter $f$ defining the error distribution. Hence, the theoretical autocovariances $\Gamma(h; \theta, f)$ are functions of both $\theta$ and $f$. \textcolor{black}{Suppose that $\theta_0, f_0$ are identifiable from the condition: $\Gamma_0(h; \theta, f)=0, \; h=1,...,H$ \footnote{For example, in the SVAR model, identification can be ensured by assuming that the components of errors $u_t$ are cross-sectionally independent and non-Gaussian. A detailed discussion of identification is out of the scope of this paper.}.} Then, the null hypothesis becomes:
$$H_0 : \{\Gamma_0 (h) =0, \; h=1,...,H\},$$
\nin where $\Gamma_0(h)$ is the true autocovariance. In terms of parameters $\theta, f$, it corresponds to the parameter set $\Theta_0= \{ \theta, f: (\Gamma(h; \theta, f) = 0, \; h=1,...,H \}$. 
So far, the transformation $a$ has not been introduced to simplify the notation.  When these transformations are accounted for, the null hypothesis and the associated parameter set become:
$$H_{0,a} \approx \Theta_{0,a}= \{ \theta, f: \Gamma_{0,a} (h; \theta, f) =0, \; h=1,...,H\}.$$
\nin where $\Gamma_{0,a} (h; \theta, f)$ is the true autocovariance of the transformed errors at lag $h$. When the error terms $u_t$ are serially independent, the nonlinear autocovariances $\Gamma_{0,a} (h; \theta_0, f_0) = 0,\; \forall h$. Hence, if the model is correctly specified, 
%and $H \geq p$, \textcolor{black} {where $p$ is the highest lag of $Y_{t-h}$ in $\tilde{Y}_t$ in model (3.1), Section 3.1}, 
the GCov test does not reject the null hypothesis $H_{0,a}$ of the absence of linear and nonlinear dependence in the errors $u_t$. 
%Since one cannot consider all the lags and nonlinear transformations, the null hypothesis $H_{0,a}$ is not equivalent to the serial independence condition, but is arbitrarily close to it, depending on lag $H$ and the set of nonlinear transformations considered. 

The test of hypothesis $H_{0,a}$ at level $\alpha$ is performed as follows: the null hypothesis $H_{0,a}$ is rejected when $\hat{\xi}_T (H)_a > \chi^2 _{1-\alpha}(K^2H-dim(\theta))$ and $H_{0,a}$ is not rejected otherwise. The asymptotic size tends to the nominal size when $T\rightarrow \infty$:
$$\lim_{T \rightarrow \infty} P_{\theta,f} [ \hat{\xi}_T(H)_a >  \chi^2 _{1-\alpha}(K^2H-dim(\theta))] = \alpha, \; \forall \theta, f \in \Theta_0, \; \forall \alpha.$$

\subsubsection{Local Alternatives and Local Power}

\textcolor{black}{Rather than evaluating the performance of our test against a fixed alternative, we can examine a sequence of local alternatives that drift towards the null at rate $\frac{1}{\sqrt{T}}$.
The drifting sequence of local alternatives makes it harder and harder to reject the null as the sample grows. The evaluation of the test under several types of local alternatives  reveals the direction of alternative to consider when the null hypothesis is rejected [Davidson and MacKinnon (2006), Section 3, Escanciano (2007), Section 2.3, Dovonon, Hall, and Kleibergen (2020), Section 4].}

Because the semi-parametric model depends not only on the vector of dynamic parameters $\theta$ but also on the functional parameter $f$, the general alternative hypothesis is difficult to formulate.
In particular, the alternative could contain models in which $u_t$'s are serially dependent, and we would need to introduce additional parameters to accommodate these cases.

For clarity, we omit in this Section the subscripts $a$ of transformation, and assume a semi-parametric alternative model defined by:
\begin{equation}
 g^*(\tilde{Y}_t; \theta, \gamma) = u_t,  \; \mbox{where} \; u_t \; \mbox{are i.i.d.}, 
\end{equation}
\nin with an additional (scalar) parameter $\gamma$, a known function $g^*$, such that $g^*(\tilde{Y}_t; \theta,0) = g(\tilde{Y}_t; \theta)$ and dimension $dim(g^*) = dim(u_t)$. Under this alternative model, the autocovariances depend on $\theta, \gamma$ and the marginal distribution $f$ of errors $u_t$, and are denoted by $\Gamma(h; \theta, \gamma, f)$.

The alternative hypothesis based on autocovariances is parametrized by $\theta, \gamma, f$. It is written, in terms of the parameters as:
$$H_1 \approx \Theta_1= \{\theta, \gamma, f: \Gamma(h; \theta, \gamma, f) = 0, \; h=1,...,H \}$$
\nin and the null hypothesis is:
$$H_0= \{\theta, \gamma, f: \Gamma(h; \theta, 0, f) = 0, \; h=1,...,H \} = \{ \gamma=0\} \cap H_1.$$

\textcolor{black}{Let us examine the fixed and local alternatives in the context of the DAR(1) model. The alternatives concern the presence of serial dependence at higher lags in 1. the conditional mean, and 2. the conditional variance.}

\medskip
\nin \textcolor{black}{{\bf Example 4: DAR(1) model cont.}} The (fixed) alternative models are:

\nin 1. $y_t = \phi_1 y_{t-1} + \gamma y_{t-2} + u_t \sqrt{w + \alpha y_{t-1}^2}$, or $u_t = (y_t- \phi_1 y_{t-1} - \gamma y_{t-2})/\sqrt{w + \alpha y_{t-1}^2} = g^*(\tilde{Y}_t; \theta, \gamma) $

\nin 2. $y_t = \phi_1 y_{t-1} + u_t \sqrt{w + \alpha y_{t-1}^2 + \gamma y_{t-2}^2  }$, or $u_t = (y_t- \phi_1 y_{t-1})/\sqrt{w + \alpha y_{t-1}^2 + \gamma y_{t-2}^2} = g^*(\tilde{Y}_t; \theta, \gamma) $

%\nin $y_t = \phi_1 y_{t-1} + \gamma y_{t-2} + u_t \sqrt{w + \alpha y_{t-1}^2  + \gamma y_{t-2}^2}$, or 
%$u_t = (y_t- \phi_1 y_{t-1} - \gamma y_{t-2})/\sqrt{w + \alpha y_{t-1}^2 + \gamma y_{t-2}^2} = g^*(\tilde{Y}_t; \theta, \gamma) $

%\nin In contrast, we will not be able to test against this alternative model:
%\nin 4. $y_t = \phi_1 y_{t-1}  + (u_t + \gamma u_{t-1}) \sqrt{w + \alpha y_{t-1}^2}$, or $u_t = (\frac{1}{1-\gamma L}) \left[(y_t- \phi_1 y_{t-1})/\sqrt{w + \alpha y_{t-1}^2 }\right]$
%for $\gamma \neq 1$. We see that $u_t \neq g^*(\tilde{Y}_t; \theta, \gamma) $
%with $\tilde{Y}_t = Y_{t-1},...,Y_{t-p}$ where $p$ is a fixed lag.
\textcolor{black}{The above alternative models 1-2 depend on an additional parameter $\gamma$. They are equivalent when $\gamma=0$, or extend the model in various directions when $\gamma \neq 0$. Thus, we have a set of directional alternatives.}

\medskip
The local alternatives are defined in a neighborhood of the true value $\theta_0, f_0$ satisfying the null hypothesis. We consider parametric directional alternatives where:
$$\theta_T \approx \theta_0 + \mu/\sqrt{T}, \;\; \gamma_T \approx \nu/\sqrt{T}, \;\; f_T \approx f_0. $$

\nin\textcolor{black}{ {\bf Example 5: DAR(1) model cont.}}  The local alternatives of the DAR model are obtained by the first-order Taylor expansion about $\gamma=0$ \textcolor{black}{of alternative models 1 to 2 given above.} We get:

\nin 1. $y_t = \phi_1 y_{t-1} + \gamma y_{t-2} + u_t \sqrt{w + \alpha y_{t-1}^2}$, or $u_t = (y_t- \phi_1 y_{t-1} - \gamma y_{t-2})/\sqrt{w + \alpha y_{t-1}^2} = g^*(\tilde{Y}_t; \theta, \gamma) $

\nin 2. $y_t \approx \phi_1 y_{t-1} + u_t \left[ \sqrt{w + \alpha y_{t-1}^2 } + \frac{\gamma}{2} \frac{y_{t-2}^2}{\sqrt{w+\alpha y_{t-1}^2}} \right] $, or $u_t \approx (y_t- \phi_1 y_{t-1})/\left[ \sqrt{w + \alpha y_{t-1}^2 } + \frac{\gamma}{2} \frac{y_{t-2}^2}{\sqrt{w+\alpha y_{t-1}^2}} \right] \approx g^*(\tilde{Y}_t; \theta, \gamma) $

%\nin 3. $y_t \approx \phi_1 y_{t-1} +  \gamma y_{t-2} + u_t \left[ \sqrt{w + \alpha y_{t-1}^2 } + \frac{\gamma}{2} \frac{y_{t-2}^2}{\sqrt{w+\alpha y_{t-1}^2}} \right] $, 
%or $u_t \approx (y_t- \phi_1 y_{t-1} - \gamma y_{t-2})/\left[ \sqrt{w + \alpha y_{t-1}^2 } \right.$ 
%$ \left. + \frac{\gamma}{2} \frac{y_{t-2}^2}{\sqrt{w+\alpha y_{t-1}^2}} \right]  $
%$\approx \frac{y_t - \phi_1 y_{t-1}}{\sqrt{w+\alpha y_{t-1}^2}} - \gamma \left[ \frac{y_{t-2}}{\sqrt{w+\alpha y_{t-1}^2}} + 0.5 
%\frac{(y_t - \phi_1 y_{t-1})y_{t-2}^2}{(w+\alpha y_{t-1}^2)^{3/2}}\right]
%\approx g^*(\tilde{Y}_t; \theta, \gamma) $

Under the sequence of local alternatives, we consider doubly indexed sequences $(y_{T,t})$, i.e., a sequence of processes indexed by $T$ (triangular array).
In this framework, what matters is the local impact on autocovariances, i.e.,:
$$ \Gamma(h; \theta_T, \gamma_T, f_0) \approx \Gamma(h; \theta_0, 0, f_0) + \frac{\partial \Gamma(h; \theta_0, 0,f_0)}{\partial \theta'} ( \theta_T - \theta_0) + \frac{\partial \Gamma(h; \theta_0, 0, f_0)}{\partial \gamma'} ( \gamma_T - \gamma_0),$$

\nin with $\Gamma(h; \theta_0, 0, f_0)=0$. This leads to local alternatives written on the autocovariances:
\begin{equation}
\Gamma(h; \theta_T, \gamma_T, f_0) = \Delta(h; \theta_0, f_0, \mu, \nu)/\sqrt{T},    
\end{equation}
\nin with: 
\begin{equation}
\Delta(h; \theta_0, f_0, \mu, \nu) = \frac{\partial \Gamma(h; \theta_0, 0, f_0)}{\partial \theta'} \mu +
\frac{\partial \Gamma(h; \theta_0, 0, f_0)}{\partial \gamma'} \nu .
\end{equation}
\nin whose vec representation is denoted by $\delta(h; \theta_0, f_0, \mu, \nu) = vec \Delta(h; \theta_0, f_0, \mu, \nu)$. 
Then, the asymptotic local power of the test, given $f_0$ fixed, is
$$ \lim_{T \rightarrow \infty }P_{\theta_T, \gamma_T, f_0}[ \hat{\xi}_T(H) > \chi^2 _{1-\alpha}(K^2H-dim(\theta))] = \beta (\theta_0, f_0, \mu, \nu; \alpha), $$
\nin for any $\mu, \nu, \alpha$ and $(\theta_0, f_0) \in \Theta_0$.

%When $T \rightarrow \infty$, one can consider smaller values of $\alpha$, and one can distinguish the elements of $\Theta_1$ closer to $\Theta_0$.
%Hence we can study $\beta(\theta_T, \alpha_T)$ either a) when $\theta_T \rightarrow \Theta_1$, or b) when $\alpha_T \rightarrow 0$. 

%a) This can be done by examining the sequence of local alternatives, as follows. We consider a fixed value $\alpha_T = \alpha$, $ \theta_T \rightarrow \theta_0 \in \Theta_0$ at the speed $\theta_T = \theta_0 + \frac{1}{\sqrt{T}} \delta$ and study the power.

%b)  This can be done by considering a fixed  $\theta_T = \theta_1 \in \Theta_1 $ and $\alpha_T \rightarrow 0$, according to the Bahadur approach.
%The approximate slope of this test is $ L_T( \hat{\theta}_T)$ [Geweke( 1981)].

\medskip

\textbf{Proposition 1: }
Under the sequence of local alternatives, and the regularity conditions given in Appendices B.1-B.2:

\nin i) The autocovariance estimator:
$$\hat{\Gamma}_T(h; \theta) = \frac{1}{T} \sum_{t=1}^T g(y_{T,t}; \theta) g'(y_{T,t-h}; \theta) - \frac{1}{T} \sum_{t=1}^T g(y_{T,t}; \theta)' \frac{1}{T} \sum_{t=1}^T g(y_{T,t}; \theta)$$
\nin converges in probability: 
$$ \hat{\Gamma}_T (h; \theta) \textcolor{black}{\stackrel{p}{\rightarrow}} \Gamma(h; \theta_0, 0, f_0),$$
\nin for all $\theta, h$ where the limit is computed under the null hypothesis. This convergence is uniform in $\theta$.

\nin ii) The GCov estimator converges in probability to $\theta_0$:
$$\hat{\theta}_T \textcolor{black}{\stackrel{p}{\rightarrow}} \theta_0.$$

\nin {\it Proof: }
i) The proof is based on the Law of Large Numbers (LLN) for doubly indexed sequences [see e.g.,  Andrews (1988), Newey (1991) and Appendices B.1-B.2 for regularity conditions]. The LLN implies the convergence of the estimated autocovariances.

\nin ii) \textcolor{black}{Under the sequence of local alternatives}, the objective function:
 \begin{equation}
L_T(\theta) = \sum\limits_{h=1}^{H} Tr[ \hat{\Gamma}_T(h; \theta) \hat{\Gamma}_T(0;\theta)^{-1}
        \hat{\Gamma}_T(h;\theta)'
        \hat{\Gamma}_T(0;\theta)^{-1}],
    \end{equation}
\nin tends to
\begin{equation}
L_{\infty} (\theta)= \sum\limits_{h=1}^{H} Tr[ \Gamma(h; \theta_0,0, f_0, \theta) \Gamma(0;\theta_0, 0 , f_0, \theta)^{-1}
        \Gamma(h;\theta_0, 0, f_0, \theta)'
        \Gamma(0;\theta_0, 0, f_0, \theta)^{-1}]. 
    \end{equation}
\nin This limit is the same as the limit of the objective function under the null hypothesis. Then, the consistency is proven as in Gourieroux and Jasiak (2023).
\hfill QED

\medskip
\textcolor{black}{Let us now show that the GCov specification test statistic converges in distribution under local alternatives to a non-central chi-square distributed variable with a non-centrality parameter involving the \textcolor{black}{direction of the alternative}.}
The distribution of the test statistic computed from the residuals of the model $g(y_t, \theta)=u_t$ estimated by the GCov estimator under the sequence of local alternatives is given in Proposition 2 below. 

\medskip

\textbf{Proposition 2: } Let us consider the specification test of the null hypothesis:
$$H_0 = \Theta_0  =   \{\theta, f: \Gamma(h; \theta,f) =0  \; \; \forall h=1,...,H\},$$
against the sequence of local alternatives:
$$H_{1,T} = \Theta_{1,T}  =  \{ \theta = \theta_0 + \mu/\sqrt{T}, \gamma = \nu/\sqrt{T}, f=f_0, \; \mbox{with} \; (\theta_0, f_0) \in \Theta_0\}.$$
The expansion of the test statistic under the sequence of local alternatives is:
\begin{equation}
\hat{\xi}_T(H) = T \sum_{h=1}^H \{vec [\sqrt{T} \hat{\Gamma}_T (h; \theta_T, \gamma_T, f_0)] \Pi(h; \theta_0, f_0)
vec [\sqrt{T} \hat{\Gamma}_T (h; \theta_T, \gamma_T, f_0)] \} + o_p(1),
\end{equation}
\nin where
\begin{eqnarray*}
   \Pi (h; \theta_0, f_0) & = & [\Gamma_0 (0,\theta_0, f_0)^{-1}\otimes \Gamma_0 (0,\theta_0, f_0)^{-1}] - [\Gamma_0 (0,\theta_0, f_0)^{-1}\otimes \Gamma_0 (0,\theta_0, f_0)^{-1}] \frac{\partial vec \Gamma(h,\theta_0, f_0)}{\partial \theta'} \\
 & &   \left\{\frac{\partial vec \Gamma(h,\theta_0, f_0)'}{\partial \theta} [\Gamma_0 (0,\theta_0, f_0)^{-1}\otimes \Gamma_0 (0,\theta_0, f_0)^{-1}]\frac{\partial vec \Gamma(h,\theta_0, f_0)}{\partial \theta} \right\}^{-1}\\
& & \times   \frac{\partial vec \Gamma(h,\theta_0, f_0)'}{\partial \theta'} [\Gamma_0 (0,\theta_0, f_0)^{-1}\otimes \Gamma_0 (0,\theta_0, f_0)^{-1}].
\end{eqnarray*}
\nin Then, under the sequence of local alternatives,  $ \hat{\xi}_T(H)  \overset d \sim \chi^2 (K^2 H-dim(\theta), \lambda(\theta_0, f_0, \mu, \nu)),$ where the non-centrality parameter is:
$$\lambda(\theta_0, f_0, \mu, \nu)= \Sum_{h=1}^H \delta(h; \theta_0, f_0, \mu, \nu)' \Pi(h; \theta_0, f_0) \delta(h; \theta_0, f_0, \mu, \nu)),$$

\nin with $\delta(h; \theta_0, f_0, \mu, \nu) = \frac{\partial \Gamma(h; \theta_0, 0, f_0)}{\partial \theta'} \mu +
\frac{\partial \Gamma(h,\theta_0, 0, f_0)'}{\partial \gamma'} \nu$.

\nin {\it Proof: }The proof of Proposition 2 is based on the Central Limit Theorem (CLT) for doubly indexed sequences [see, e.g., Wooldridge and White (1988)] and given in Appendices B.5-B.6.

\medskip

Let the cumulative distribution function (c.d.f.) of the non-central chi-square distribution be denoted by $F(x; \kappa, \lambda)$, where $\kappa$ denotes the degrees of freedom and $\lambda$ is the non-centrality parameter.
Then, $F(x; \kappa,\lambda) = 1 - Q_{\kappa/2} (\sqrt{\lambda}, \sqrt{x})$, where $Q_{\delta}(a,b)$ is a Marcum Q-function. For positive integer values of $\delta$, the Marcum Q-function is defined by:
$$Q_{\delta}(a,b) = \left\{ \begin{array}{ll} H_{\delta} (a,b) & a < b, \\
                                        0.5+ H_{\delta}(a,a), & a=b, \\
                                        1 + H_{\delta}(a,b), & a>b, \end{array} \right.$$

\nin where
$H_{\delta}(a,b) = \frac{\zeta^{1-\delta}}{2 \pi} exp( - \frac{a^2 + b^2}{2}) \int_0^{2\pi} \frac{cos (\delta-1)w - \zeta cos\, \delta\, w}{1-2 \zeta cos w + \zeta^2} exp(ab\, cos\, w) dw$
and $\zeta = a/b$. Then, we get:

\medskip

\textcolor{black}{
{\bf Corollary 1:} } The local asymptotic power of the GCov specification test is defined as:
$$\beta(\theta_0, f_0, \mu, \nu; \alpha) = Q_{(K^2H - dim(\theta))/2} [\sqrt{\lambda(\theta_0, f_0, \mu, \nu)}, \sqrt{\chi^2_{1-\alpha} (K^2H - dim(\theta))}],$$
\nin where $\lambda(\theta_0, f_0, \mu, \nu)$ is given in Proposition 2.

\medskip
From the monotonicity property of the Q-function, it follows that the local asymptotic power function is strictly decreasing in the non-centrality parameter $\lambda(\theta_0, f_0, \mu, \nu)$.

\setcounter{equation}{0}\def\theequation{4.\arabic{equation}}

\color{black}

\section{Bootstrap-Adjusted GCov Test}

\textcolor{black}{Even though the asymptotic distribution of the GCov test statistic is known to be a $\chi^2$ distribution \textcolor{black}{[resp. non-central $\chi^2$ distribution] under the null hypothesis [resp. local alternatives]}, and asymptotically valid critical values are available, one may be interested in finding critical values for hypothesis testing in finite samples.}
This section describes the bootstrap for GCov test statistic \textcolor{black}{under the regularity conditions given in Appendix B.7}. 
%The application of the bootstrap GCov test to noncausal processes and the bootstrap GCov test based on the AML estimator are illustrated through simulations in Section 6.3. 

\subsection{Bootstrap Under the Null Hypothesis}

We approximate the distribution of $\tilde{\xi}_T$ by computing the test statistic $\xi(H, \hat{\theta}_T^s, \hat{f}_T^s)$ with $\hat{\theta}^s_T$ and $\hat{f}^s_T$ obtained from the bootstrapped residuals, and the bootstrapped values $y_1^s,...,y_T^s$. We assume $\tilde{Y}_t = (Y_t, Y_{t-1},..., Y_{t-p})$ and denote by $c$ the inverse of function $g$ with respect to $y_t$. 
Then, the critical value of the test statistic $\tilde{\xi}_T(H)$ can be found by parametric bootstrap along the following steps:
\medskip

\nin 1. Draw randomly with replacements $T$ residuals $\hat{u}_{T,t}^s, t=1,...,T$ from residuals $\hat{u}_{T,t} = g(\tilde{Y}_t, \hat{\theta}_T), t=1,...,T$. 

\medskip
\nin 2. Build the bootstrapped time series of length $T$: $y_{T,t}^s = c(y^s_{T,t-1},...,y^s_{T,t-p},
\hat{u}_{T,t}^s, \hat{\theta}_T), t=1,...,T, \; s=1,...,S$ \footnote{with the starting values $y_{T,-1}^s = y_{-1},...,y_{T,-p}=y_{-p}$, if the stationary process is causal and such that $u_t$ is independent of $y_{t-1},...,y_{t-p} \forall t$. If the process is a mixed VAR, the bootstrapped values $y_1^s,...,y_T^s$ can be computed from $\hat{u}_{T,t}^s$ using the formulas in Gourieroux and Jasiak (2017). For a VAR model in a multiplicative representation, see Lanne and Saikkonen (2013). For univariate MAR(r,s) processes, see Gourieroux and Jasiak (2016) and Section 6.1.}.

\medskip
\nin 3.  Re-estimate the parameter vector $\theta$ by GCov from $y_{T,t}^s, \; t=1,...,T$, providing $\hat{\theta}_T^s, \; s=1,...,S$. Under the regularity conditions discussed in \textcolor{black}{Appendix B.7}, the asymptotic distribution of $\sqrt{T} (\hat{\theta}_T^s - \hat{\theta}_T)$ conditional on the sample $y_t, t=1,...,T$ is the same as the asymptotic distribution
of $\sqrt{T} (\hat{\theta}_T - \theta_0) $. 

\medskip
\nin 4. Compute the test statistic $\hat{\xi}_T^s(H)$ from $\hat{u}_{T,t}^s, t=1,...,T$, their nonlinear transforms and $\hat{\theta}_T^s$.

\medskip
\nin 5. Rank the test statistics $\hat{\xi}^s_T(H), \; s=1,...,S$, and use the 95th percentile $\hat{q}_{T,95\%}$, say, as the critical value for testing 
the null hypothesis of absence of nonlinear or linear serial dependence.
Then, the null hypothesis is rejected if:

$$ \hat{\xi}_T(H) > \hat{q}_{T,95\%},$$

\nin and it is not rejected, otherwise. Henceforth, the bootstrap size-adjusted test is called the GCov bootstrap test.

\subsection{Bootstrap Analysis of the Local Power of Test}

The bootstrap can also be used to obtain an approximation of the local power of the size-adjusted GCov bootstrap test (see Escanciano (2007), Sections 2.3 and 3). \textcolor{black}{The approach involves two types of bootstrap performed under the null and under the alternative, respectively.} The local power approximation is obtained as follows:
  
 \medskip
\nin Step 1: Estimate the model $g(\tilde{y}_t, \theta, \gamma)=u_t, \; t=1,..,T$ under the alternative to get $\hat{\theta}_{1,T}, \hat{\gamma}_{1,T}$.

\medskip

\nin Step 2: Compute the residuals under the alternative:
$\hat{u}_{1,T,t} = g(y_t, \hat{\theta}_{1,T}, \hat{\gamma}_{1,T})$.

\medskip
\nin Step 3: Get the bootstrapped residuals $\hat{u}_{1,b,T,t}^s$ by drawing in the sample distribution of $\hat{u}_{1,T,t}, t=1,...,T$ for $s=1,...,S$.

\medskip
\nin Step 4: Calculate the bootstrapped pseudo-observations $y_{1,b,T,t}^s$ from 
$$g(\tilde{y}_{1,b,T,t}^s; \hat{\theta}_{1,T}, \hat{\gamma}_{1,T}) = \hat{u}_{1,b,T,t}^s, \; s=1,...,S.$$ 
$$\iff y_{1,b,T,t}^s = c(y_{1,b,T,t-1}^s,...,y_{1,b,T,t-p}^s, \hat{u}_{1,b,T,t}^s, \hat{\theta}_{1,T}, \hat{\gamma}_{1,T}), \; s=1,...,S.$$

% Let $\theta_{1,0}, \gamma_{1,0}$ denote the true of the parameters under the alternatives.

\medskip
\nin Step 5: From $y_{1,b,T,t}^s, t=1,...,T$ bootstrapped under the alternative, compute the values of the estimates $\hat{\theta}_{0,b,T}^s$  of $\theta$ under the null and the associated test statistic $\hat{\xi}_{0,b,T}^s, s=1,...,S$. Let the test statistic computed under the null be denoted by $\hat{\xi}_{0,T}$.
Then, the empirical distribution of $\hat{\xi}_{0,b,T}^s- \hat{\xi}_{0,T}, s=1,...,S$ approximates the distribution of $\hat{\xi}_{0,T}$ under the local alternative.

\medskip
\nin \textcolor{black}{ Step 6: Evaluate the bootstrapped local power of the (size-adjusted) GCov bootstrap test as:
$1- \hat{F}_{b,T}^s (\hat{q}_{T, 95\%} )$, where $\hat{F}_{b,T}^s(\alpha)$ denotes
the bootstrap c.d.f. under the alternative and $\hat{q}_{T, 95\%}$ is defined in Section 4.1}.

%The regularity conditions for the validity of the bootstrap under the null and local alternatives are discussed in Appendix B.7. 

\subsection{An Extended Continuous Updating GMM}

To discuss the relationship with the current literature on bootstrap applied to overidentification tests based on moment conditions, we \textcolor{black}{consider a comparable extension of the Continuous Updating GMM (CUGMM) estimator. This extended CUGMM estimates an additional vector of parameters $\beta$ that is needed to center each moment condition based on nonlinear transformations in order to obtain moment conditions equivalent to nonlinear autocovariance conditions of the GCov.}

For ease of exposition, we assume that $H=1$ and enlarge the parameter set to $\theta, \beta$, where $dim(\beta) = K$. Then the set of centered moment conditions \textcolor{black}{equivalent to zero nonlinear autocovariance conditions} is:
\begin{eqnarray}
E[a_k[g(\tilde{Y}_t, \theta)] - \beta_k] & = & 0, \; k=1,...,K, \\
E\{ [a_j [g(\tilde{Y}_t, \theta)] - \beta_j] [a_k [g(\tilde{Y}_{t-1}, \theta)] - \beta_k]\} & = & 0, \; j,k =1,...,K. \nonumber
\end{eqnarray}
\nin This set of moment conditions jointly identifies the parameter $\theta$ and the additional parameter $\beta$.

\medskip
{\bf Proposition 3:}

i) Under the condition of just-identification, $dim(\theta)=K^2$ and $H=1$, the extended CUGMM concentrated in $\beta$ and applied to the set of moment conditions (4.1) yields the estimators $\hat{\theta}_T,
\hat{\beta}_T$ and the objective function $\hat{\xi}^*_T(\hat{\theta}_T,
\hat{\beta}_T)$ such that $\hat{\theta}_T$ is equal to the GCov estimator of $\theta$, and the minimum of $\hat{\xi}^*_T(\hat{\theta}_T,
\hat{\beta}_T)$ is equal to the minimum value of zero of the GCov objective function.

ii) Under overidentification, $dim(\theta)<K^2$, the GCov and the concentrated extended CUGMM estimators of $\theta$ are equivalent at order $1/\sqrt{T}$ and differ at order $1/T$. The associated test statistics have the same asymptotic chi-square distributions at the first order, but differ at higher orders.

{\it Proof:} i) If $dim(\theta)=K^2$, due to the just identification of $\beta$ for a given $\theta$, we can concentrate $\beta$ out of the CUGMM objective function by using:
$$\hat{\beta}_T(\theta) = \frac{1}{T} \sum_{t=1}^T a[g(\tilde{Y}_t, \theta)].$$
\nin Next, after substituting $\hat{\beta}_T(\theta)$ into the CUGMM objective function, we find that at the minimum, the concentrated objective function of the extended CUGMM  is equal to the minimum of the GCov objective function. The result follows.

ii) Under the overidentification, the asymptotic distribution of $\sqrt{T}(\hat{\theta}_T - \theta_0)$ is normal, and both estimators are semi-parametrically efficient for the given set of autocovariance conditions and have the same asymptotic variance. This implies a common chi-square distribution of test statistics at the first order.  The asymptotic Taylor expansions of The GCov and extended concentrated CUGMM estimators show a difference in asymptotic biases at order $1/T$ due to different weighting, which implies that these estimators differ at order $1/T$.

\hfill QED.

When the lag $H$ is increased to $H>1$ and higher, the dimensions of the set of moment conditions and of the weighting matrix in the objective function of the extended CUGMM increase by a factor $KH$. In practice, this may cause an invertibility issue, which will not arise in the GCov where the dimensions of the objective function remain unchanged when $H$ increases.  

Since the GCov and extended concentrated CUGMM estimators differ asymptotically by the bias at order $1/T$, and the bootstrap methods applied to the above estimators are expected to adjust for these biases, sufficient regularity conditions for the validity of the bootstrap for the GCov and extended concentrated CUGMM, respectively, will be the same [see Dovonon and Goncalves (2017)]. \textcolor{black}{ An important difference arises when the set of moment conditions increases to infinity [see Section 5]. Then, the number of parameters $\beta$ in the extended CUGMM
tends to infinity too, which is at odds with the results established in the literature on CUGMM with an infinite set of moment conditions and a parameter of fixed dimension.}

\setcounter{equation}{0}\def\theequation{5.\arabic{equation}}

\section{Increasing the Set of Autocovariance Conditions}
The GCov estimator has been applied so far using a finite set of zero nonlinear autocovariance conditions, that is, a given order $H$ and a given set of $K$ nonlinear error transformations.
In this section, 
we discuss how the asymptotic performance of the GCov specification test can be improved by increasing the set of nonlinear transformations denoted by $\mathcal{A}$ and possibly by taking into account an infinite set of autocovariance conditions.

\subsection{Implicit Null and Alternative Hypotheses}

Let us compare the implicit null and alternative hypotheses considered with those of potential interest. We distinguish the following hypotheses concerning the parameter $\theta$ and the distribution $f$ of the process $(u_t)$\footnote{\textcolor{black}{As in previous sections, we assume a well-specified model (3.1): $u_t=g(\tilde{Y}_t, \theta)$ and the true value $\theta_0$ of $\theta$. To simplify the notation, the same symbol  $u_t$ is used for errors that depend on either $\theta$ or $\theta_0$.}}:

\nin i) $H_{0,\mathcal{A}} = \{ \theta, f: Cov[a(u_t), \textcolor{black}{\tilde{a}}(u_{t-h})]=0, \; \forall h, \; \forall a, \textcolor{black}{\tilde{a}} \in \mathcal{A}\}$ and the associated alternative $H_{1,\mathcal{A}}$. These hypotheses depend on the selected set of scalar transformations $\mathcal{A}$.

\nin ii) $H_{0,pair} = \{ \theta, f: u_t,\;\mbox{and}\; u_{t-h} \; \mbox{are independent}, \; \forall h\}$ and its alternative $H_{1,pair}$.

\nin iii) $H_{0,ind} = \{ \theta, f: u_t, u_{t-1},...,u_{t-h} \; \mbox{are independent} \;  \forall t,h \}$ and  its alternative $H_{1,ind}$.

\nin iv) $H_{0,iid} = \{ \theta, f: u_t's \; \mbox{are iid}\}$ and its alternative $H_{1,iid}$.

\nin  We have $H_{0,\mathcal{A}} \supset H_{0,pair} \supset H_{0,ind} \supset H_{0,iid}$ and
$H_{1,\mathcal{A}} \subset H_{1,pair} \subset H_{1,ind} \subset H_{1,iid}$.

\nin The test of $H_{0,\mathcal{A}}$ is consistent and of asymptotic power of 1 against $H_{1,\mathcal{A}}$. However, this test is not of asymptotic power 1 against $H_{1,pair}, H_{1,ind}$ and $H_{1,iid}$.
By increasing the set of nonlinear transformations $a$ in $\mathcal{A}$ appropriately, we expect to improve the asymptotic efficiency of the GCov estimator and to increase the set of alternatives against which the GCov specification test is asymptotically consistent.

\subsection{The Semi-Parametric Efficiency Bound}

The asymptotic performance of the GCov specification test depends on the asymptotic properties of the GCov estimator of parameter $\theta$. 
We can distinguish  the semi-parametric models and the associated semi-parametric efficiency bounds, which are:

i) The semi-parametric efficiency bound that accounts for the pairwise independence between $u_t$ and $u_{t-h}$ for any $h$.

ii) The semi-parametric efficiency bound that takes into account the joint independence of $u_t, u_{t-1},...,u_{t-h}$, since the pairwise independence does not imply the joint independence.

iii) The semi-parametric efficiency bound that takes into account the fact that the $u_t$'s are identically distributed. It coincides with the parametric efficiency bound, as shown
in Example 6.

%In the framework of nonlinear autocovariance-based test statistics with a one-step
The GCov estimator attains the first semi-parametric efficiency bound i), but not the parametric efficiency bound under $H_{0,iid}$, which requires a two-step estimator illustrated below.

\medskip

\nin \textbf{Example 6: Adaptive Estimator}

Let $\hat{\theta}_T$ denote a GCov estimator based on a finite set $\mathcal{A}$ of nonlinear transformations. Given this consistent estimator, we can compute the residuals $\hat{u}_{T,t}, t=1,...,T$ of the model and then approximate nonparametrically the true density $f_0$ of $u_t$ by the kernel-based density $\hat{f}_T(u)$ based on the residuals $\hat{u}_{T,t}, t=1,...,T$.
Next, in the second step, parameter $\theta$ can be estimated by a pseudo-maximum likelihood method, where the true density $f_0(u)$ is replaced by $\hat{f}_T(u)$
[see e.g., Bickel (1982) and Newey (1988)]. Under standard regularity conditions, this leads to a two-step estimator of $\theta$ that reaches the parametric efficiency bound. Due to this improvement of the asymptotic properties of the GCov estimator, the pseudo-likelihood ratio test based on this two-step estimator has better asymptotic power properties than the GCov test based on a finite set $\mathcal{A}$. 

\subsection{How to Choose an Infinite Set of Transformations}

The asymptotic properties of the GCov estimator and of the associated test statistics can be improved by increasing the finite set of nonlinear autocovariance conditions to a larger finite or infinite set.
The extension to an infinite set is easy when these conditions correspond to an orthonormal basis of the Hilbert space $L^2 (u_t, u_{t-1},...)$. In our framework, we can increase the set of nonlinear autocovariance conditions by increasing the maximum lag $H$, or the set of transformations $\mathcal{A}$. We saw that, under the null hypothesis, the orthogonality of nonlinear autocovariance conditions with respect to the lag is satisfied. This explains the simplified form of the test statistic written as a sum of terms associated with lags $h=1,...,H$. 
In contrast, the set of nonlinear transformations does not necessarily correspond to an orthonormal basis, and inverting the variance matrix of a large dimension can become a problem from both the theoretical and computational perspectives.

We follow the approach of Bierens (1990) and build a sequence of orthonormal bases in $L^2(u_t, u_{t-1},...,u_{t-H})$ of square-integrable transformations of $(u_t, u_{t-1},...,u_{t-H})$ (or, equivalently in $L^2(u_t)$). We proceed as follows:

1. Consider a countable subset of transformations $\mathcal{A}$, called a system of generators, that allows identifying the unknown distribution $f_0$ of $u_t$;

2. Select from that set $\mathcal{A}$ an increasing sequence of finite subsets of transformations $\mathcal{A}_n$, such that $\bigcup_n \mathcal{A}_n$ is dense;

3. Orthonormalize within each subset;

\nin (See Bierens (1990), Corollary 1 for the system of generators based on exponential transforms). In practice, it is difficult to find a system of generators that is informative about $\theta$.
An important feature of the causal-noncausal models considered in our paper is the presence of extreme risks and persistence, creating locally explosive patterns, including spikes and bubbles. This implies that the model errors do not necessarily have higher power moments. Moreover, some parameters driving the tails of their distributions, including those distinguishing between the negative and positive tails, may be non-identifiable
from the transformations $u_t u_t'$ only.

In this respect, some standard linear systems of generators are not convenient. For example, the polynomial transformations of $u_t$ cannot be used if $u_t$ has no moments of order greater than 3. Moreover, the sine-cosine transformations used in the ADCV literature may not be informative of the tail driving parameters [see, e.g., Wan and Davis (2022), Fokianos and Pitsillou (2018) for the ADCV test of the independence hypothesis based on a joint characteristic function]. The same remark applies to the standard spline bases.

A preferred system of generators would assign weights to the power transformations, ensuring their square integrability. For ease of exposition, consider a univariate process with positive errors $u_t$ and with trajectories admitting only positive and increasing bubbles caused by a large positive shock to $u_t$\footnote{The extension to a multivariate $u_t$ is done by finding the products of generators for univariate $u_t$'s.}. Then, a countable linear system of generators is:

$\mathcal{A} = \{ a_{t,p}(u) = u^p \exp(-tu), \; p \in  \textcolor{black}{\mathbb{N}}, \; t \in \mathcal{Q} \subset [0,1]\},$

\nin where $\mathcal{Q}$ is the set of rational numbers, with an increasing sequence of finite subsets given by: $\mathcal{A}_n = \{ a_{t_j,n,\textcolor{black}{p}}(u) = u^p \exp(-t_{j,n} u), \; p \in \textcolor{black}{ \mathbb{N}}, \; t_{j,n} \in \mathcal{Q}, \; j=1,...n\},$

\nin where $(t_{1,n},...,t_{j,n})$ are the elements of a grid that is getting finer as $n$ increases.

We observe that the sequence $\{\mathcal{A}_n\}$ is such that $\bigcup_n \mathcal{A}_n$ is dense in $L^2$ and allows identifying the distribution of $u$, based on the inversion formula of the real Laplace transform called Post's inversion formula [Post (1930), and Feller (1971), Chapter 13, for the modern proof]. We make the following assumption:

\medskip

\nin {\bf Assumption 1:}
\nin $U$ is positive and has a distribution with continuous density $f_0(u)$ such that: 
$\sup_{u>0} f_0(u)/\exp(bu) < \infty$, for some $b>0$.

\medskip
\nin This condition on the density function implies that the distribution cannot have a large probability mass at zero, and its right tail can be of any size. 

Under Assumption 1, the Laplace transform: $\Psi(t) = E[\exp(-tU)]$, $t \in [0,1]$ characterizes the distribution of $U$.
However, as mentioned earlier $\mathcal{A}^*= \{ a_t(u) = \exp(-tu),\; t \in [0,1]\}$
is not a convenient system of generators, because we may not be able to write $f_0$ as the limit of linear combinations of decreasing exponentials. A better system of generators follows from the Post's inversion formula.

\bigskip
 {\bf Proposition 4: Post's inversion formula [Post (1930)] }

\nin  Under Assumption 1, we have:
$$f_0(v) = \lim_{n \rightarrow \infty} \frac{1}{n!} \left(\frac{n}{v} \right)^{n+1} E [ U^n \exp \left(- \frac{n}{v} U \right)], \; \forall v, \; a.e.$$
\nin This suggests the linear system of generators:
$$\mathcal{A} = \{ a_{t,p}(u) = u^p \exp(-tu), \; p \in \mathbb{N}, t \in [0,1] \},$$
\nin where the power transforms are weighted by decreasing exponentials to ensure their square integrability, and the associated sequence given by:
$$\mathcal{A}_n = \{ a_{t_{j,n},p}(u) = u^p \exp(-t_{j,n}u), \; p \in \{ 1,...,n \}, t_{j,n} \in \mathcal{Q} \subset [0,1], \; j=1,...,n \}.$$

\medskip

In practice, only weights with $t$ close to 0 are useful. To see that, let us recall that the real Laplace transform $\Psi$ of a positive variable is characterized by its Taylor series expansion:
$$E[\exp (-tU)] = \sum_{j=0}^{\infty} \frac{t^j}{j!} \mu_j.$$
\nin If $U$ admits power moments of any order, we have $\mu_j = E(U^j), \; \forall j$. In our framework of heavy-tailed $U$, such moments may not exist, but regardless, this series expansion exists and Taylor's coefficients $\mu_3, \mu_4$ define new notions of skewness and kurtosis measures. 

\medskip
\nin \textbf{Remark 5:} In practice, the Post's inversion formula does not provide a tractable means of inverting the Laplace transform, and it is an ill-posed problem [see Bryan (2006), Conclusion]. In this respect, the regularized GCov estimator introduced later in this section can be seen as a solution to this ill-posed problem when the goal is to estimate $\theta$. 
We use Post's inversion formula only to show that $\mathcal{A}$ is an adequate linear system of generators.  

\medskip
\nin \textbf{Remark 6:} Similar results can be obtained if additional information is available on the size of the tail of the distribution of $U$, like, for example, the tails are Pareto. Then, we can apply the Hardy-Littlewood Tauberian theorem with a Pareto weighting function to define an alternative system of generators [Feller (1971)].

\medskip
\nin \textbf{Remark 7:} The causal-noncausal MAR(r,s) processes are often applied to positive time series, such as commodity prices. Then, their solutions have a two-sided moving average representation $y_t = \sum_{j=-\infty}^{\infty} c_j u_{t-j}$ in errors taking values in $(0, \infty)$. The positivity restriction on process $(y_t)$ implies that $c_j \geq 0, \forall j$, and $u_t \geq 0, \forall t$ (up to sign identification), validating the assumption $U>0$. If the error $u_t$ takes both positive and negative values, we can 
observe "positive" and "negative" bubbles. Then, the weights have to be replaced by $\exp(-t|U|)$, and the even and odd powers of $U$ have to take positive or negative signs, respectively. If $u_t$ has a symmetric distribution, the generators become $a_{t,p} = |u|^p \exp(-t|u|)$.

%\medskip

%In practice, it may be beneficial to limit the number of transformations by considering only the informative transformations. Some parameters can be estimated 
%either from well-chosen transformations, or 
%from a few power transformations along with an increased lag $H$ to capture the persistence. 
%It may be useful to consider quadratic transformations along with some decreasing exponential transformations.

\subsection{Regularized Orthonormalization of the System of Generators}

It is impossible to construct an exact orthonormal basis from a system of generators because the true distribution of $U$ under the i.i.d. hypothesis is unknown. Hence, a two-step approach is required. We derive below the orthonormalization in the multidimensional case and discuss the form of the portmanteau statistic and its asymptotic distribution in the following subsection. The empirical orthonormalization in the spirit of the Gram-Schmidt forward regression [Chen et al. (2025)] proceeds in the following steps:

\nin step 1. Consider a finite set $\mathcal{A}_0$ of transformations from which $\theta$ is identifiable. Estimate $\theta$ by 
a consistent estimator $\tilde{\theta}_{n,T}$ and compute $\hat{u}_{T,t}, \; t=1,...,T$.

\medskip

\nin step 2. Consider an infinite set $\mathcal{A} = \{ a(u)= u_1^{p_1},...,u_J^{p_J} \exp(- \tau'u),\; p_1,...,p_J \in \mathcal{N}, \; \tau \in [0,1]^J \}$ and the increasing sequence of finite sets $\mathcal{A}_n$:

$\mathcal{A}_n = \{ a(u)= u_1^{p_1},...,u_J^{p_J} \exp(- \tau_{j,n}'u),\; p_1,...,p_J \in (0,1,...,P_n), \; \tau_{1,n},...,\tau_{n,n} \in [0,1]^J \},$ 

\nin where $\bigcup_n(\tau_{1,n},...,\tau_{n,n})$ is dense in $[0,1]^J$ when $n \rightarrow \infty$.

\nin The number of transformations in $\mathcal{A}_n$ is $P_n^J J^n$, where $K_n \equiv P_n^J J^n$ with the elements $a_{k,n},k=1,...,K_n$ arranged in an increasing order:
$\mathcal{A}_n = \{ a_{k,n}, \; k=1,...,K_n=P_n^J J^n\}$.

\medskip

\nin step 3. (Regularized) Orthonormalization 

The mapping of $a_{k,n}, \; k=1,...,P_n^J J^n$ into an orthonormal basis $a_{k,n}^*, \; k=1,...,K_n$ is obtained as follows. We run regressions to get at step $n$ the orthonormal functions $a_{k,n}^*, \; k=1,...,K_n$, with zero mean.

\nin a) We start from regressing $a_{1,n} (\hat{u}_{T,t})$ on the constant:
$$a_{1,n}  (\hat{u}_{T,t}) = \alpha_{1,n,T} + \hat{w}_{1,n,T,t},$$
\nin $t=1,...,T$ with the residuals $\hat{w}_{1,n,T,t}$. Next, compute:
$$a_{1,n,T}^*(u) = \hat{w}_{1,n,T} (u)/||\hat{w}_{1,n,T}||_T,$$
\nin where $\hat{w}_{1,n,T}(u) = a_{1,n}(u) - \alpha_{1,n,T}$. Let $R^2_{1,n,T}$ denote the R-square in the above regression. Then $a_{1,n,T}^*$ is used if $1-R^2_{1,n,T}$ is sufficiently different from 0, i.e., $ 1-R^2_{1,n,T} > \epsilon_{n,T}$, where $\epsilon_{n,T}$ is an appropriately chosen regularization tuning parameter  [The regularization through tuning parameters is an empirical analogue of the theoretical condition in the proof of Corollary 1 in Bierens (1990)]. Otherwise, disregard $a_{1,n}$ and start from  $a_{2,n}$.

\nin b) In the next step, $a_{2,n}$ is projected on $a_{1,n,T}^*$ as follows. We run the regression:
$$a_{2,n}(\hat{u}_{T,t}) = \alpha_{2,n,T} + \beta_{2,n,T} a_{1,n,T}^*(\hat{u}_{T,t})  + \hat{w}_{2,n,T,t},$$
\nin $t=1,...,T$ with the residuals $\hat{w}_{2,n,T,t}$. Next, compute:
$$a_{2,n,T}^*(u) = \hat{w}_{2,n,T}(u)/||\hat{w}_{2,n,T}||_T,$$
\nin where $\hat{w}_{2,n,T}(u) = a_{2,n}(u) - \alpha_{2,n,T} - \beta_{2,n,T} a^*_{1,n,T}(u)$. This can be done if $||\hat{w}_{2,n,T}||_T$ is sufficiently different from 0, which is the case when $1-R^2_{2,n,T} >\epsilon_{n,T}$. Otherwise, proceed with $a_{3,n}$ instead.

\nin c) The third step is 
$$a_{3,n}(\hat{u}_{T,t}) = \alpha_{3,n,T} + \beta_{3,1,n,T} a_{1,n,T}^*(\hat{u}_{T,t}) + \beta_{3,2,n,T} a_{2,n,T}^*(\hat{u}_{T,t})  + \hat{w}_{3,n,T,t},$$
\nin and so on, up to $k=K_n$. 
%Since the regressors are orthogonal, multivariate regressions can be replaced by simple regressions in practice.

\nin We end up with a selected set of transformations  $\mathcal{A}_n^* = \{a_{k,n}^*, k=1,...,K_n^* \}$, which are zero mean and orthonormal with respect to the sample distribution of $\hat{u}_{T,t}, \; t=1,...,T$, and a random $K_n^*$, such that $K_n^* \leq K_n$.
Since the autocovariance conditions concern pairs $u_t, u_{t-h}$ and  transformations of type $a(u_t), \textcolor{black}{\tilde{a}(u_{t-h})}$, we need to consider transformations in 
$\mathcal{A}_n^2$, where $dim(\mathcal{A}_n^2) = P_n^{2J} J^{2n} =K_n^2$.

The above regularized orthonormalization depends on the estimator $\tilde{\theta}_{n,T} $ of $\theta$ selected in the first step. In practice, we can consider either: a) $\tilde{\theta}_{n,T} =
\hat{\theta}_{n,T}$ equal to the GCov estimator, or b) $\tilde{\theta}_{n,T} $, which is a Diagonal GCov estimator with $\hat{\Gamma}(0)$ replaced by a matrix containing only the diagonal elements
of $\hat{\Gamma}(0)$ in the objective function [Gourieroux and Jasiak (2017)]. For a large number of transformations, the Diagonal GCov estimator can be easily computed, while the GCov estimator becomes computationally challenging.

\subsection{Two-step Portmanteau Statistic and its Asymptotic Distribution}
\color{black}
The orthonormalization simplifies the expression of the objective function and/or facilitates the derivation of the asymptotic distribution of the GCov specification test statistics.  The orthonormalization approach outlined in Section 5.4 provides an identity weighting matrix, so that the resulting objective function has the following expression:
\begin{equation}
L_{n,T}(\theta) = \sum_{h=1}^H \sum_{j=1}^{K_n} \sum_{k=1}^{K_n} \left( \frac{1}{T}
\sum_{t=1+h}^T a_{j,n,T}^* [g(\tilde{y}_t; \theta)] a_{k,n,T}^*[g(\tilde{y}_{t-h}; \theta)] \right)^2,
\end{equation}
\nin and $\hat{\xi}_{n,T} = T L_{n,T}(\theta)$, since the inverse of the variance-covariance matrix is equal to the identity matrix. \textcolor{black}{Note that the transformations $a_{j,n,T}^*$ depend on the starting values $\tilde{\theta}_{n,T}$.}

\textcolor{black}{Below, we examine two types of GCov test statistics, 
depending on the initial value $\hat{\theta}_{n,T}$, or $\tilde{\theta}_{n,T}$
used in the orthonormalization algorithm. Then, the GCov test statistics based on the minimizers of (5.1) are:}
$$\hat{\xi}_{n,T}(Arg min_{\theta} \,  L_{n,T} (\theta| \hat{\theta}_{n,T})) \equiv \xi_{n,T}(\hat{\hat{\theta}}_{n,T}| \hat{\theta}_{n,T}) \; \mbox{and} \;
\tilde{\xi}_{n,T}(Arg min_{\theta} \,  L_{n,T} (\theta| \tilde{\theta}_{n,T})) \equiv \xi_{n,T}(\tilde{\tilde{\theta}}_{n,T}| \tilde{\theta}_{n,T}),$$

\nin where $\hat{\hat{\theta}}_{n,T}$ and $\tilde{\tilde{\theta}}_{n,T}$ denote the minimizers of the objective function (5.1). Note that $\hat{\hat{\theta}}_{n,T}$ can be computationally challenging for large $K$ and different from $\tilde{\tilde{\theta}}_{n,T}$, although they are both asymptotically equivalent.

\bigskip
\textcolor{black}{\bf Proposition 5}
\color{black}

Under $H_{0,iid}$ and the regularity conditions given in Appendix B.8, we have:
$$(\hat{\xi}_{n,T} - H K_n^{*2})/\sqrt{2H K_n^{*2}} \stackrel{d}{\rightarrow} N(0,1)
\; \mbox{and} \; 
(\tilde{\xi}_{n,T} - H K_n^{*2})/\sqrt{2H K_n^{*2}} \stackrel{d}{\rightarrow} N(0,1),$$

\nin when $T$ and $K_n$ tend to infinity, \textcolor{black}{$K_n^6/T$ tends to 0, and the regularization tuning parameter tends to 0 at a rate such that there exist $\delta_1, \delta_2$, $0 < \delta_1 \leq \delta_2 < 1$ with $\delta_1 < E K_n^*/K_n < \delta_2$.}

{\it Proof}: See Koenker and Machado (1999) and Appendix B.8.

\medskip

The conditions on the number of transformations are sufficient conditions for the asymptotic normality [see Koenker and  Machado (1999)]. They imply that $K^{*}_n$ has to increase with $T$, but at a rate slow enough to allow for solving the ill-posed problem due to inversion of the sample autocovariance matrix at lag 0.

The change of asymptotic distribution from $\chi^2(K^2H-dim(\theta))$ for a fixed set of transformations to the normal distribution given above is easy to explain. If the dimension of the parameter is fixed and $K_n^*$ (and/or $H_n$) tend to infinity when $T$ increases to infinity, the $\chi^2$ distribution with an infinite number of degrees of freedom is well approximated by a Gaussian distribution. 
%$N(K^2H, 2K^2H)$. 
Hence, Proposition 5 applies this approximation to a centered and standardized test statistic.

As mentioned earlier, our discussion is focused on the increase of the set $\mathcal{A}_n$ of transformations. It would also be possible to increase the maximum lag $H$ with the number of observations and get a similar result when $T, H, K_n$ tend to infinity at a rate such that $(H K_n^2)/T$ tends to zero.

The asymptotic results given above are valid for any choice of sequence $\{ \mathcal{A}_n \}$ such that $\bigcup_n \mathcal{A}_n$ is dense and identifies the true distribution $f_0$, and for any ordering of transformations in  $\mathcal{A}_n$. However, the choice of sequence $\mathcal{A}_n$ and transformation ordering will have an effect at higher orders and in finite samples.

\color{black}
\setcounter{equation}{0}\def\theequation{6.\arabic{equation}}
\section{Finite Sample Performance of NLSD, GCov Specification and Bootstrap Tests }

This section examines
the finite sample performance of proposed test statistics in selected causal-noncausal autoregressive processes. We perform simulations to study the size and power of a) the NLSD
test (eq. 2.5) to test for the absence of (non)linear serial dependence in time series, b) the GCov specification test (eq. 3.3) in Section 6.1,  and c) GCov bootstrap test [Section 4.1] in Section 6.2. Section 6.3 shows an example of the GCov specification test with many nonlinear transformations.

\subsection{Simulation Study}

We consider univariate causal-noncausal MAR(r,s) processes with i.i.d. errors from a uniform distribution $\cal{U}$ $_{[-1,1]}$, a Laplace distribution with mean zero and variance one, and a t-student distribution with $\nu$ degrees of freedom, called t($\nu$), with mean zero and variance $\nu/(\nu-2)$, $\nu>2$. We generate these processes in samples of size  $T = 100,200,500$.

\subsubsection{Data Generating Process}

The simulation method introduced in Gourieroux and Jasiak (2016) is used to generate the MAR(r,s) process (eq. \ref{mar(r,s)})  where $r$ and $s$ denote the orders of causal and noncausal polynomials, respectively. For $r=0$ and $s=1$, we generate MAR(0,1), i.e., the noncausal autoregressive process of order 1:  
\begin{equation}
y_t= \psi y_{t+1} + u_t \;\;, |\psi| <1.
\end{equation}
\color{black}
\nin By setting $r=1$ and $s=1$, we generate MAR(1,1) processes defined in equation (3.7).
It follows from Lanne and Saikkonen (2011) that its unobserved components $v_{1,t}, v_{2,t}$ are defined by:
\begin{equation}
  v_{1,t}  \equiv (1- \phi L)  y_t \leftrightarrow (1-\psi L^{-1})  v_{1,t} = u_t,\;\;
  v_{2,t} \equiv  (1-\psi L^{-1})  y_t \leftrightarrow (1- \phi L)   v_{2,t} = u_t,
\end{equation}
\nin and can be interpreted as the "causal" and "noncausal"
components. 
%Gourieroux, Jasiak (2016) show that  i) $v_{1,t}$ is $u$-noncausal (i.e. a function of present and future values of $u_t$) and $y$-causal (i.e. a function of present and past values of $y$) and ii) $v_{2,t}$ is $u$-causal (i.e. a function of present and past values of $u_t$) and $y$-noncausal (i.e. a function of present and future values of $y$). 
Then, the MAR(1,1) process has the following deterministic representations based on these unobserved components, used for the simulation and bootstrapping of $y_t$:
 \begin{equation}
i)\, y_t = \Frac{1}{1-\phi \psi} (\phi v_{2,t-1} + v_{1,t}), \;\;ii) \, y_t = \Frac{1}{1-\phi \psi} (v_{2,t} + \psi v_{1,t+1}),
\end{equation}
\noindent where in i) $y_t$ is a linear function of the
first lag of $v_{2,t}$ and of the current value of $v_{1,t}$, and in ii)
$y_t$ is a linear function of the
current value of $v_{2,t}$ and of the 
first lag of $v_{1,t}$. \color{black}  

%When $\psi= 0$, we get the MAR(1,0), a purely causal autoregressive process and if $\phi= 0$, we get a purely noncausal process. If both autoregressive polynomials have non-zero coefficients, then equation (4.2) describes a mixed causal-noncausal MAR(1,1) process. The mixed process contains both leads and lags of $y_t$, and admits a two-sided moving average representation [Gourieroux, Zakoian (2015)]. 

\begin{figure}[ht!]
    \centering
    
    % \begin{subfigure}[b]{0.49\textwidth}
    %     \centering
    %     \includegraphics[width=\textwidth]{MAR01U02.png}
    %     \caption{MAR(0,1), $u_t \sim U$, $\psi_1=0.2 $  }
    %     \label{ar1c1200}
    % \end{subfigure}
    % \hfill
        \begin{subfigure}[b]{0.31\textwidth}
        \centering
        \includegraphics[width=\textwidth]{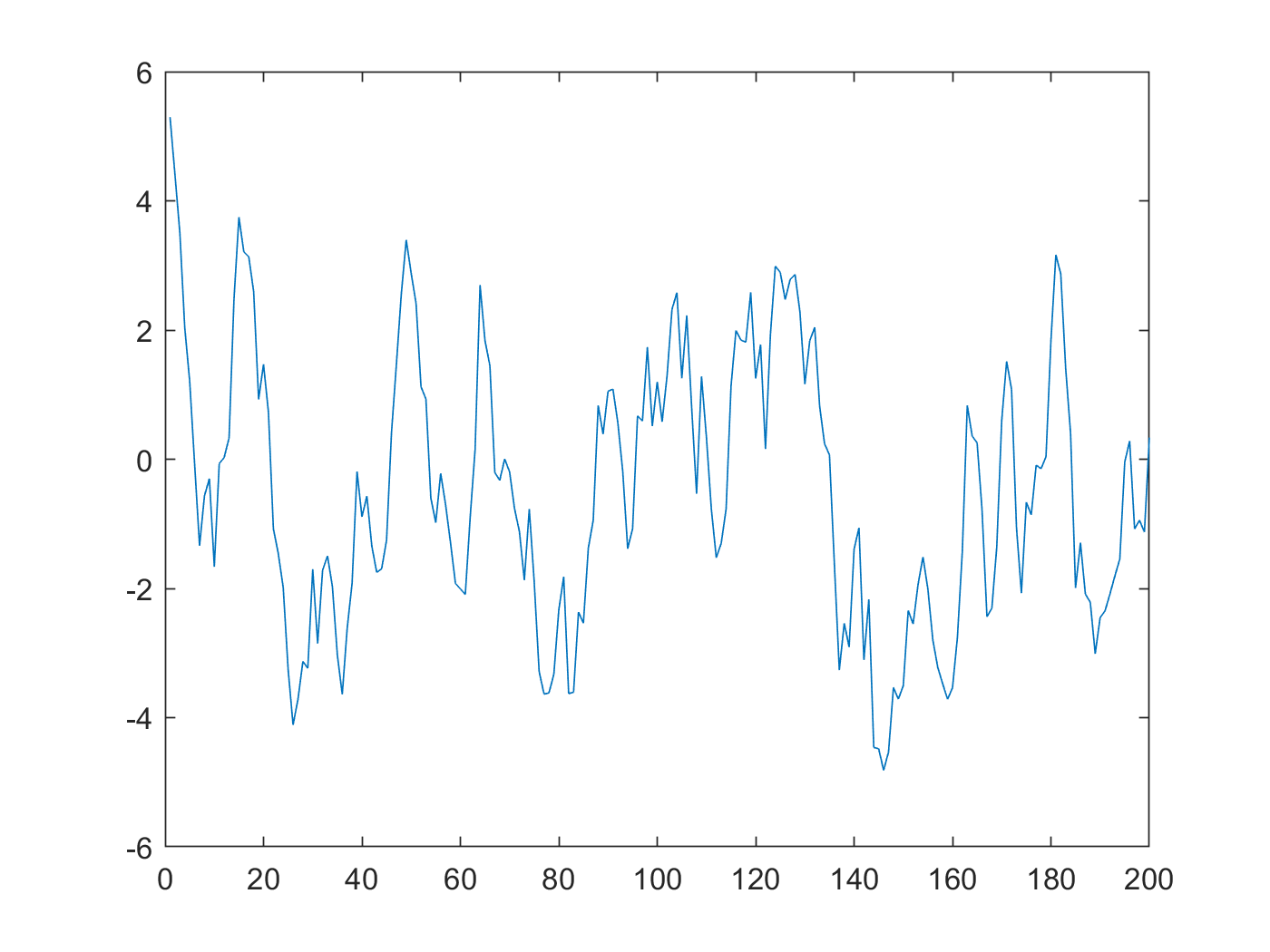}
        \caption{ $u_t \sim U$}
        \label{mar11c1200}
    \end{subfigure}
    \hfill
        \begin{subfigure}[b]{0.31\textwidth}
        \centering
        \includegraphics[width=\textwidth]{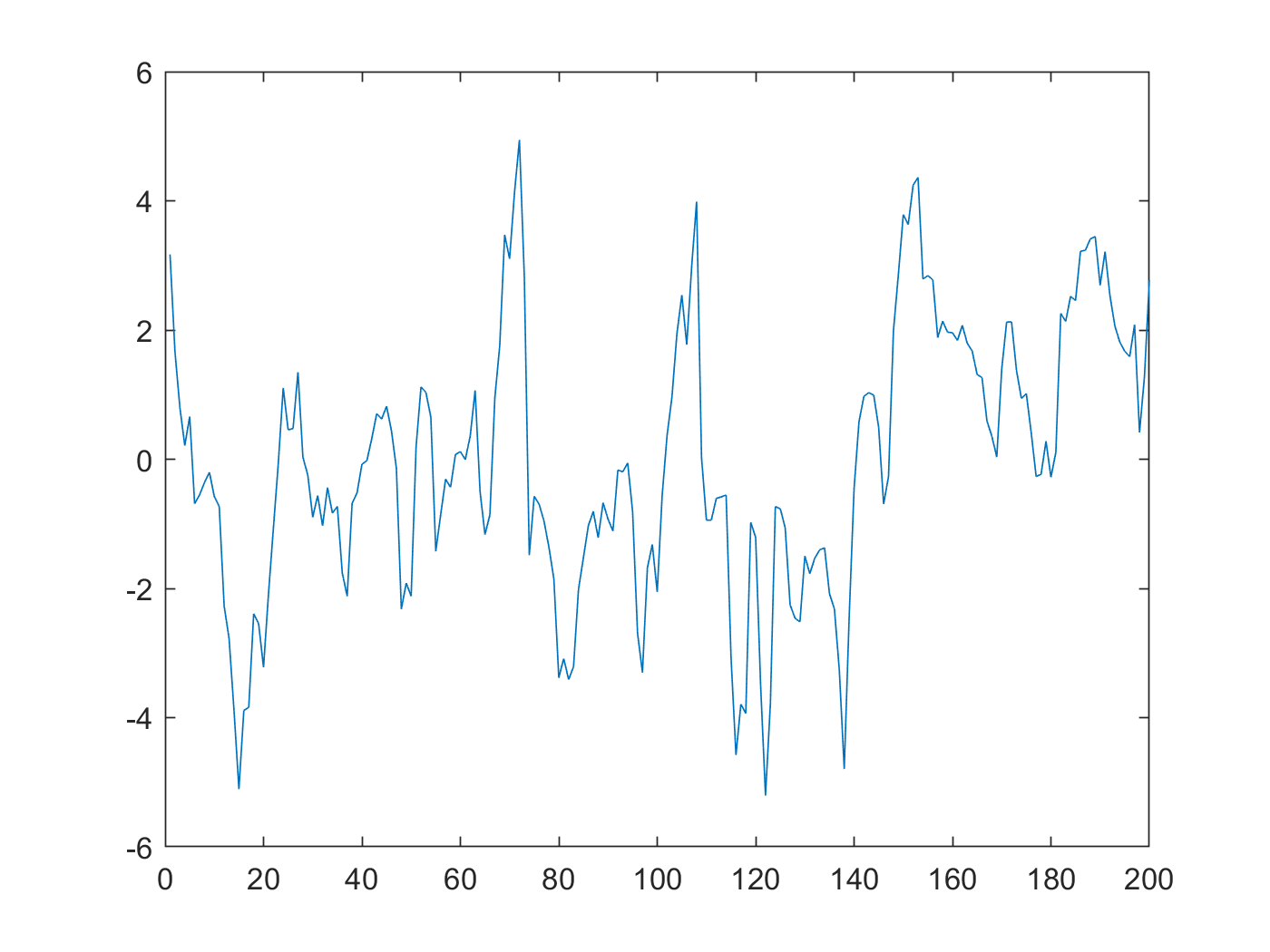}
        \caption{ $u_t \sim L$}
        \label{mar11t21200}
    \end{subfigure}
    \hfill
    \begin{subfigure}[b]{0.31\textwidth}
        \centering
        \includegraphics[width=\textwidth]{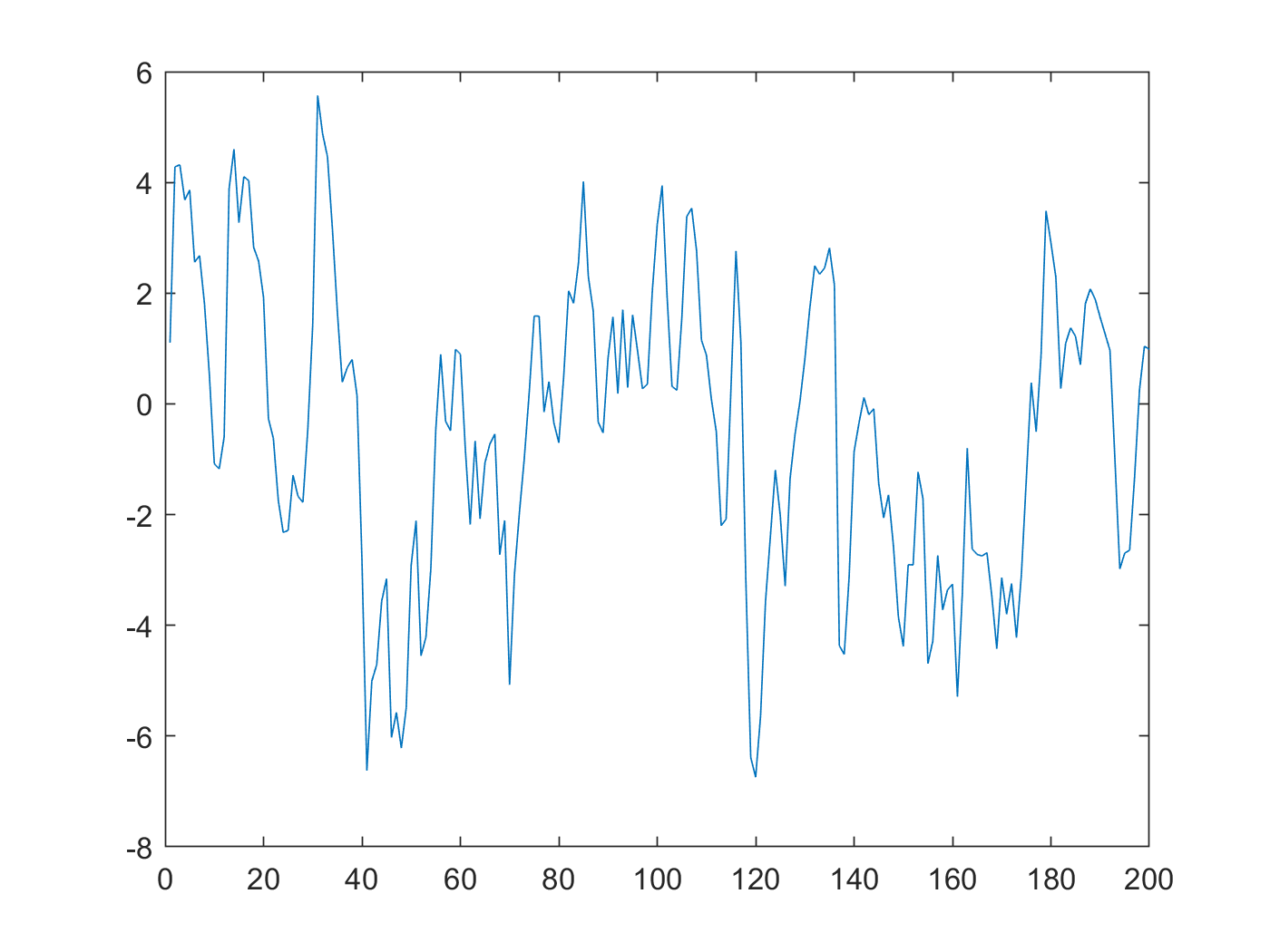}
        \caption{ $u_t \sim t(5)$ }
        \label{mar11t51200}
    \end{subfigure}
    
    \caption{MAR(1,1) processes with $\phi=0.2$ and $\psi=0.8$, and with error densities: L.:Laplace, U.:Uniform, t(5): t-Student with d.f.=5, over a period of $T=200$}
    \label{figure 1}
\end{figure}

 Figure \ref{figure 1} shows examples of trajectories of the MAR(1,1) processes simulated with the three error distributions given above and the coefficients $\phi=0.2$ and $\psi=0.8$  satisfying the strict stationarity condition.
We observe that a large error value creates a spike in the trajectory of MAR(1,1) with explosion and collapse rates determined by coefficients $\psi$ and $\phi$, respectively. 
%The simulated trajectories of the MAR(0,1) processes are displayed in Figure ... Appendix C. %In pure processes, we observe a jump if $\psi = 0$ and $\phi>0$, and an explosive bubble if $\psi$ is small, positive, and $\phi=0$. 

\subsubsection{Nonlinear Transformations}
\textcolor{black}{
In practice, we choose a finite set of nonlinear autocovariances, depending on the dynamics of the process. For example, we consider the sample autocovariances to capture linear serial dependence in the data, or residuals. To account for nonlinear serial dependence, we use the nonlinear autocovariances, which can be selected based on graphically displayed nonlinear autocorrelograms [Gourieroux, Jasiak (2001)]. This approach provides additional insights on the choice of $H$. A priori, we can set $H \geq p$, where $p$ is the lag length of the model. Specifically, processes with local explosive patterns and/or time-varying volatility are characterized by conditional heteroscedasticity, which can be captured by the autocovariances of squared observations. The third powers can accommodate skewed distributions, common in financial data. 
Logarithms are useful for capturing nonlinearities and transforming positive price series into variables that take both positive and negative values. 
%The sign transformation can separate the volatility dynamics from the bid-ask bounce in the financial data. 
For models with Cauchy distributed errors, we may employ fractional transformations, such as the square root of absolute value, since the unconditional expectation of a fractional power of a Cauchy distributed variable exists, while the unconditional expectation of an integer order does not. 
%When choosing multiple transformations, it is important to avoid combinations that could result in a reduced rank of matrix $\Gamma(0,.)$\footnote{See Giancaterini et al. (2024)}. 
Section 5 shows that combinations of power transformations weighted by exponential functions can also be used and that by increasing the set of associated autocovariance conditions, one can improve the performance of the GCov estimator and test. }

\subsubsection{NLSD Test for Time Series}

This section examines the finite-sample size and power of the (non)linear serial dependence (NLSD) test of $y_t$  computed from transformations $a$ of a univariate time series $y_t$. 
%where the true error distribution $f_0$ is either Uniform, Laplace, or t(5),
The null hypothesis:
$H_{0,a} = \{\Gamma_{0,a} (h) =0, \; \forall  h=1,...,H\},$
%$$H_{0} = (\gamma=0) \equiv (y_t = u_t) $$
\nin is tested against the fixed alternative hypothesis:
$H_{1,a} = \{ \exists h: \mbox{such that }\; \Gamma_{0,a} (h) \neq 0\}.$
Under this fixed alternative, the process is a MAR(0,1) model with an autoregressive coefficient denoted by $\gamma$ that varies between 0.1 and 0.9. Thus, the alternative can be written in terms of the parameters as:
$H_{1,a} = \{\gamma, f: \Gamma(h; \gamma,f) =0 \}.$
%$(1-\gamma L^{-1}) y_t = u_t$
For each value of $\gamma$ and each error density $f_0$, we simulate the strong white noise series $y_1^s,...,y_T^s, s=1,...,S$ with S=5000 replications and compute the NLSD test statistics, based on $K=2$ transformations of time series: $y_t, y_t^2$,  and lag $H=1$. The results are reported in Table \ref{table:dependenceMAR(0,1)}, for the nominal size of the test of $0.05$.
%We first consider a fixed alternative, assuming a given error density.
The first row of Table \ref{table:dependenceMAR(0,1)} with the zero values of coefficients
$\gamma$ provides the rejection rates corresponding to the size. The remaining rows illustrate the size-adjusted\footnote{The size adjustment is based on the 95th percentile of the sample distribution of simulated test statistic.} power of the test against the fixed alternative of a MAR(0,1) process with coefficients $\gamma=0.3,0.7$.
The columns of Table \ref{table:dependenceMAR(0,1)} show the outcomes for different sample sizes and the Uniform, Laplace, and t(5) error distributions. Additional results for other values of $\gamma$ are provided in Table 6 of Appendix C.1.

\begin{table}
\footnotesize{
\centering
\caption{NLSD Test of the absence of (non)linear dependence against the fixed alternative of MAR(0,1) at 5\% significance level: empirical size and power }
\begin{tabular}{|c|ccc|ccc|ccc|}
\hline
\multirow{2}{*}{$\gamma$} & \multicolumn{3}{c|}{T=100} & \multicolumn{3}{c|}{T=200} & \multicolumn{3}{c|}{T=500} \\
\cline{2-10} 
& Uniform       & Laplace      & t(5)      & Uniform       & Laplace      & t(5)      & Uniform       & Laplace      & t(5)     \\ \hline
    0.0 & 0.0414 & 0.0484 & 0.0512 & 0.0450 & 0.0502 & 0.0518 & 0.0496 & 0.0540 & 0.0480 \\
    % \hline
    % 0.1 & 0.0878 & 0.0706 & 0.0684 & 0.1502 & 0.1360 & 0.1220 & 0.3760 & 0.3390 & 0.3572 \\
    % \hline
    % 0.2 & 0.2682 & 0.2202 & 0.2240 & 0.5618 & 0.5390 & 0.5322 & 0.9586 & 0.9546 & 0.9570 \\
    \hline
    0.3 & 0.6014 & 0.5742 & 0.5674 & 0.9202 & 0.9266 & 0.9256 & 1 & 0.9998 & 1 \\
    % \hline
    % 0.4 & 0.8754 & 0.8814 & 0.8754 & 0.9966 & 0.9988 & 0.9976 & 1 & 1 & 1 \\
    % \hline
    % 0.5 & 0.9796 & 0.9822 & 0.9856 & 1 & 1 & 1 & 1 & 1 & 1 \\
    % \hline
    % 0.6 & 0.998 & 0.9986 & 0.9998 & 1 & 1 & 1 & 1 & 1 & 1 \\
    \hline
    0.7 & 1 & 1 & 1 & 1 & 1 & 1 & 1 & 1 & 1 \\
    % \hline
    % 0.8 & 1 & 1 & 1 & 1 & 1 & 1 & 1 & 1 & 1 \\
    % \hline
    % 0.9 & 1 & 1 & 1 & 1 & 1 & 1 & 1 & 1 & 1 \\
    \hline
\end{tabular}
\label{table:dependenceMAR(0,1)}
\caption*{The first row ($\gamma=0$) shows the size of the test and the remaining rows show the size-adjusted power against fixed alternatives.}
}
\end{table}

The results reported in Table \ref{table:dependenceMAR(0,1)} show close to the nominal empirical size and good power of the test against fixed alternatives, given each error density.
When the sample size increases, the size converges to 0.05 and the power converges to 1.  We also observe that the power of the test increases in the values of the autoregressive coefficient $\psi$ of MAR(0,1).

\begin{figure}[ht!]
    \centering
    
    \begin{subfigure}[b]{0.31\textwidth}
        \centering
        \includegraphics[width=\textwidth]{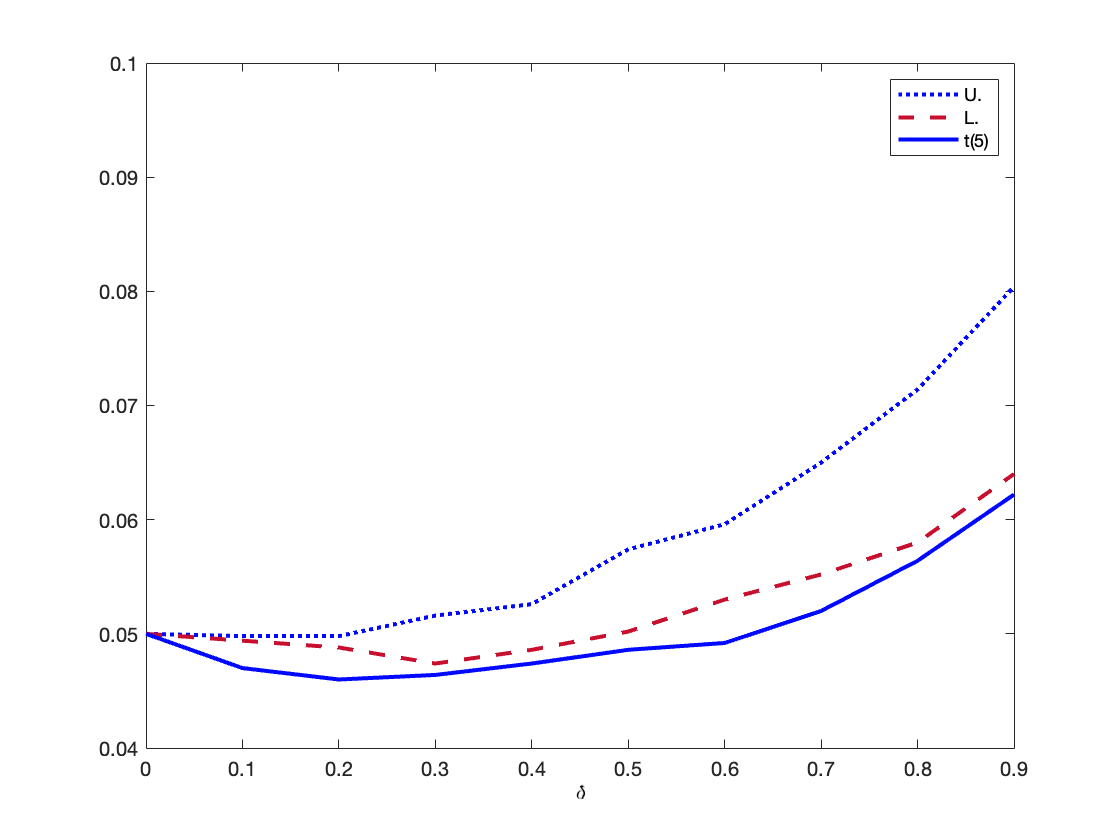}
        \caption{$T=100$ }
        \label{local-100}
    \end{subfigure}
    \hfill
        \begin{subfigure}[b]{0.31\textwidth}
        \centering
        \includegraphics[width=\textwidth]{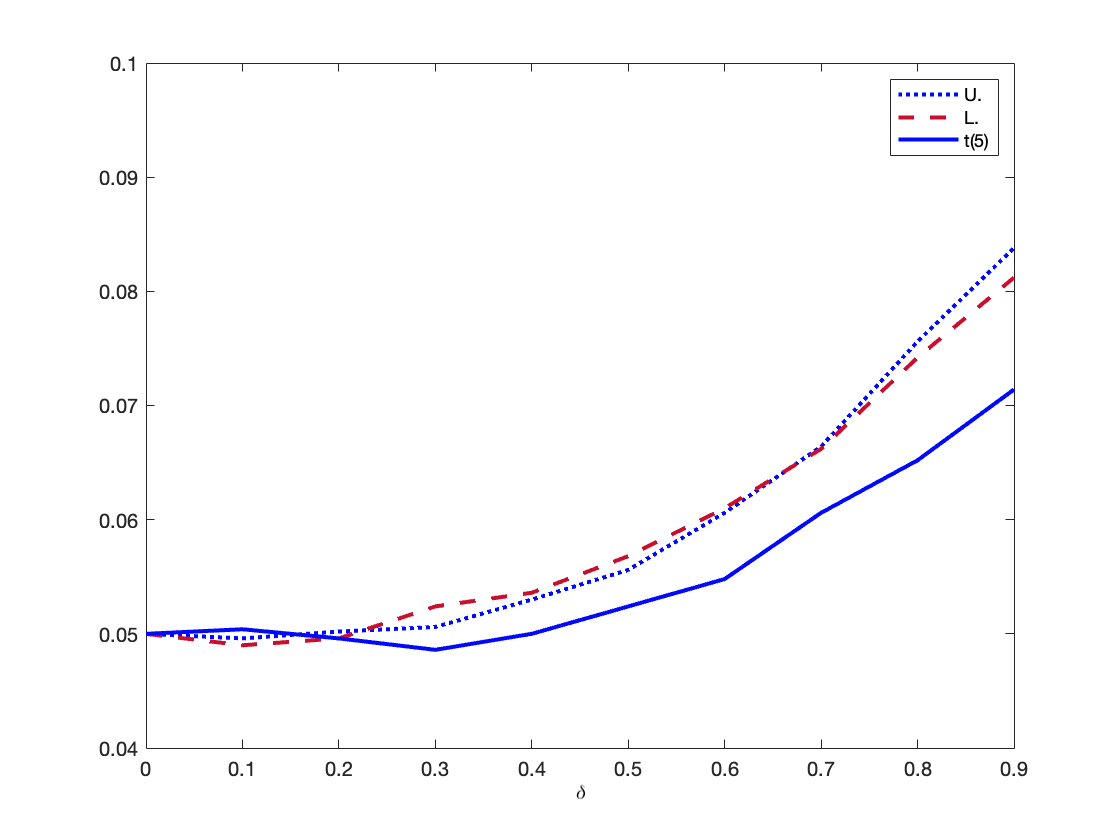}
        \caption{$T=200$ }
        \label{local-200}
    \end{subfigure}
        \hfill
        \begin{subfigure}[b]{0.31\textwidth}
        \centering
        \includegraphics[width=\textwidth]{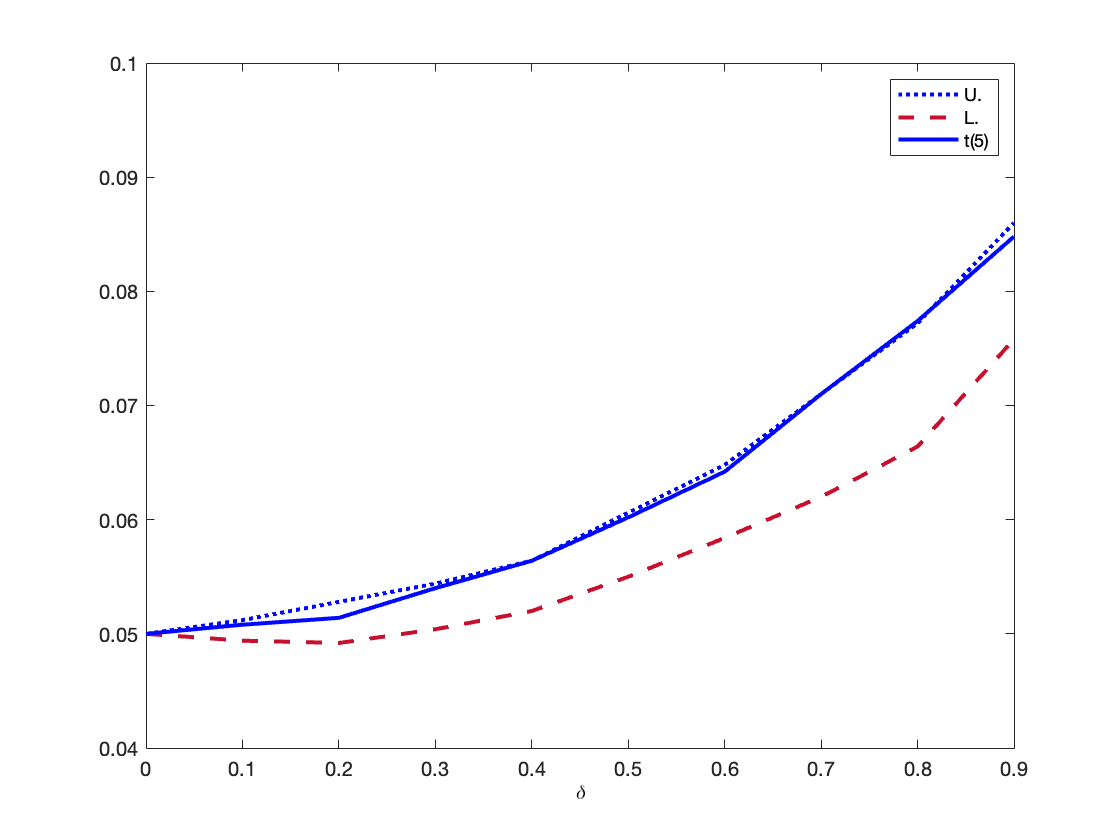}
        \caption{$T=500$ }
        \label{local-300}
    \end{subfigure}
 
    \caption{Local asymptotic power of the NLSD test.}
    \label{Figure local power}
\end{figure}

Furthermore, we investigate the power under the local alternatives by generating MAR(0,1) models with autoregressive coefficients equal to $\frac{\delta}{\sqrt{T}}$, where $\delta$ varies between $0$ and $0.9$. The results are provided in Figure \ref{Figure local power}. Since we consider local alternatives, we expect asymptotically the powers to be close to the size for small $\delta$, while for bigger $\delta$, we deviate further away.

%Figure \ref{Figure local power} illustrates the local asymptotic power of the test computed from the last column of Table 2 for T=500 and size-adjusted. The local asymptotic power is displayed for the time series with the Uniform, Laplace and t(5) distributions by three functions of $\delta=0,0.1,\dots,0.9$, where $\psi_T=\gamma_T=\frac{\delta}{\sqrt(T)}$.

%\begin{figure}[ht!]
%    \centering
%    \includegraphics[width=\textwidth]{local power.png}
% \includegraphics[width=9cm]{Figure1.png}
%    \caption{Local asymptotic power of the test of the absence of (non)linear dependence.}
%    \label{local power}
%\end{figure}

\nin  We observe that the test has good local power, and the size-adjusted power functions in the neighborhood of the null hypothesis increase fast and nonlinearly in $\delta$. The local asymptotic properties 
%be slightly worse for the Laplace distributed time series in small samples. 
\textcolor{black}{depend on the sample size and the error distribution. When the sample size increases from 100 to 500, we observe a greater improvement for the Uniform and t(5) distributions compared to the Laplace distribution, which has the least heavy tail among these three distributions.}

\subsubsection{GCov Specification Test for Semi-Parametric Models}

This section examines the finite-sample size and power of the GCov specification test applied to nonlinear transforms of the residuals of a model. The true model is a MAR(0,1) with errors $u_t = (1-\psi L^{-1})y_t$. It is estimated by the GCov estimator with lag length  H=3. We consider K=2 with the residuals $\hat{u}_{T,t}$ and their squares $\hat{u}_{T,t}^2$ as non-linear transformations, where  $\hat{u}_{T,t} = (1-\hat{\psi}_T L^{-1})y_t$, for the MAR(0,1) process.

%\nin {\bf Empirical size}

 To study the size, we compute the test statistics from the autocovariance matrices of nonlinear transformations of the residuals, denoted by $\Gamma_a(h; ., f)$. The null hypothesis of "correct specification" is: $H_{0,a} = \{\psi, f: \Gamma_{0,a} (h; \psi, f) =0, \; \forall  h=1,...,H\}$, and it is tested against $H_{0,a} = \{\psi, f: \exists h: \mbox{such that }\; \Gamma_{0,a} (h; \psi, f) \neq 0 \}$.

We generate the noncausal MAR(0,1) processes with the autoregressive coefficients $\psi$  
\textcolor{black}{equal to $0.3$ and $0.7$}, using S=5000 replications.
The results are reported in Table \ref{Table22}  for the nominal size of 0.05.
\textcolor{black}{The top two rows of Table \ref{Table22}  illustrate the rejection rates corresponding to the empirical size of the GCov specification test applied to the  MAR(0,1)  model, and the columns display the results for the Uniform, Laplace, and t(5) error distributions and different sample sizes}.  
%The columns of Table \ref{Table22} summarize the results for different sample sizes and error distributions.
Table \ref{Table22} shows that the GCov specification test is conservative at $T=100$ in all cases.
When the sample size increases, the size approaches the nominal size for all distributions. The power increases in $\psi$, similar to the pattern in Table 1, and quickly converges to one.

\begin{table}[]
\footnotesize{
\centering
\caption{Empirical size and power of GCov specification test of MAR(0,1) at 5\% significance}

\begin{tabular}{|c|cc|ccc|ccc|ccc|}
\hline
\multirow{2}{*}{S./P.} & \multirow{2}{*}{$\phi$} & \multirow{2}{*}{$\psi$} & \multicolumn{3}{c|}{T=100}                                  & \multicolumn{3}{c|}{T=200}                              & \multicolumn{3}{c|}{T=500}                              \\ \cline{4-12} 
                       &                      &                      & \multicolumn{1}{c|}{Uniform} & \multicolumn{1}{c|}{Laplace}     & t(5)  & \multicolumn{1}{c|}{Uniform} & \multicolumn{1}{c|}{Laplace} & t(5)  & \multicolumn{1}{c|}{Uniform} & \multicolumn{1}{c|}{Laplace} & t(5)  \\ \hline
\multirow{2}{*}{S.}    &            0          & 0.3                  & 0.0224 & 0.0454 & 0.0386 & 0.0348 & 0.0566 & 0.0534 & 0.0406 & 0.0544 & 0.0560  \\ \cline{2-12} 
                       &          0            & 0.7                  & 0.0200 & 0.0400 & 0.0338 & 0.0298 & 0.0528 & 0.0468 & 0.0408 & 0.0552 & 0.0528  \\ \hline
\multirow{2}{*}{P.}    & 0.8                  & 0.3                  & 0.1016  & 0.1724  & 0.1788  & 0.3404  & 0.4468  & 0.4672  & 0.9092  & 0.9282  & 0.9300 \\ \cline{2-12} 
                       & 0.8                  & 0.7                  & 0.9680  & 0.9882  & 0.9896  & 1       & 1       & 1       & 1       & 1       & 1      \\ \hline
\end{tabular}
\label{GCov test size and power}
\caption*{ S.: size, P.: power (size-adjusted) }
\label{Table22}
}

\end{table}

An interesting way to investigate the power of the GCov specification test is to consider the null hypothesis of MAR(0,1) and deviate from it by adding a causal autoregressive coefficient $\gamma = \phi$. This transforms the MAR(0,1) model under the null hypothesis into a  MAR(1,1) model under the alternative. Then, the fixed  alternative hypothesis is:
 $H_{1,a} = \{ \psi, \gamma, f: \exists h: \mbox{such that }\; \Gamma_{0,a} (h; \psi, \gamma, f) \neq 0\}.$

\textcolor{black}{Rows 3 and 4 of Table \ref{Table22} report the size-adjusted power of the test applied to a noncausal MAR(0,1) model with coefficients $\psi=0.3$ and $\psi = 0.7$, when an additional causal coefficient $\gamma=\phi=0.8$ of MAR(1,1) is present under the fixed alternative, for each given error density and sample size reported in the columns. By comparing the results for $\psi=0.7$ and $\psi=0.3$, we can conclude that the empirical power increases in $\psi$ for all sample sizes and distributions. Additional results for other values of the causal persistence coefficient $\gamma=\phi$}
are given in Table \ref{newwtable:power}, Appendix C.1.
%By comparing the results for $\psi=0.7$ and $\psi=0.3$ in Table \ref{newwtable:power}, we can conclude that the empirical power increases in $\psi$.}

\begin{figure}[ht!]
    \centering
    
    \begin{subfigure}[b]{0.49\textwidth}
        \centering
        \includegraphics[width=\textwidth]{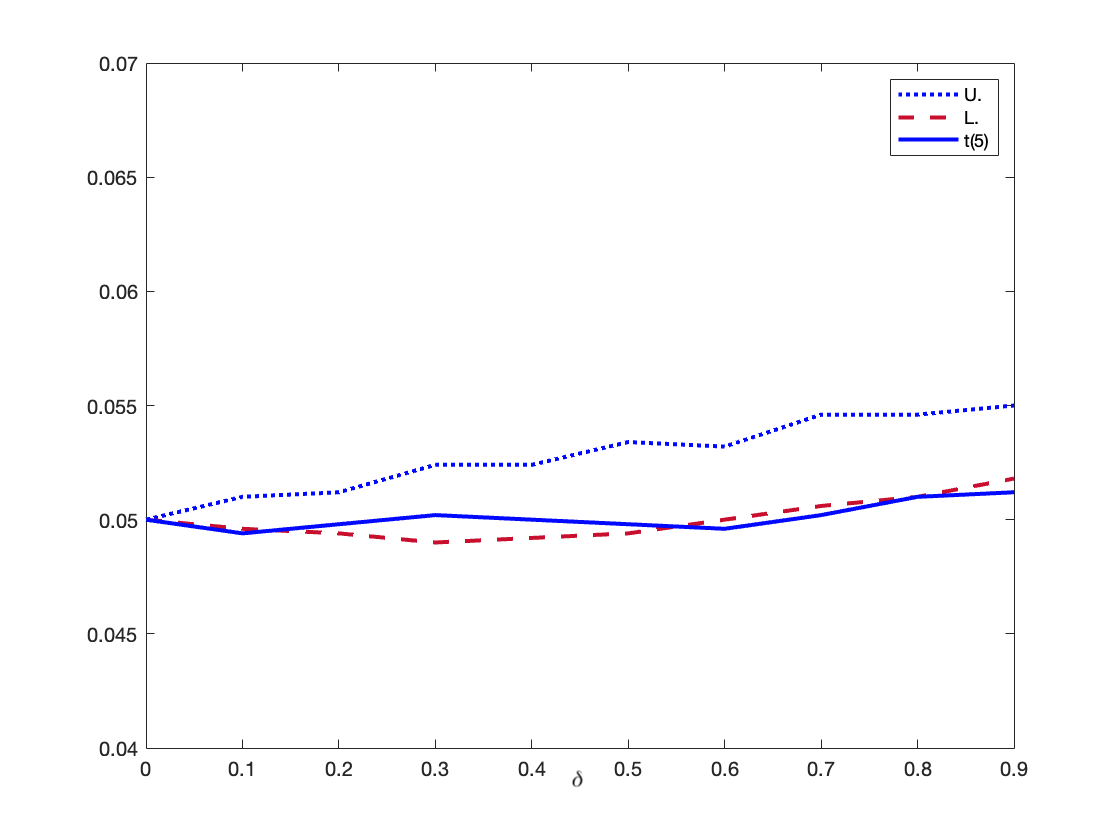}
        \caption{MAR(0,1) with $\psi=0.3$ }
        \label{local03}
    \end{subfigure}
    \hfill
        \begin{subfigure}[b]{0.49\textwidth}
        \centering
        \includegraphics[width=\textwidth]{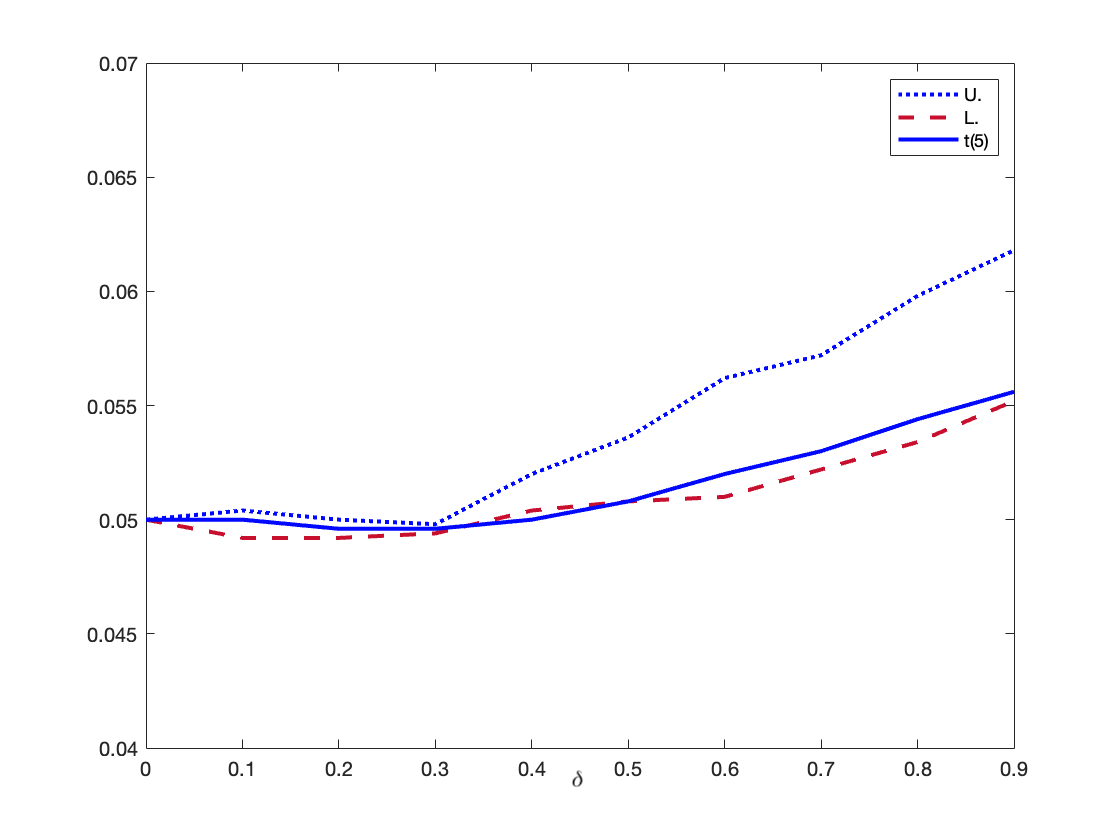}
        \caption{MAR(0,1) with $\psi=0.7$ }
        \label{local07}
    \end{subfigure}

    \caption{Local asymptotic power of GCov specification test} 
    \label{local power MAR(0,1)}
\end{figure}

Figure 3 illustrates the power of the GCov specification test against local alternatives, computed from the sample of size $T=500$ and size-adjusted. The three local power functions of the MAR(0,1) model with Uniform, Laplace and t(5) distributed errors are plotted against $\delta=0,0.1,\dots,0.9$, where $\phi_T=\gamma_T=\frac{\delta}{\sqrt{T}}$.
Panel (a) displays the results for the MAR(0,1) model with $\psi=0.3$ and panel (b)  shows the results for $\psi=0.7$. We observe that the test has good local power in each case.  \textcolor{black}{The best local power is displayed in both panels by the Uniform distributed errors with the heaviest tails.} The local power functions increase faster in panel (b) 
for the model with higher noncausal persistence than in panel (a). Overall, the rate of increase of local power functions is lower compared with the local power of the (non)linear dependence NLSD test displayed in Figure 2. Hence, larger sample sizes are recommended for the specification test. 

\subsubsection{Infinite Variance Errors}
\textcolor{black}{So far, we have considered only error distributions with finite variances. The size and power properties of the GCov specification test of the MAR(0,1) model with Cauchy distributed errors are illustrated in Table \ref{Csizenew}, Appendix C.2. We use K=3, with the square roots and logarithms of the absolute values of residuals as nonlinear transformations, and set the maximum lag equal to $H=3$. 
%For size, we generate MAR(0,1) with $\psi=0.3$ and $\psi=0.7$ and fit the correct model by using the GCov estimator. For power, we consider the alternative of MAR(1,1) with the coefficient $\phi=0.8$. 
We observe that the empirical size is close to the nominal for T=200, and the power quickly converges to one for both values of $\psi$ as the number of observations increases. Compared to Table \ref{Table22}, the power is higher for $T=100$ and the convergence in $T$ is faster. }

%Furthermore, we consider the MAR(0,1) process and study the power under the local alternatives. To do that we deviate from the null hypothesis of MAR(0,1) by adding parameter $\gamma=\phi$ and increasing $\phi$  gradually with $\phi=\frac{\delta}{\sqrt{T}}$.

% We use a similar approach to investigate the empirical power of the GCov test in the multivariate framework. We use the same models at error distributions as those investigated in the previous section on empirical size and illustrated in Table 5. 
 %To study the power under a fixed alternative the GCov test is computed for coefficients $\Phi^*$ that deviate  from the true coefficient $\Phi$. More specifically, we keep three coefficients  $\phi_{12}, \phi_{21}, \phi_{2}$ of the $\Phi$ matrix fixed, and perturb $\phi_{11}$ by adding matrix P (2*2):

% Table \ref{multipower} indicates the results of fixing $\phi_{12}$,$\phi_{21}$ and $\phi_{22}$ and just changing the $\phi_{11}$ by $p_{12}=p_{21}=p_{22}=0$ and $p_{11}=-0.3,-0.2,-0.1,0,0.1,0.2,0.3$. 

%Furthermore, to investigate the empirical power of the GCov test in the VAR(1) model under the local alternatives: $\phi_{11} = \phi_{11} + \delta/\sqrt{T}$. We use the same values of matrix $\Phi^*$ but instead of adding $P$  we add $\frac{P}{\sqrt{T}}$. Table \ref{localmultipower} demonstrates the local empirical power results in multivariate framework. 

\subsection{GCov Bootstrap Test}
\color{black}
 
In the first part of this section, we consider the bootstrap approximation of finite-sample size and power of the GCov specification test based on the GCov estimator. In the second part, we examine the GCov test and GCov bootstrap test, both with the AML estimator.

\subsubsection{GCov Bootstrap Test with GCov Estimator}

The performance of GCov bootstrap test depends on the quality of the bootstrap approximation of finite sample properties of a GCov test.
We consider a MAR(0,1) model under the null hypothesis, and a MAR(1,1) under the fixed alternative with the same error distributions and parameter values as those in Section 6.1.4. The GCov test statistic is based on $K=2$ with residuals and their squares, and $H=3$. The bootstrap with $S=100$ is replicated 1000 times.

Table \ref{boostrapapp}, Appendix C.3 shows the bootstrap approximation of finite sample size and power of GCov specification test obtained from rejection rates with a chi-square quantile used as the critical value and two sampling schemes: with and without replacement. By comparing these results with Table 2, we observe that the bootstrap provides a reliable approximation of finite-sample size, and the quality of the approximation improves with increasing $T$. Moreover, we find that sampling with replacement provides more accurate results. An additional insight is obtained for the MAR(0,1) process with t(5)-distributed errors and $\psi=0.7$. Figure \ref{boot MAR(0,1)}, Appendix C.3 shows the convergence of the distribution of the GCov bootstrap test statistic to the theoretical chi-square distribution when the sample size increases from $T$=100 to $T$=500.

% \medskip
% TO BE COMPLETED

%\medskip

%\subsubsection{GCov Bootstrap Test with the AML Estimator}
%For diagnostic checking of parametric models, the GCov bootstrap test can be applied to the residuals $\hat{u}_{T,t} = g(\tilde{Y}_t, \tilde{\theta}_T)$, where $\tilde{\theta}_T$ is not the GCov estimator, but instead a maximum likelihood estimator, for example. 
%In this section, we illustrate by simulations the size and power of this GCov bootstrap test in finite sample. 

\subsubsection{Parametric Estimation of Causal-Noncausal Models} 

For diagnostic checking of parametric models, the GCov specification test can be applied to residuals $\hat{u}_{T,t} = g(\tilde{Y}_t, \tilde{\theta}_T)$, where $\tilde{\theta}_T$ is not the GCov estimator.
%but instead a maximum likelihood estimator, for example. 

When a causal-noncausal model is fully parametric, and the errors are assumed to follow a parametric density, the Approximate Maximum Likelihood (ML) estimator can be used.
The Approximate Maximum Likelihood (AML) estimator of univariate MAR(r,s) processes defined in equation (3.6) and introduced by Lanne and Saikkonen (2011) is: 
% The AML disregards the first $r$ state
%variables
%%that summarize the effect of shocks before time $r$ and the last $s$ state variables that summarize the
%effect of shocks after time $T-s$ and is therefore  constructed from 
%errors $u_{r+1},...u_{T-s-1}$ only.  Hence, the sample is reduced to
%$T-(r+s)$ observations:
%the first error to be included in the log-likelihood 
%function  is $u_{r+1}$ and he last one is
%$u_{T-s-1}$. %The Approximate Maximum Likelihood (AML) estimator is defined as:

$$
(\hat{\Psi},\, \hat{\Phi},\, \hat{\nu}) = Arg max_{\Psi, \Phi, \nu}\, \sum_{t=r+1}^{T-s} \ln \, f[\Psi( 
L^{-1}) \Phi (L) y_t ;  
\nu ],
$$
\nin where $f[.; \nu]$ denotes the non-Gaussian probability density function of $u_t$, such as a t($\nu$)-student density, for example.
Davis and Song (2020) discuss the ML estimator for the multivariate causal-noncausal VAR process given in Example 3, Section 3.1, with a parametric error density.

\subsubsection{GCov Test and GCov Bootstrap Test with AML Estimator}

To illustrate the performance of GCov bootstrap test with the AML estimator, we consider the simulated processes with errors from the t(4), t(5) and t(6) distributions. Like in the previous Section, the model under the null hypothesis is a MAR(0,1), and under the fixed alternative is a MAR(1,1).  First, we estimate the MAR(0,1) model from the AML estimator based on a t-student log-likelihood function. Next, we plug the AML residuals into the formula of the GCov specification test statistic in eq. (3.3), i.e., we calculate the sample autocovariances from the AML residuals and squared residuals (K=2) and their lags up to H=3.

%with Uniform, Laplace, and t(5) distributed errors. Hence, there is a misspecification of the models with Uniform and Laplace error distributions, resulting in the Quasi-AML (QAML) estimators. 

\begin{table}[H]
\footnotesize{
\centering
\caption{ \textbf{A:} Empirical size and power of GCov test with plug-in AML residuals and \textbf{B:} Empirical size and power of GCov bootstrap test of MAR(0,1) with AML, at 5\% significance}

\begin{tabular}{|c|c|cc|ccc|ccc|ccc|}
\hline
\multirow{2}{*}{Panel} & \multirow{2}{*}{S./P.} & \multirow{2}{*}{$\phi$} & \multirow{2}{*}{$\psi$} 
& \multicolumn{3}{c|}{T=100} & \multicolumn{3}{c|}{T=200} & \multicolumn{3}{c|}{T=500} \\ \cline{5-13} 
& & & & \multicolumn{1}{c|}{t(4)} & \multicolumn{1}{c|}{t(5)} & t(6) 
& \multicolumn{1}{c|}{t(4)} & \multicolumn{1}{c|}{t(5)} & t(6) 
& \multicolumn{1}{c|}{t(4)} & \multicolumn{1}{c|}{t(5)} & t(6) \\ \hline

\multirow{4}{*}{A} 
& \multirow{2}{*}{S.} & 0 & 0.3 & 0.065 & 0.061 & 0.061 & 0.071 & 0.050 & 0.054 & 0.070 & 0.051 & 0.061 \\ \cline{3-13} 
& & 0 & 0.7 & 0.064 & 0.064 & 0.061 & 0.072 & 0.048 & 0.050 & 0.072 & 0.056 & 0.062 \\ \cline{2-13} 
& \multirow{2}{*}{P.} & 0.8 & 0.3 & 0.431 & 0.400 & 0.391 & 0.735 & 0.697 & 0.645 & 0.988 & 0.990 & 0.986 \\ \cline{3-13} 
& & 0.8 & 0.7 & 0.994 & 0.996 & 0.993 & 1 & 1 & 1 & 1 & 1 & 1 \\ \hline

\multirow{4}{*}{B} 
& \multirow{2}{*}{S.} & 0 & 0.3 & 0.060 & 0.061 & 0.067 & 0.064 & 0.048 & 0.051 & 0.058 & 0.045 & 0.056 \\ \cline{3-13} 
& & 0 & 0.7 & 0.061 & 0.064 & 0.063 & 0.066 & 0.050 & 0.049 & 0.053 & 0.044 & 0.055 \\ \cline{2-13} 
& \multirow{2}{*}{P.} & 0.8 & 0.3 & 0.430 & 0.410 & 0.388 & 0.704 & 0.664 & 0.633 & 0.982 & 0.980 & 0.975 \\ \cline{3-13} 
& & 0.8 & 0.7 & 0.991 & 0.993 & 0.911 & 1 & 1 & 1 & 1 & 1 & 1 \\ \hline
\end{tabular}

\label{bootstrap1}
\caption*{ S.: size, P.: power (size-adjusted) }
}
\end{table}

  \nin This approach is a one-step procedure proposed as an alternative to two separate Ljung-Box tests applied to the residuals and their squares. The experiment is replicated 1000 times. \textcolor{black}{ Then, the finite sample size and power are evaluated from rejection rates based on the theoretical critical value and given in part A of Table \ref{bootstrap1}, where we observe a slight size distortion of the test in finite samples. Nevertheless, the test has reasonably good performance.} 
  
  Instead of this simple procedure, one can use the GCov bootstrap test to ensure better finite sample properties. We examine the size and power of the GCov bootstrap test of MAR(0,1) against MAR(1,1) with $S=100$ and AML-estimated residuals based on 1000 replications. The results are reported in Table \ref{bootstrap1}-B based on sampling with replacement. We find that the size of the GCov bootstrap is close to the nominal level of 0.5 for all sample sizes and error distributions. The power of the GCov bootstrap test is high and increases to 1 with sample size\footnote{The size adjustment is based on the bootstrapped critical value.}. Moreover, we report the size and power of the bootstrap test based on sampling without replacement in Table \ref{comparetable3} of Appendix C.3, providing additional evidence based on test size that sampling with replacement proposed in Section 4 is preferred.

\subsubsection{Infinite Variance Errors}

\textcolor{black}{ Table \ref{bootCSizeOLS}, Appendix C.2 presents additional results on the GCov bootstrap test of MAR(0,1) with Cauchy error distribution and OLS estimator. The size of our test compares favorably with the size of the bootstrap test of coefficient $\psi$ in  MAR(0,1) based on the (unconstrained) OLS estimator and sampling without replacement introduced in Cavaliere et al. (2020), who consider the test of the null hypothesis $\rho=\psi=0.5$ by unrestricted bootstrap for Stable error distribution with parameters $\alpha =1.0$ and $\beta=0.0$ based on the test statistic $r_T$. The size of our GCov bootstrap test is between 5.3 and 5.5, depending on $\psi$, which is close to 5.4 reported in Table 2, Cavaliere et al. (2020).  The results from sampling with and without replacement are both provided in Table\ref{bootCSizeOLS}. We observe that, as in Table 3, sampling with replacement provides a size closer to the nominal. }

%\textcolor{black}{ In Appendix C.2 we also provide in Table 11 a bootstrap approximation of the size and power of the GCov test applied to a Cauchy MAR(0,1). It shows that for T=500 we obtain size and power close to those reported in Table \ref{Csizenew} in the case of sampling with replacement. }

\subsection{GCov Specification Test with Many  Transformations}

\color{black}

We examine the finite sample performance of the GCov specification test with many transformations applied to the MAR(0,1) processes with t(5) distributed errors, coefficients $\psi=0.3$ and  $\psi=0.7$ generated in samples of size: $T=100, 200, 500$. The models are estimated by GCov with $K=7,8,9$ transformations, which are the absolute values of consecutive powers of the residuals weighted by exponential functions with $t=0.01$, and the maximum lags are $H=3,4,5$. Thus, the number of nonlinear autocovariance conditions used in this study ranges from 48 to 80. The size of the GCov specification test based on rejection rates from 1000 replications is reported in Table \ref{Manytransformations}. We observe that the size  approaches the nominal value as $K$, $H$ and $T$ increase. For both values of $\psi$ and all values of $H$ we observe that the size is closer to the nominal for $K=9$, in comparison to $K=7,8$. 
For $\psi=0.7$ the nominal size is reached for $K=7$ and $T=500$, for all $H$ considered. For $\psi=0.3$ more transformations $K=9$ and $T=500$ are needed, for all $H$ considered.

\begin{table}[H]
\footnotesize{
\centering
\caption{ Empirical size of GCov test with many transformations at  5\%  level, $T=500$.}
\begin{tabular}{|c|c|c|c|c|c|c|c|}
\hline
\multirow{2}{*}{S./P.} & \multirow{2}{*}{$H$} 
& \multicolumn{2}{c|}{$K=7$} & \multicolumn{2}{c|}{$K=8$} & \multicolumn{2}{c|}{$K=9$} \\ \cline{3-8}
& & $\psi=0.3$ & $\psi=0.7$ & $\psi=0.3$ & $\psi=0.7$ & $\psi=0.3$ & $\psi=0.7$ \\ \hline

\multirow{3}{*}{S.} 
& $H=3$ & 0.033 & 0.051 & 0.035 & 0.051 & 0.054 & 0.050 \\ \cline{2-8}
& $H=4$ & 0.031 & 0.049 & 0.034 & 0.059 & 0.050 & 0.056 \\ \cline{2-8}
& $H=5$ & 0.031 & 0.053 & 0.029 & 0.052 & 0.053 & 0.060 \\ \hline

% \multirow{3}{*}{P.} 
% & $H=3$ & 0.703 & 0.897 & 0.535 & 0.529 & 0.567 & 0.877 \\ \cline{2-8}
% & $H=4$ & 0.664 & 0.889 & 0.528 & 0.514 & 0.571 & 0.872 \\ \cline{2-8}
% & $H=5$ & 0.673 & 0.891 & 0.493 & 0.486 & 0.552 & 0.863 \\ \hline
\end{tabular}

\label{Manytransformations}

}
\end{table}

The power functions are illustrated in Appendix C 4. We find that the power of test converges quickly to 1 in sample size $T$, and the convergence is faster for higher values of $\psi$.

% \begin{figure}[H]
% %    \centering
%  %   \begin{subfigure}[b]{0.4\textwidth}
%         \centering
%         \includegraphics[width=0.49\textwidth]{SizeUK6H10.png}
%         \caption{Size of the GCov test with large K, H as a function of sample size}
% %   \end{subfigure}
% %    \hfill
%  %       \begin{subfigure}[b]{0.49\textwidth}
% %        \centering
% %        \includegraphics[width=\textwidth]{SizeMt5.png}
% %        \caption{t(5) distribution }
% %    \end{subfigure}
%     %\caption{Size  of GCov specification test with many transformations of noncausal AR(1) } 
%     \label{Csizenew}
% \end{figure}

% %\clearpage

% \\nin As expected, the size-adjusted power is 1 for both parameter values and all combinations of $K, H$ and $T$ given in Table 4. 

\color{black}
\section{Empirical Application}

In this section, we apply the GCov specification test to a univariate causal-noncausal model fitted to the series of aluminum prices in U.S. Dollars per metric ton. This approach is motivated by the presence of spikes and bubbles in aluminum prices, and the recent literature on causal-noncausal modeling of commodity prices [Hecq, Lieb, and Telg  (2016), Fries and Zakoian (2019), Gourieroux and Jasiak (2023)]. First, we apply the (non)linear serial dependence NLSD test to the data, and next use the GCov specification test to examine the goodness of fit of the estimated causal-noncausal processes.

\begin{figure}
    \centering
    
    \begin{subfigure}[b]{0.49\textwidth}
        \centering
        \includegraphics[width=\textwidth]{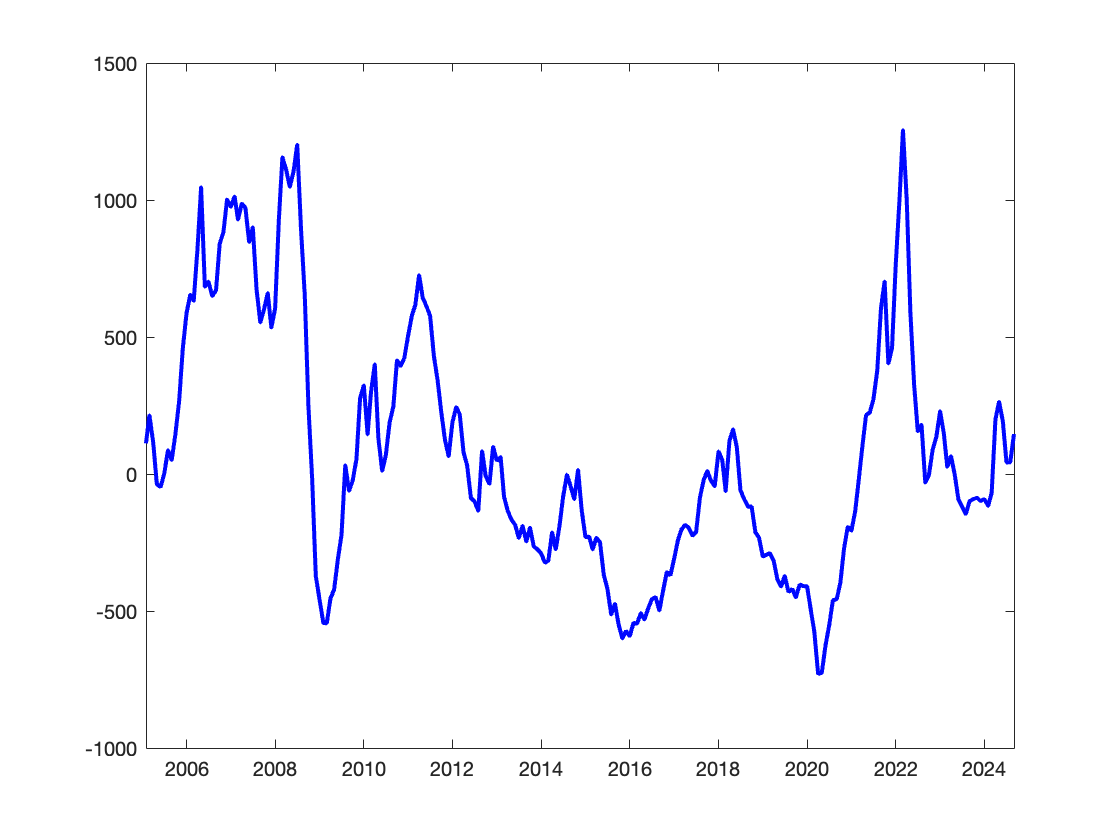}
        \caption{Detrended aluminum price }
        \label{2-a}
    \end{subfigure}
    \hfill
        \begin{subfigure}[b]{0.49\textwidth}
        \centering
        \includegraphics[width=\textwidth]{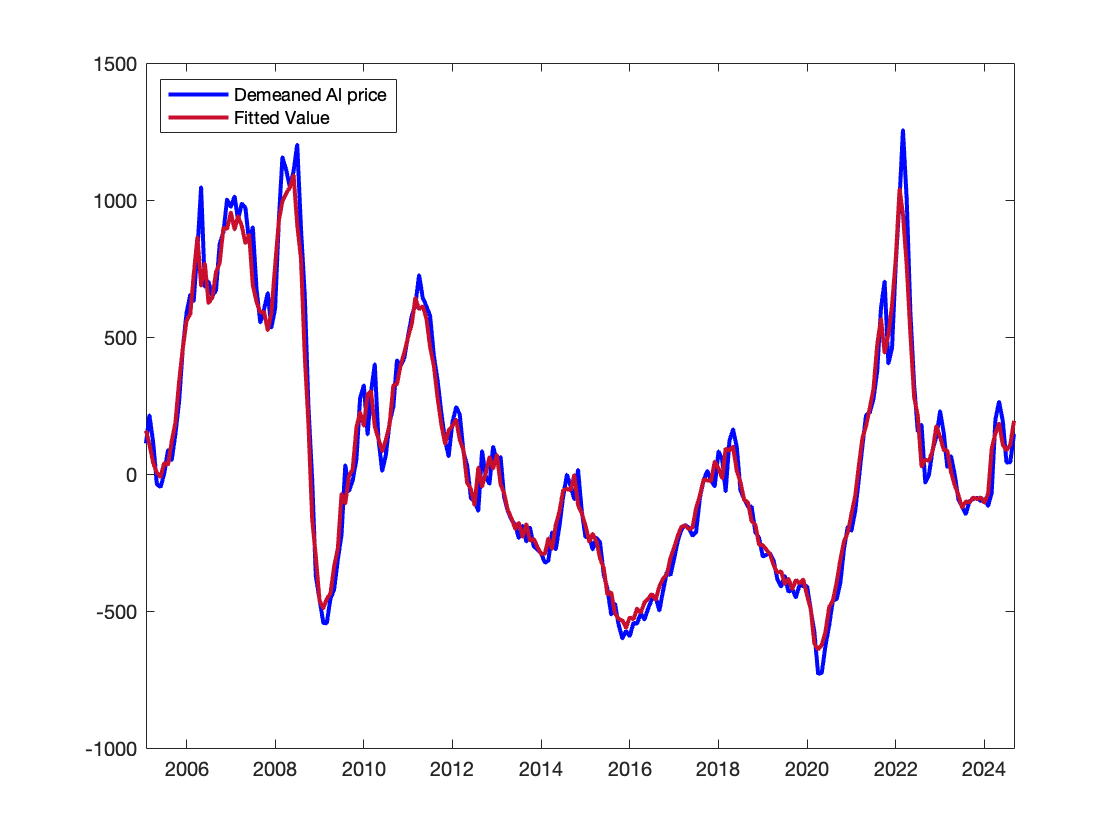}
        \caption{ MAR(1,1) fitted values}
        \label{2-b}
    \end{subfigure}
    % \hfill
    % \begin{subfigure}[b]{0.49\textwidth}
    %     \centering
    %     \includegraphics[width=\textwidth]{Jan17-residuals.png}
    %     \caption{MAR(1,1) residuals}
    %     \label{2-c}
    % \end{subfigure}
    % \hfill
    % \begin{subfigure}[b]{0.49\textwidth}
    %     \centering
    %     \includegraphics[width=\textwidth]{Jan17-MAR11utvt-AL.png}
    %     \caption{MAR(1,1) causal-noncausal components}
    %     \label{2-d}
    % \end{subfigure}
    
    \caption{Detrended aluminum price,  MAR(1,1) fitted values }
    \label{figure 2}
\end{figure}

Our sample consists of $T=228$ monthly average prices recorded between January 2005 and October 2024 and known as the Global price of Aluminum \footnote{International Monetary Fund, Global price of Aluminum [PALUMUSDM], retrieved from FRED, Federal Reserve Bank of St. Louis; https://fred.stlouisfed.org/series/PALUMUSDM, December 20, 2024.}. \textcolor{black}{We detrend the series of prices by regressing it on time (polynomial of degree one).}  The detrended prices are plotted in Figure \ref{2-a}, where we observe multiple spikes and a sudden drop in aluminum prices during the 2008 recession when the commodity prices fell due to weak demand. In addition, in 2020, we see a spike in the price of aluminum, which is due to the weak supply of commodities at the beginning of the Covid period. 

% \begin{table}[ht!]
% \centering
% \caption{ Aluminum price: summary statistics}
% \begin{tabular}{|c|c|c|c|c|c|c|c|}
% \hline
%                    & Count & Mean    & Min     & Max     & SD     & Skewness & Kurtosis \\ \hline
% aluminum price & 336   & 1820.40 & 1040.00 & 3498.40 & 464.12 & 0.80     & 3.20     \\ \hline
% \end{tabular}
% \label{stats}
% \end{table}
To ensure the identification of causal and noncausal dynamics, we test the data for normality by applying the Kolmogorov-Smirnov normality test. The test statistic of \textcolor{black}{0.50} exceeds the critical value of 0.08. Hence, the null hypothesis of the normality of the aluminum price distribution is rejected. Figure \ref{figure 5} (a) in  Appendix C.4 provides the sample density plot of demeaned aluminum price and compares it to the Gaussian density to confirm that the aluminum prices are non-Gaussian.

We test for nonlinear serial dependence in the aluminum prices using the NLSD test introduced in Section 2. We compute the test statistic from the series using \textcolor{black}{H=9} and K=2. The value of the NLDS test is \textcolor{black}{1675.4} and exceeds the critical value of \textcolor{black}{50.99}, showing (non)linear serial dependence in the data. This finding is confirmed by the ACF of the series and their squares in Figures \ref{3-a} and \ref{3-b} in Appendix C.4.

Next, we use the semi-parametric GCov estimator and
explore several specifications of the causal-noncausal MAR(r,s) models for varying autoregressive orders $r$ and $s$  without any distributional assumptions on the errors. In Table \ref{applications} we report the GCov estimated parameters with \textcolor{black}{H=9} and K=2, where we use the residuals and the logarithm of the second power of the residuals \textcolor{black}{to capture the linear and nonlinear serial dependencies, respectively. By considering nonlinear autocovariances at higher lags, we accommodate possible long-range serial dependence, with the total of 18 transformations.}  Under the strict stationarity assumption, all models have autoregressive polynomials with roots outside the unit circle. The estimated roots of the polynomials $\hat{\Phi}(L)$ and $\hat{\Psi}(L^{-1})$  are given in the last column of Table 4.

% \begin{table}[ht!]

% \centering
% \caption{ Estimated NLSD test on Aluminum price series with critical values at $5\%$ significance level, and Ljung-Box test of series in levels and squares}
% \begin{tabular}{|cc|cc|cc|}
% \hline
%                                 NLSD test       &  $\chi^2_{0.95}(16)$           & LB($y_t$)           & $\chi^2_{0.95}(20)$  & LB($y^2_t$)           & $\chi^2_{0.95}(20)$      \\ \hline
%     $1391.40$ & $21.03$ & $2228.4$ & $31.41$ & $1155.8$ & $31.41$ \\ \hline
% \end{tabular}
% \label{NLSD TEST}
% \end{table}

 We apply the GCov specification test to each model to assess the fit. The test statistics and the associated critical values are given in column 3 of Table \ref{applications}. We find that the GCov specification test does not reject the null hypothesis of correct specification of the MAR(1,1) model, indicating the absence of (non)linear serial dependence in the residuals. The in-sample fitted values of the MAR(1,1) model are shown in
Figure \ref{2-b}. Figure \ref{figure 5} (b) in Appendix C.5 illustrates the non-Gaussian distribution of the residuals. The ACFs of the residuals and their squares given in Figures \ref{3-c} and \ref{3-d} \textcolor{black}{of Appendix C.4} are not statistically significant, which confirms the results of the GCov specification test.  Furthermore, we plot the fitted causal and noncausal components of MAR(1,1), defined in equation (6.2), $\hat{v}_{1,t} =  (1-\hat{\psi} L^{-1})  y_t$ and $\hat{v}_{2,t}  = (1- \hat{\phi} L)  y_t$ in Figure \ref{2-d} \textcolor{black}{ of Appendix C.5}. 
%The components $v_{1,t}, v_{2,t}$ are discussed in Section 6.1.1 and defined in equation (6.2), with $v_{2,t}$ capturing the locally explosive patterns.

\begin{table}[ht!]

\centering
\caption{ GCov estimated parameters, GCov specification test with $\chi^2$ critical values at $5\%$ significance level, and roots of $\hat{\Phi}(L^{-1})$  and $\hat{\Psi}(L)$ }

\begin{tabular}{|c|ccc|cc|ccc|}
\hline
                  & $\phi_1$          & $\psi_1$  & $\psi_2$            & test statistic       & $\chi^2_{0.95}$          & $L_1^{\phi}$             & $L_1^{\psi}$             & $L_2^{\psi}$     \\ \hline
MAR(0,1)          &          &          $0.93^*$  &                    & 57.53         & 49.80          & 1.07          &                   &           \\ \hline
\textbf{MAR(1,1)} & \textbf{0.41}$^*$ & \textbf{0.87}$^*$ &   & \textbf{22.32} & \textbf{48.60} & \textbf{2.43} & \textbf{1.14} & \textbf{} \\ \hline
MAR(1,2)       & $0.62^*$   & $0.76^*$          & 0.02  & 22.60        & 47.4      & 1.60 & 1.27   & -36.93               \\ \hline
\end{tabular}
\label{applications}
\caption*{* indicates statistical significance at 5\%}
\end{table}

In addition, we estimate the MAR(1,1) model using the AML estimator with a log-likelihood function based on the fitted t-student error distribution with 3.9 degrees of freedom\footnote{The degrees of freedom are an additional AML parameter estimated in this model.} and apply the GCov bootstrap test. \textcolor{black}{Since AML suffers from a bi-modality issue [Hecq, Lieb, and Telg  (2016) and  Bec, Nielsen, and Sa\"idi (2020)], we employ a grid search of initial values as suggested in Bec, Nielsen, and Sa\"idi (2020)
%. For the sake of computational time, we only apply the grid search on initial values of
for parameters $\phi$ and $\psi$ with a grid size of $0.01$ between 0 and 1. The AML estimates are $\hat{\phi}=0.42$ and $\hat{\psi}=0.90$. The GCov bootstrap test does not reject the null hypothesis of i.i.d. errors, based on comparing the value of the test statistic $42.30$ with the critical value of $57.78$.}

% \begin{table}[ht!]

% \centering
% \caption{ AML estimated parameters, GCov bootstrap test with critical values at $5\%$ significance level, and Ljung-Box test of residuals and residuals square}
% \begin{tabular}{|c|cc|cc|cc|cc|}
% \hline
%                   & $\phi$          & $\psi$              & bootstrap test       &  CV           & LB($\hat{\epsilon}_t$)           & $\chi^2_{0.95}(20)$  & LB($\hat{\epsilon}^2_t$)           & $\chi^2_{0.95}(20)$      \\ \hline
% MAR(1,1) & $0.42^*$ & $0.90^*$ &    $42.30$ & $57.78$ & $13.14$&  $31.41$ & $26.06$ & $31.41$ \\ \hline
% \end{tabular}
% \label{applications-AML}
% \caption*{* indicates statistical significance at 5\%}
% \end{table}

 \textcolor{black}{The parameters of MAR(1,1) estimated by the GCov and AML estimators are very close. In both cases, we do not reject the null of correct specification, based on the GCov specification test (Table \ref{applications}) and the GCov bootstrap test, respectively. By comparing the approaches used to estimate the parameters, we can argue that the GCov method provides an estimator and a test with a known asymptotic distribution in one step, while the AML estimator needs great attention to the initial values of optimization and grid search. Moreover, the GCov bootstrap test replaces two separate Liung-Box tests with one testing procedure of a joint hypothesis.}

% By comparing Tables \ref{applications} and \ref{applications-AML}, we find that the AML estimate of the causal coefficient of the MAR(1,1) process is closer to the unit root. Based on the GCov bootstrap test results, we reject the null hypothesis of i.i.d. errors in the AML estimated model. However, the Ljung-Box test does not reject the absence of dependence in residuals at the level of 5\% and can only detect the existence of dependence in the squared residuals. The AML results could suffer from misspecification of the parametric likelihood function, as we see from the portmanteau test results in Tables \ref{applications} and \ref{applications-AML} that the GCov-estimated model has a satisfactory fit, while the model based on the AML does not. Moreover, the GCov bootstrap test is advantageous, compared to the Ljung-Box test, since it rejects the null hypothesis of i.i.d. errors in one step.

\section{Conclusion}

This paper examined nonlinear serial dependence testing in non-Gaussian time series and specification testing in semi-parametric models with non-Gaussian i.i.d. errors. We introduced a new test of the null hypothesis of the absence of (non)linear dependence in time series, called the NLSD test. We studied analytically the asymptotic properties of the GCov specification test under the local alternative hypotheses to provide convincing empirical evidence of its potential as a widely applicable diagnostic tool for testing the goodness of fit.
%of semi-parametric dynamic models with i.i.d. non-Gaussian errors. 
We showed that under the local alternatives, the GCov specification test statistic converges to a non-central chi-square distribution with a non-centrality parameter depending on the direction of the alternative.
We also 
%described the local asymptotic power of the GCov specification test and
compared the GCov specification test with a concentrated and extended CUGMM-based overidentification test.

For finite sample size adjustment of GCov specification tests in dynamic non-Gaussian models estimated by GCov or maximum likelihood, we introduced the GCov bootstrap test and provided regularity conditions ensuring its validity.

We also explored how the asymptotic performance of the GCov specification test can be improved by increasing the set of nonlinear autocovariance conditions. We showed that the nonlinear transformations that help identify the unknown error distribution are power functions weighted by decreasing exponential functions of errors. The GCov estimator based on that enlarged set of nonlinear transformations is asymptotically more efficient, which increases the set of alternatives against which the GCov test has asymptotic power of 1. In addition, by appropriately choosing the nonlinear transformations and orthonormalizing them in the Hilbert space, we obtained an optimally weighted test statistic with an asymptotic normal distribution.

%We examined the finite sample performance of the GCov specification test analytically under a sequence of local alternatives to study the hypotheses about the parameters of a semi-parametric model. 

We examined through simulations the finite sample properties of the 
 GCov specification, bootstrap and NLSD tests under the fixed and local alternative hypotheses. Our study shows that these tests perform well in detecting nonlinear serial dependence in mixed causal-noncausal processes 
and residuals of nonlinear models. 
%The GCov specification test successfully detects the correct fit in models estimated by  maximum likelihood estimators. 
%and is a reliable diagnostic tool.  
For illustration, we applied the NLSD test to the aluminum price and used the GCov specification and bootstrap tests to select the optimal fit of a causal-noncausal MAR model.

\textbf{Acknowledgements}

The authors thank A. Djogbenou, C. Gourieroux, V. Zinde-Walsh, and the participants of the Canadian Economic Association (CEA) 2023 meetings, 2023 European Winter Meeting of the Econometric Society, \textcolor{black}{ Canadian Econometrics Study Group (CESG) 2024 meeting, CFE-CMStatistics 2024 Conference, Statistics Society of Canada  2025 Annual Meeting, 2025 Annual Conference of the International Association for Applied Econometrics (IAAE), and the workshop in honor of James Mackinnon at Queen's University for their helpful comments. Moreover, we thank the Associate Editor and two anonymous referees for their helpful comments and suggestions.}

\textbf{Funding}

This work was financially supported by the Natural Sciences and Engineering Research Council of Canada (NSERC).

\bigskip

\begin{center}
REFERENCES
\end{center}

\color{black}

\small{

\nin Anderson, T. (1999): "Asymptotic Theory for Canonical Correlation Analysis", Journal of Multivariate Analysis, 70, 1-29.

\nin Anderson, T. (2002): "Canonical Correlation Analysis and Reduced Rank Regression in Autoregressive Models", Annals of Statistics, 30, 1134-1154.

\nin Andrews, D. (1988): "Laws of Large Numbers for Dependent Non-Identically Distributed Random Variables", Econometric Theory, 4, 458-467.

%\medskip
%\nin Andrews, D. (1992): "Generic Uniform Convergence", Econometric Theory, 8, 241-257. 

%\medskip
%\nin Bartlett, M. (1947): "The General Canonical Correlation Distribution", Annals of Mathematical Statistics, 18, 1-17. 

\nin \textcolor{black}{Bec, F., Nielsen, H. B., and S.Sa\"idi (2020): "Mixed Causal-Noncausal Autoregressions: Bimodality Issues in Estimation and Unit Root Testing", Oxford Bulletin of Economics and Statistics, 82(6), 1413-1428.}

\nin \textcolor{black}{Bickel, P. (1982): "On Adaptive Estimation", Annals of Statistics, 10, 647-671.}

\nin \textcolor{black}{Bierens, H. (1990): "A Consistent Conditional Moment Test of Functional Form", Econometrica, Vol 58, 1443-1458.}

\nin Box, G. and D. Pierce (1970): "Distribution of Residual Autocorrelations in Autoregressive-Integrated Moving Average Time Series Models", JASA, 65, 1509-1526.

%\medskip
%\nin Brown, B. (1971): "Martingale Central Limit Theorem", Annals of Mathematical Statistics, 42, 59-66

% \color{black}
% \medskip
% \nin Brown, B., and W. Newey (2002): "Generalized Method of Moments, Efficient Bootstrapping and Improved Inference", Journal of Business and Economic Statistics, 20, 507-513.

\color{black}

\nin Bryan, K. (2006) "Elementary Inversion of the Laplace Transform", Research Paper, Department of Mathematics, Rose-Hulman Institute of Technology.

%\medskip
%\nin Boudjellaba, H., Dufour, J.M. and R. Roy (1994): "Simplified Conditions for Noncausality Between Vectors in Multivariate ARMA Models", Journal of Econometrics, 63, 271-287.

%\color{black}
%\medskip
\nin Cavaliere, G., Nielsen, H, and A. Rahbek (2020): "Bootstrapping Noncausal Autoregressions with Applications to Explosive Bubble Modelling", Journal of Business and Economic Statistics, 38, 55-67.
\color{black}

\nin Chan, K., Ho, L., and H. Tong (2006): "A Note on Time-Reversibility of Multivariate Linear
Processes", Biometrika 93, 221-227.

\nin Chen, J., Gao, Z., and R. Tsay (2025): "Forward Variable Selection in Ultra-High Dimensional Linear Regression Using Gram-Schmidt Orthogonalization", working paper, University of Chicago.

\nin Chu, B. (2023): "A Distance-Based Test of Independence Between Two Multivariate Time Series", Journal of Multivariate Analysis, 195, 105-151.

% \medskip
% \nin Chitturi, R. (1974): "Distribution of Residual Autocorrelations in Multiple Autoregressive Schemes", JASA, 69, 928-934.

\nin Chitturi, R. (1976): "Distribution of Multivariate White Noise Autocorrelations", JASA, 71, 223-226.

% \medskip
% \nin Cubbada, G. and A. Hecq (2011): "Testing for  Common Autocorrelation in Data-Rich Environments", Journal of Forecasting, 30, 325-335.

\color{black}

\nin Davidson, R., and J. MacKinnon (2006): "The Power of Bootstrap and Asymptotic Test", Journal of Econometrics, 133, 421-441.
\color{black}

\nin Davis, R., Matsui, M., Thomas Mikosch, T. and P. Wan (2018): "Applications of Distance Correlation to Time Series", Bernoulli 24(4A), 3087-3116.

\nin Davis, R. and L. Song (2020): "Noncausal Vector AR Processes with Application to Economic Time Series", Journal of Econometrics, 216, 246-267.

% \medskip
% \nin Davis, R and P. Wan (2022): "Goodness of Fit Testing for Time Series Models via Distance Covariance", Journal of Econometrics, 227, 4-24.

% \medskip
% \nin Davis, R and P. Wu (1997): "Bootstrapping M-Estimates in Regressions and Autoregressions with Infinite Variances, Statistica Sinica, 7, 1135-1156.

% \medskip
% \nin De Jong, R. (1995): "Laws of Large Numbers for Dependent Heterogeneous Processes", Econometric Theory, 11, 347-358.

% \medskip
% \nin De Jong, R. (1998): "Weak Laws of Large Numbers for Dependent Random Variables", Annales d'Economie et de Statistique, 51, 210-225.

\nin De Gooijer, J.G. (2023): "On Portmanteau-Type Tests for Nonlinear Multivariate Time Series", Journal of Multivariate Analysis, 195, 105-157.

\color{black}

\nin Dovonon, P., and S. Goncalves (2017): "Bootstrapping the GMM Overidentification Test Under First-Order Underidentification," Journal of Econometrics, 201, 43-71.
\color{black}

\nin \textcolor{black}{ Dovonon, P., Hall, A., and F. Kleibergen (2020): "Inference in Second-Order Identified Models", Journal of Econometrics, 218, 346-372 }

% \medskip
% \nin \textcolor{black}{ Dovonon, P. and N. Gospodinov (2024): "Specification Testing for Conditional Moment Restrictions Under Local Identification Failure",  Quantitative Economics, forthcoming}

%\medskip
%\nin Duchesne, P. and R. Roy (2003): "Robust Tests for Independence of Two Time Series", Statistica Sinica, 13, 827-852. 

%\medskip
%\nin Duchesne, P. and R. Roy (2004): "On Consistent Testing for Serial Correlation of Unknown Form in Vector Time Series Models" Journal of Multivariate Analysis, 89, 148-180. 

% \medskip
% \nin Dvoretski, A. (1970): "Asymptotic Normality for Sums of Dependent Random Variables", Proc. of the Sixth Berkeley Symposium on Math. Stat. and Prob., Vol II, 513-535.

%\medskip
%\nin El Himdi, H. and R. Roy (1997): "Tests for Noncorrelation of Two Multivariate ARMA Time Series", Canadian Journal of Statistics, 25, 233-256.

%\medskip
%\nin Engle, R., Lilien, D. and R. Robbins (1987): "Estimating Time Varying Risk Premia in the Term Structure: The ARCH-M Models", Econometrica, 55, 391-407.

\color{black}

\nin Escanciano, M. (2007): "Model Checks Using Residual Marked Empirical Processes", Statistica Sinica, 17, 115-138
\color{black}

\nin \textcolor{black}{ Feller, W. (1971): "Probability Theory and Its Applications", Vol 3, 2nd Ed, Wiley, Chapter 13: "Laplace Transforms, Tauberian Theorems, Resolvents".}

\nin \textcolor{black}{ Fokianos, K. and M. Pitsillou (2018): "Testing Independence for Multivariate Time Series via the Autodistance Correlation Matrix", Biometrika, 105, 337-352.}

% \medskip
% \nin Francq, C., Roy, R. and J.M. Zakoian (2005): "Diagnostic Checking in ARMA Models with Uncorrelated Errors", JASA, 100, 532-544.

% \medskip
% \nin Francq, C., and J.M. Zakoian (2005): " A Central Limit Theorem for Mixing Triangular Arrays of Variables Whose Dependence is Allowed to Grow with the Sample Size", Econometric Theory, 21, 1165-1171.

\nin Fries, S., and  J. M. Zakoian (2019):" Mixed Causal-Noncausal AR Processes and the Modelling of Explosive Bubbles", Econometric Theory, 35(6), 1234-1270.

%\medskip
%\nin Geweke, J. (1981): "The Approximate Slopes of Econometric Tests", Econometrica, 49, 1427-1442.

% \medskip
% \nin \textcolor{black}{Giancaterini, F., Hecq, A., Jasiak, J., \& Neyazi, A. M. (2025). "Regularized generalized covariance (rgcov) estimator", arXiv preprint arXiv:2504.18678.}

\nin Giancaterini F, Hecq, A., Jasiak, J. and A. Manafi Neyazi (2024): "Regularized Generalized Covariance (RGCov) Estimator", ArXiv 6390395.

\nin  Gourieroux, C. and J. Jasiak (2001): "Nonlinear Autocorrelograms; an Application to Inter-Trade Durations", Journal of Time Series Analysis, Vol. 23, No 2, 1-28

\nin  Gourieroux, C. and J. Jasiak (2016): "Filtering, Prediction and Simulation Methods for Noncausal Processes",  Journal of Time Series Analysis, Vol 37, 405-430

\nin Gourieroux, C. and J. Jasiak (2017): "Noncausal Vector Autoregressive Process: Representation, Identification and Semi-Parametric Estimation", Journal of Econometrics, 200, 118-134.

%\medskip
%\nin Gourieroux, C. and J. Jasiak (2018): "Misspecification of Noncausal Order in Autoregressive Processes", Journal of Econometrics, 205(1), 226-248

%\medskip
%\nin Gourieroux, C. and J. Jasiak (2019): "Robust Analysis of the Martingale Hypothesis", Econometrics and Statistics, Vol 9, 17-41 

\nin Gourieroux, C. and J. Jasiak (2023): "Generalized Covariance Estimator", Journal of Business and Economic Statistics, 41, 1315-1327.

% \medskip
% \nin Gourieroux, C. and A. Monfort (1995): " Statistics and Econometric Models", Vol 2, Cambridge Univ. Press.

\nin Gourieroux, C., A. Monfort, and J.P. Renne (2017): "Statistical Inference for Independent
Component Analysis: Application to Structural VAR Models", Journal of Econometrics 196, 111-126.

\nin  Gourieroux, C., A. Monfort, and J.P. Renne (2018): "Identification and Estimation in Non-
Fundamental Structural Models", Review of Economic Studies, 2020, 87 (4), pp.1915-1953.

\nin  Gourieroux, C. and J.M. Zakoian (2017): "Local Explosion Modelling by Non-causal Process", Journal of the Royal Statistical Society, Series B, 79,  737-756.

% \color{black}
% \medskip
% \nin Hahn, J. (1996): "A Note on Bootstrapping Generalized Method of Moments Estimators", Econometric Theory, 12, 187-197.
% \color{black}

% \color{black}
% \medskip
% \nin Hall, P., and J. Horowitz (1996): "Bootstrap Critical Values for Tests Based on Generalized Method of Moments Estimators", Econometrica, 64, 831-916.
% \color{black}

% \medskip
% \nin \textcolor{black}{Han, C. and P.C.B. Phillips (2006): "GMM with Many Moment Conditions", Econometrics, 74, 147-192.}

% \medskip
% \nin Hannan, J. (1967): "Canonical Correlation and Multiple Equation Systems in Economics", Econometrica, 35, 123-138.

% \medskip
% \nin Hannan, J. (1976): "The Asymptotic Distribution of Serial Covariances", Annals of Statistics, 4, 396-399. 

% \medskip
% \nin Haugh, L. (1976): "Checking the Independence of Two Covariance-Stationary Time Series: A Univariate Residual Cross-Correlation Approach", JASA, 71, 378-385.

% \medskip
% \nin Hecq, A., Issler, J. V., and S. Telg (2020): "Mixed Causal-Noncausal Autoregressions with Exogenous Regressors", Journal of Applied Econometrics, 35(3), 328-343.

\nin  Guay, A. (2021): "Identification of Structural Vector Autoregressions through Higher Unconditional Moments", Journal of Econometrics, 225, 27-46.

\nin Hecq, A., Lieb, L., and S. Telg (2016): "Identification of Mixed Causal-Noncausal Models in Finite Samples", Annals of Economics and Statistics, (123/124), 307-331.

% \medskip
% \nin Hecq, A., and L. Sun (2020): "Selecting Between Causal and Noncausal Models with Quantile Autoregressions", Studies in Nonlinear Dynamics and Econometrics, 25(5), 393-416.

\nin Hecq, A. and E. Voisin (2021): "Forecasting Bubbles with Mixed Causal-Noncausal Autoregressive Models", Econometrics and Statistics, 20, 29-45.

% \medskip
% \nin Hosking, J. (1980): "The Multivariate Portmanteau Statistic", JASA, 75, 602-608.

% \medskip
% \nin Hosking, J. (1981)a: "Equivalent Forms of the Multivariate Portmanteau Statistic", JRSS B, 43, 261-262.

% \medskip
% \nin Hosking, J. (1981)b: "Lagrange Multiplier Tests of Multivariate Time Series Models", JRSS B, 43, 219-230. 

\nin Hotelling, H. (1936): "Relation Between Two Sets of Variants", Biometrika, 28, 321-377.

% \medskip
% \nin Inoue, A. and M. Shintani (2006): Bootstrapping GMM Estimators for Time Series", Journal of Econometrics, 133, 531-555.
%\medskip
%\nin Hurn, R. and C. Johnson (1999): "Topics in Matrix Analysis", Cambridge University Press

% \medskip
% \nin Jennrich, R. (1969): "Asymptotic Properties of Nonlinear Least Squares", Annals of Mathematical Statistics, 40, 633-643.

\nin  Keweloh, S. (2020): "A Generalized Method of Moments Estimator for Structural Autoregressions Based on Higher Moments", Journal of Business \& Economic Statistics, 39(3), 1-29.

\nin \textcolor{black}{Koenker, R. and J.A.F. Machado (1999) "GMM Inference when the Number of Moment Conditions is Large", Journal of Econometrics, 93, 327-344}

% \medskip
% \nin Kundu, S., Majumdar, S., and K. Mukherjee (2000): "Central Limit Theorems Revisited", Statistics and Probability Letters, 47, 265-275.

\nin  Lanne, M., and J. Luoto (2019), "GMM Estimation of Non-Gaussian Structural Vector Autoregression", Journal of Business \& Economic Statistics, 69-81

\nin Lanne, M., M. Meitz, and P. Saikkonen (2017), "Identification and Estimation of Non-Gaussian
Structural Vector Autoregressions", Journal of Econometrics 196, pp.288-304.

\nin Lanne, M., and P. Saikkonen, (2011): "Noncausal Autoregressions for Economic Time Series", Journal of Time Series Econometrics, 3(3).

\nin Lanne, M., and P. Saikkonen, (2013): "Noncausal Vector Autoregression", Econometric Theory, 29(3), 447-481.

% \medskip
% \nin Leucht, A., and M. Neumann (2003): " Consistency of General Bootstrap Methods for degenerate U-Type and V-Type Statistics", Journal of Multivariate Analysis, 100, 1622-1633.

% \color{black}
% \medskip
% \nin Li, W. and A. McLeod (1981): "Distribution of the Residual Autocorrelations in Multivariate ARMA Time Series Models", JRSS B, 43, 231-233. 

% \color{black}
% \medskip
% \nin Li, M. and Y. Zhang (2022): "Bootstrapping Multivariate Portmanteau Tests for Vector Autoregressive Models with Weak Assumptions on Errors", Computational Statistics and Data Analysis, 165, 107321.
% \color{black}

\nin Ling, S. (2007):  "A Double AR(p) Model: Structure and Estimation", Statistica Sinica, 17,161-175.

%\medskip
%\nin Magnus, J. and H. Neudecker (2019): " Matrix Differential Calculus with Applications in Statistics and Econometrics", Wiley.

% \medskip
% \nin Mahdi, E., and A. McLeod (2012): "Improved Multivariate Portmanteau Test", Journal of Time Series Analysis, 33(2), 211-222.

%\medskip
%\nin Mahdi, E., and Fisher, T. J. (2022): "Bootstrapping a powerful mixed portmanteau test for time series", Journal of Applied Statistics, 1-26.

\nin \textcolor{black}{Newey, W. (1991): "Uniform Convergence in Probability and Stochastic Equicontinuity", Econometrica, 59, 1161-1167.}

\nin \textcolor{black}{Newey, W. (1988): "Adaptive Estimation of Regression Models Via Moment Restrictions", Journal of Econometrics, 38, 301-334.}

\nin \textcolor{black}{Post, E. (1930): "Generalized Differentiation", Transactions of the American Mathematical Society, 32, 723-781.}

\nin Robinson, P. (1973): " Generalized Canonical Analysis for Time Series", Journal of Multivariate  Analysis, 3, 141-160.

%\medskip
%\nin \textcolor{black}{Royden, H.L., and P.M. Fitzpatrick (2010): "Real Analysis", 4th ed., Prentice Hall}

\nin 
Swensen, A. (2022): "On Causal and Non-Causal Cointegrated Vector Autoregressive Time Series", Journal of Time Series Analysis, 43(2), 178-196.

% \medskip
% \nin Szroeter, J. (1983): "Generalized Wald Methods for Testing Nonlinear Implicit and Overidentifying Restrictions", Econometrica, 51, 335-353.

\nin Velasco, C. and I. Lobato (2018): "Frequency Domain Minimum Distance Inference for Possibly Noninvertible and Noncausal ARMA Models", Annals of Statistics, 46, 555-579.

\nin Wan, P., and R. Davis (2022): "Goodness-of-Fit Testing for Time Series Models via Distance Covariance", Journal of Econometrics, 227(1), 4-24.

\nin Wooldridge, J. and H. White (1988): "Some Invariance Principles and Central Limit Theorems for Dependent Heterogeneous Processes", Econometric Theory, 4, 210- 230. 
}

\newpage

\setlength{\baselineskip}{.26in}
\thispagestyle{empty}
\vspace*{0cm}
\begin{center}

\setlength{\baselineskip}{.32in}
{\bbf GCov-Based Portmanteau Test \\ Online Appendices}

\vspace{0.4in}

\vspace{0.4in}

\large{Joann Jasiak}\footnote{York University, 
e-mail:{\it jasiakj@yorku.ca}},
\large{and Aryan Manafi Neyazi}\footnote{York University, e-mail: {\it aryanm@yorku.ca}.\\}

\setlength{\baselineskip}{.26in}
\vspace{0.8in}

\today\\
\end{center}
\medskip
\begin{center}
Abstract \\
\end{center}

This supplementary material to "GCov-Based Portmanteau Test" contains the appendices and is organized as follows. Appendices A and B present the theoretical results and regularity conditions.
Appendix C provides additional results based on simulations and an empirical application of the GCov specification test to the causal-noncausal model of aluminum prices.

\renewcommand{\thefootnote}{\arabic{footnote}}

\newpage
\setcounter{equation}{0}\def\theequation{A.\arabic{equation}}
\begin{center}
\textbf{Appendix A}
\end{center}

\nin The following notation is used:

$m$ - dimension of $Y_t$

$u_t$ is of dimension $J = dim(g)$ 

K - dimension of transformations $a$

$a(u_t) = g_a$ is of dimension $K$

$\Gamma$ or $(\Gamma^a$) is of dimension $K \times K$  

$dim(\theta)$ - dimension of $\theta$

$dim(\gamma)=1$,  $\gamma$ is a scalar

$Id$ is the Identity matrix 

\medskip

\begin{center}
{\bf Asymptotic Behavior of the Portmanteau Statistic Under the Independence Hypothesis} 
\end{center}

\nin This Appendix reviews the results that already exist in the literature and are used in the proofs of new results in Appendix B.

\nin {\bf A.1 Asymptotic Behavior of Sample Autoregressive Coefficients }

\nin Suppose that process ($Y_t$) is strictly stationary and follows a VAR(1) model:

\begin{equation}
Y_t =  \alpha + B Y_{t-1} + u_t,
\end{equation}

\nin where $u_t$ is a square integrable strong white noise, $E(u_t)= 0$, $V(u_t) = \Sigma$, where $\Sigma$ is invertible and the coefficient matrix $B$ has no eigenvalues of modulus 1. This VAR model is a SUR model with identical regressors $X_t = Y_{t-1}$ in all equations. In this case, the OLS estimators applied equation by equation are equal to the GLS estimator
of $B$ \footnote{In this Appendix, the index T of the estimators is omitted to simplify the notation.}. The estimator
$\hat{B}' =  \hat{\Gamma}(0)^{-1} \hat{\Gamma}(1)' $ is asymptotically
normally distributed:

$$
\sqrt{T} [ vec (\hat{B}') - vec B'] \approx N[0, \Sigma \otimes \Gamma(0)^{-1}].
$$

\nin Under the null hypothesis: $H_0 = ( \Gamma(1) = 0) = (B=0)$, we have $\Sigma= \Gamma(0)$ and

$$
\sqrt{T} vec (B') \sim N(0, \Gamma(0) \otimes [\Gamma(0)^{-1}]).
$$

\nin where the $\otimes$ denotes the Kronecker product [see Chitturi (1974), eq. (1.13)]. 

\nin {\bf A.2 Portmanteau Statistic as a Lagrange Multiplier test}

\nin It can be shown that the Lagrange Multiplier test statistic
%\footnote{This is a Lagrange Multiplier test statistic as the asymptotic covariance matrix of $vec(\hat{B}')$ is estimated under the null hypothesis [see Hosking (1981)].} 
for testing $H_0 = (\Gamma(1) =0) = (B=0)$ is:

%\begin{eqnarray*}
%\hat{\xi}_T (1) & = & T vec [\hat{\Gamma}(0)^{-1} \hat{\Gamma}(1)']' [\hat{\Gamma}(0)^{-1} \otimes \hat{\Gamma}(0)] vec [\hat{\Gamma}(0)^{-1} \hat{\Gamma}(1)'] \\
%& = & T vec [\hat{\Gamma}(0)^{-1} \hat{\Gamma}(1)']' [\hat{\Gamma}(0)^{-1/2} \otimes \hat{\Gamma}(0)^{1/2}]
%[\hat{\Gamma}(0)^{-1/2} \otimes \hat{\Gamma}(0)^{1/2}]  vec [\hat{\Gamma}(0)^{-1} \hat{\Gamma}(1)']\\
%& = & T vec [\hat{\Gamma}(0)^{-1/2} \hat{\Gamma}(1)'\hat{\Gamma}(0)^{-1/2} ]'vec [\hat{\Gamma}(0)^{-1/2} \hat{\Gamma}(1)'\hat{\Gamma}(0)^{-1/2} ], 
%\end{eqnarray*}

%\nin by using the equality: $vec(ABC) = (C'\otimes A) vec B$ [see Lemma 4.3.1 in Horn, Johnson (1999) or Magnus, Neudecker (2019), ch. 18, p. 440-441]. Moreover, we have $[vec C]'[vec C] = Tr \;CC'$. Therefore,

%\begin{eqnarray}
%\label{2.9}
%\hat{\xi}_T(1) & = & T \;Tr [\hat{\Gamma}(0)^{-1/2} \hat{\Gamma}(1)'\hat{\Gamma}(0)^{-1} \hat{\Gamma}(1) \hat{\Gamma}(0)^{-1/2}] \nonumber\\
%& = & T \;Tr [ \hat{\Gamma}(1)' \hat{\Gamma}(0)^{-1} \hat{\Gamma}(1) \hat{\Gamma}(0)^{-1}] \nonumber\\
%& = & T \;Tr \hat{R}^2(1). 
%\end{eqnarray}

\begin{equation}
\label{A.2}
\hat{\xi}_T(1) = T \;Tr [ \hat{\Gamma}(1)' \hat{\Gamma}(0)^{-1} \hat{\Gamma}(1) \hat{\Gamma}(0)^{-1}]  = T \;Tr \hat{R}^2(1). 
\end{equation}

\nin [See, e.g., Gourieroux and Jasiak (2023), Supplementary Material].

\nin {\bf A.3 Asymptotic Behavior of Sample Autocovariance}

\nin The asymptotic distribution of $\sqrt{T} vec[\hat{\Gamma}(1)'- \Gamma(1)']$ for model (A.1) is given in Gourieroux and Jasiak (2023), Supplementary Material [see also Chitturi (1976), Hannan (1976)]. We have:

$$\sqrt{T} [\hat{\Gamma}(1)'- \Gamma(1)'] = \hat{\Gamma}(0) \sqrt{T} [\hat{B}'-B'] = \Gamma(0) \sqrt{T} [\hat{B}'-B'] + o_p(1), $$

\nin and 
 $$  vec[\sqrt{T} [\hat{\Gamma}(1)'- \Gamma(1)']] = vec [ \Gamma(0) \sqrt{T} [\hat{B}' -B']] = [Id \otimes \Gamma(0)] vec (\sqrt{T} [\hat{B}'-B'] )  + o_p(1).$$

\nin Under the null hypothesis  $H_0: = ( \Gamma(1) = 0) = (B=0)$ of independently and identically distributed (i.i.d.) process $(Y_t)$ with finite fourth order moment:

\begin{eqnarray*}
 vec[\sqrt{T} [\hat{\Gamma}(1)'- \Gamma(1)']] & \sim & N[ 0, [Id \otimes \Gamma(0)] [ \Gamma(0) \otimes \Gamma(0)^{-1}] [Id \otimes \Gamma(0)]] \\
& = & N[ 0, \Gamma(0) \otimes \Gamma(0)].
\end{eqnarray*}

\nin It follows that under this null hypothesis, the statistic 
(\ref{A.2}) 
has asymptotically a chi-square distribution $\chi^2(K^2)$, where $K=m$ is the dimension of $(Y_t)$.

\bigskip
\nin {\bf A.4 Statistic Based on Several Autocovariances}

The interpretation as a SUR regression can be extended to any lag H. Then, under the stationarity assumption, the VAR model becomes: 

\begin{equation}
 Y_t = \alpha + B_1 Y_{t-1} + \cdots + B_H Y_{t-H} + u_t,   
\end{equation}

\nin where $(u_t)$ is a square integrable strong white noise and the companion matrix of autoregressive coefficients has no eigenvalues of modulus 1. Under the null hypothesis of the independence of $Y_t$, or equivalently under $H_0= \{ B_1 = \cdots =B_H =0 \}$, the explanatory variables are orthogonal, and the OLS estimators of $B_1,...,B_H$ are such that $\hat{B}_h$ coincides with the OLS estimator in the simple SUR model $Y_t = \alpha_h + B_h Y_{t-h} +v_t$. It follows that, under this null hypothesis, the estimators $\sqrt{T} \hat{B}_h, \; h=1,...,H$ are independent, normally distributed with the same distribution $N(0, \Gamma(0) \otimes \Gamma(0))$.
Then, the test statistics: 
\begin{equation}
  \hat{\xi}_T (H) \approx T \Sum_{h=1}^H vec[\sqrt{T} \hat{\Gamma}(h)']'[\hat{\Gamma}_0 (0)^{-1}\otimes \hat{\Gamma}_0 (0)^{-1}]  vec[\sqrt{T} \hat{\Gamma}(h)']
  \label{eq:3.8}
\end{equation}
\nin follows asymptotically the chi-square distribution $\chi^2(K^2H)$,  where $K=m$ is the dimension of $(Y_t)$.

\bigskip

\begin{center}
\textbf{Appendix B}

{\bf Asymptotic Distribution in the Semi-Parametric Framework}
\end{center}
\setcounter{equation}{0}\def\theequation{B.\arabic{equation}}

\nin This Appendix provides the regularity conditions and proofs of Propositions 1, 2, B1, and B2.

\nin {\bf B.1 The Law of Large Numbers (LLN) for Triangular Arrays}

As pointed out in Section 3.3.2, the proof of the consistency of estimated autocovariances and of the GCov estimator under the local alternatives is similar to the proof under the null hypothesis of independence. The difference is in the use of the LLN for empirical autocovariances of a triangular array of observations, uniform in $\theta$.

Below, we provide a sufficient set of regularity conditions.

\medskip

\nin {\bf Regularity Conditions for LLN uniform in $\theta$}. 

\nin {\it 1. Conditions on the true nonlinear dynamics}

 i) The observations satisfy the model:
 \begin{equation}
     g^*(\tilde{Y}_{T,t}; \theta_T, \gamma_T) = u_t, 
 \end{equation}

 \nin where the $u_t$'s are i.i.d. with pdf $f_0$.

 ii) The function $g^*$ is invertible with respect to $Y_{T,t}$; then we can write:

 \begin{eqnarray}
  Y_{T,t} = h(u_t, Y_{T, t-1},...,Y_{T, t-p}; \theta_T, \gamma_T).   
 \end{eqnarray}

iii) For each given $T$, ($Y_{T,t})$ with $t=1,...,T$, is a strictly stationary and ergodic solution of the autoregressive equation (B.2).

\nin {\it 2. Conditions on the parameters}

\nin Suppose that the parameter space is $\Theta \times C$, where $\theta \in \Theta \subset \mathbf{R}^{dim(\theta)}$ and $\gamma \in C \subset \mathbf{R}$. We assume that:

i) $\Theta$ and $C$ are compact sets with non-empty interiors. 
 
ii) $\theta_0$ is in the interior of $\Theta$ and 0 is in the interior of $C$. 

iii) $\theta_T = \theta_0 + \mu/\sqrt{T}, \; \gamma_T = \nu/\sqrt{T}$.

iv) The combined parameter vector $\theta,\gamma$ is identifiable in a neighborhood of $\theta_0, 0$.

\nin In particular, for $T$  sufficiently large, $\theta_T, \gamma_T$ are in the interior of $\Theta$ and $C$, respectively.

\medskip

\nin {\it 3. Regularity conditions on the function $g^*$}

i) The functions $g_k^*(y; \theta, \gamma), \; k=1,...,K$ are continuously differentiable on the interior $\Theta \times C$.

ii) Let us define: $G_k^*(\tilde{y}) = Max_{(\theta,\gamma) \in \Theta \times C} [g_k^* (\tilde{y}, \theta, \gamma)]^2$. 
We assume $E_0 G_k^*(\tilde{Y}) < \infty, \; k=1,...,K$ where $E_0$ denotes the expectation computed for the process $(\tilde{Y}_t)$ associated with the "asymptotic" parameter values $(\theta_0,0)$.

iii) Let us denote by $\mathcal{B}(\tilde{y})$ a uniform Lipschitz coefficient for functions
$g_j^*(\tilde{y}; \theta, \gamma), \; j=1,...,J$, $g_j^{*2}(\tilde{y}; \theta, \gamma), \; j=1,...,J$ and for $g_j^*(\tilde{y}; \theta, \gamma), \,g_j^*(\tilde{y}_{-h}; \theta, \gamma), \; j,k=1,...,J, h=1,...,H$. In this expression $\tilde{y}$ denotes the trajectory of the process and $\tilde{y}_{-h}$ denotes this trajectory lagged by $h$. It is assumed that
 $sup_T \frac{1}{T}
E|\mathcal{B}(Y_{T,t})] < \infty $ [Gourieroux and Jasiak (2023)], where the expectation is taken with respect to the distribution of process $Y_{T.} = (Y_{Tt}),$ with $t=1,...,T$.

\medskip

\nin {\it 4. Condition of Near Epoch Dependence [De Jong (1998)]}

The functions of the triangular array of random variables $Y_{T,t}, t \leq T, T \geq 1$ are $L_2-NED$ (near epoch dependent), i.e.,  for $\nu(r) \geq 0$ and $c_{Tt} \geq 0$ and for all $r \geq 0$ and $ t \geq 1$

\nin $ \sup_{\theta \in \Theta} E[g_j(\tilde{Y}_{T,t}, \theta) - E(g_j (\tilde{Y}_{T,t}, \theta)|Y_{T, t-r},...,Y_{T,t+r})]^2 \leq c_{T,t} \varphi(r)$

\nin $ \sup_{\theta \in \Theta} E[g_j^2(\tilde{Y}_{T,t}, \theta) - E(g_j^2 (\tilde{Y}_{T,t}, \theta)|Y_{T, t-r},...,Y_{T,t+r})]^2 \leq c_{T,t} \varphi(r)$

\nin $ \sup_{\theta \in \Theta} E[g_j(\tilde{Y}_{T,t}, \theta)g_k(\tilde{Y}_{T,t-h}, \theta) - E(g_j (\tilde{Y}_{T,t}, \theta)|Y_{T, t-r},...,Y_{T,t+r})E(g_k (\tilde{Y}_{T,t-h}, \theta)|Y_{T, t-r},...,Y_{T,t+r})]^2$
\nin $\leq c_{T,t} \varphi(r)$

\nin for all $j,k=1,...,K$, $h=1,...,H$, $\varphi(r) \rightarrow 0$ as $r \rightarrow \infty$ and 
$\lim \sup_{T \rightarrow \infty } \frac{1}{T} \sum_{t=1}^T c_{T,t} < \infty$.

From Assumption 4, it follows that the functions $g(Y_{Tt})$ are uniformly integrable mixingales [De Jong (1998)].
Because the NED condition implies a mixingale condition, the weak LLN of Theorem 2, Andrews (1988) can be applied. Then, Theorem 4 of Andrews (1992) implies the uniform weak LLN (U-WLLN).

\medskip
\nin To summarize, we get the following Proposition:

\bigskip
\nin {\bf Proposition B1:} 

Under the regularity conditions 1 to 4, we have:

$$plim_{T \rightarrow \infty } \hat{\Gamma}_T (h; \theta) = \Gamma_0 (h; \theta)$$

uniformly in $\theta \in \Theta$ for $h=1,...,H$, where $\Gamma_0(h; \theta)$ is evaluated at $\theta_0, \gamma_0$
and $f_0$ is the 
\indent true pdf of the error.

\medskip
\nin When the functions $g$ are distinguished from their transforms $g_a$, then conditions 1 and 2 concern $g$ and conditions 3 and 4 concern $g_a$.

\bigskip

\nin {\bf B.2 Central Limit Theorem (CLT) for Triangular Array}

We need to introduce additional regularity conditions to justify the expansion (3.13) and the asymptotic normality of the sample autocovariances $\sqrt{T} \hat{\Gamma}(h; \theta_T, \gamma_T, f_0)$ computed under a sequence of local alternatives. To obtain the corresponding CLT, we use the conditional Lindeberg-Feller conditions for martingale difference triangular array [Dvoretski (1970), Brown (1971)] extended to the multivariate case [Kundu et al. (2000), Th. 1.3]. To apply these conditions, we first define triangular filtration and the appropriate martingales. We denote by  ${\cal F}_{T,t}$
the information generated by the array $Y_{T ,\tau}, \; \tau \leq t$. Then, we consider the transformations $g_j^*(\tilde{Y}_{T,t}; \theta_0,0)$, $g_j^*(\tilde{Y}_{T,t}; \theta_0,0) g_k^*(\tilde{Y}_{T,t-h}; \theta_0,0), \; j,k =1,...,K, \, h=1,...,H$. They can be written as a vector $G^*(\tilde{Y}_{T,t}; \theta_0,0)$, say. Next, we transform this vector into a multivariate martingale difference array by considering:

$$X_{T,t} =  \frac{1}{\sqrt{T}} \{G(\tilde{Y}_{T,t}; \theta_0,0) - E_0 [G (\tilde{Y}_{T,t}; \theta_0,0) |  {\cal F}_{T,t-1} ] \}.$$

\nin The additional regularity conditions are the following:

\nin {\bf Regularity Conditions for the CLT}

i) The multivariate martingale difference array $X_{T,t}$ has finite second-order moments.

ii) For any vector $b$ of the same dimension as $X_{T,t}$, there exists a matrix $\Omega$ such that:

$$\sum_{t=1}^T E [ (b'X_{T,t})^2 | {\cal F}_{T,t-1}] \stackrel {P} {\rightarrow}  b' \Omega b. $$

iii) Conditional Lindeberg-Feller condition:

$$\sum_{t=1}^T E \{ (b'X_{T,t})^2 \mathbf{1}_{|b'X_{T,t}| >\epsilon }| {\cal F}_{T,t-1} \} \stackrel {P} {\rightarrow} 0, \; \mbox{for any } b \; \mbox{and} \; \epsilon >0. $$

\nin These regularity conditions ensure that the sum $S_T = \sum_{t=1}^T X_{T,t}$ tends in distribution to the multivariate Gaussian distribution $N(0, \Omega)$. Then we get the asymptotic normality of the estimated autocovariances under the sequence of local alternatives by applying the Slutsky Theorem.

\medskip
\nin {\bf Proposition B2:} 
Under the sequence of local alternatives and the regularity conditions 1-5, the vectors
$vec[ \sqrt{T} \hat{\Gamma}_T (h; \theta_T, \gamma_T, f_0)]$ are asymptotically independent, normally distributed with mean
$\Delta(h; \theta_0, f_0, \mu, \nu)$ defined in (3.10) and variance-covariance matrix $\Gamma_0(0, \theta_0) \otimes \Gamma_0(0, \theta_0)$.

\medskip
Thus, the behavior of the estimated autocovariances differs from the behavior under the null by the presence of the asymptotic bias measured by $\Delta(h; \theta_0, f_0, \mu, \nu)$.

We have introduced a set of regularity conditions to derive the asymptotic behavior of the estimated autocovariances. 
Let us now explain why this set of conditions is also sufficient to derive the asymptotic behavior of the GCov estimator and of the GCov specification test statistic. First, we review the standard expansions under the null hypothesis. Next, we derive their analogues under the sequence of local alternatives, before applying the CLT to the estimated autocovariances of a triangular array.

\bigskip

\nin {\bf B.3 First-order Expansion of the GCov Estimator under the Null Hypothesis}

\nin Below we recall the results under the null hypothesis derived in Gourieroux and Jasiak (2023). Let us consider $H=1$ for ease of exposition. The first-order conditions of the  GCov estimator are

$$
\frac{\partial Tr \hat{R}^2(1;\theta_j)}{\partial \theta_j}=0 , \;\;\; j=1,...,J=dim(\theta),
$$

\nin Let us define:
$$
A(\theta_0)= 2 \frac{\partial vec \Gamma(1; \theta_0)'}{\partial \theta} [\Gamma(0;\theta_0)^{-1}\otimes \Gamma(0;\theta_0)^{-1}],
$$
and
$$
J(\theta_0)= -2\frac{\partial vec \Gamma(1;\theta_0)'}{\partial \theta} [\Gamma(0;\theta_0)^{-1}\otimes \Gamma(0;\theta_0)^{-1}] \frac{\partial vec \Gamma(1;\theta_0)}{\partial \theta}.
$$
The first-order Taylor series expansion of the GCov estimator is: 
 \begin{equation}
    \sqrt{T}(\hat{\theta_T}-\theta_0) = J(\theta_0)^{-1} A(\theta_0)vec[ \sqrt{T}\hat{\Gamma}_T(1;\theta_0)'] + o_p(1),
 \label{fist-order}  
 \end{equation}

\bigskip

\nin \textbf{B.4 Expansion of the Portmanteau Statistic under the Null Hypothesis}

\nin The expansion of the test statistic under the null hypothesis is:
%!!!!!!!!!!!!!
\begin{equation}
    \begin{split}
    \hat{\xi}_T (H) =
    \Sum_{h=1}^H vec[\sqrt{T} \hat{\Gamma}_T(h,\theta_0, f_0)']' \Pi(h; \theta_0, f_0) vec[\sqrt{T} \hat{\Gamma}_T(h,\theta_0, f_0)'] + o_p(1),
    \end{split}
\end{equation}

\nin [See Gourieroux and Jasiak, "Generalized Covariance Estimator Supplemental Material" (2023), equation  (a.11)], where: 
\begin{eqnarray*}
   \Pi(h;\theta_0,  f_0) & = & [\Gamma_0 (0,\theta_0, f_0)^{-1}\otimes \Gamma_0 (0,\theta_0, f_0)^{-1}] - [\Gamma_0 (0,\theta_0, f_0)^{-1}\otimes \Gamma_0 (0,\theta_0, f_0)^{-1}] \frac{\partial vec \Gamma(h,\theta_0, f_0)}{\partial \theta'} \\
 & &   \left\{\frac{\partial vec \Gamma(h,\theta_0, f_0)'}{\partial \theta} [\Gamma_0 (0,\theta_0, f_0)^{-1}\otimes \Gamma_0 (0,\theta_0, f_0)^{-1}]\frac{\partial vec \Gamma(h,\theta_0, f_0)}{\partial \theta} \right\}^{-1}\\
& & \times   \frac{\partial vec \Gamma(h,\theta_0, f_0)'}{\partial \theta'} [\Gamma_0 (0,\theta_0, f_0)^{-1}\otimes \Gamma_0 (0,\theta_0, f_0)^{-1}].
\end{eqnarray*}

\nin Matrix $\Pi(h;\theta_0, f_0)$
satisfies for all $h=1,...,H$ the condition: 
$$\Pi(h;\theta_0, f_0) V_{asy} [\sqrt{T} \hat{\Gamma}_T(h,\theta_0)'] \Pi(h;\theta_0, f_0) = \Pi(h;\theta_0, f_0), $$
where $V_{asy} [\sqrt{T} \hat{\Gamma}_T(h,\theta_0, f_0)']= [\Gamma_0 (0,\theta_0, f_0)\otimes \Gamma_0 (0,\theta_0, f_0)]$.

This condition means that the matrix $\Pi(h; \theta_0, f_0)$ has an interpretation in terms of an orthogonal projector.
Therefore, under the null hypothesis, the quadratic form (A.10), where the $vec (\sqrt{T}\hat{\Gamma}_T (h; \theta_0, f_0))$ are independent identically distributed, still follows a chi-square distribution with a reduced degree of freedom.

\nin {\bf B.5 Asymptotic Behavior Under the Local Alternatives}

Under the regularity conditions 1-6, it is easy to see that expansions similar to (B.3)-(B.4) are still valid under the sequence of local alternatives by using the LLN for triangular arrays and the convergence of order $1/\sqrt{T}$ of the estimated autocovariances that follows from the CLT. For example, we still have the Taylor series expansion:

$$
\hat{\xi}_T(H) = T \sum_{h=1}^H \{vec [\sqrt{T} \hat{\Gamma}_T (h; \theta_T, \gamma_T, f_0)] \Pi(h; \theta_0, f_0)
vec [\sqrt{T} \hat{\Gamma}_T (h; \theta_T, \gamma_T, f_0)] \} + o_p(1),
$$

\nin similar to expansion (B.4) where the vectors $vec [\sqrt{T} \hat{\Gamma}_T (h; \theta_T, \gamma_T, f_0)]$, $h=1,...,H$, are now asymptotically independent with the distribution: 

$N[\delta(h; \theta_0, f_0, \mu, \nu),  \Gamma(0; \theta_0, f_0) \otimes \Gamma(0; \theta_0, f_0)],$

\nin by the CLT.

Then, under the sequence of local alternatives,  the asymptotic distribution of $\hat{\xi}_T(H)$ is a chi-square distribution with the non-centrality parameter $\lambda$:
$$\lambda(\theta_0, f_0, \mu, \nu)= \Sum_{h=1}^H \delta(h, \theta_0, f_0, \mu, \nu)' \Pi(h;\theta_0, f_0) \delta(h, \theta_0, f_0, \mu, \nu),$$
and a degree of freedom equal to the rank of matrix $\Pi(H;\theta_0, f_0) = diag [\Pi(h;\theta_0, f_0)]$, where diag denotes a diagonal matrix.

\medskip
\nin {\bf B.6 The Properties of the Independence Test under Local Alternatives}

\nin The results of Section B.5 can be applied to test the null hypothesis $H_0= (y_t=u_t)$ without the parameter $\theta$ (NLSD test) and other forms of local alternatives.

Let us consider the test of the absence of linear dependence in the time series $y_t = u_t, t=1,...,T$ against the local alternatives of an autoregressive form. More specifically, we test

$$H_0:\{\Gamma_0 (h)=0, \forall h=1,...,H\} = \{ B_1 = \cdots = B_H =0\},$$

\nin against the local alternatives. The local alternatives can be defined in terms of the autoregressive parameters $B_1,...,B_H$, or equivalently in terms of the autocovariances $\Gamma(h), h=1,...,H$. Thus, the additional parameter $\gamma$ is not necessarily a scalar. We follow the latter approach with the sequence of local alternatives:

$$H_{1,T} = \{\Gamma_T(h) = \Delta(h)/\sqrt{T}, \; h=1,...,H  \} = \{vec \Gamma_T(h) = \delta(h)/\sqrt{T}, \; h=1,...,H  \},$$

\nin with $\delta(h) = vec \Delta(h)$.

\nin Under the sequence of local alternatives, the estimated autocovariances are asymptotically independent with the asymptotic normal distributions.

$$ vec[\sqrt{T} \hat{\Gamma}_T(h)'] \overset a \sim N[\delta(h),\Gamma (0) \otimes \Gamma (0)]. $$
Hence:
$$[\Gamma (0)^{-1/2} \otimes \Gamma (0)^{-1/2}] vec[\sqrt{T} \hat{\Gamma}_T(h)'] \overset a \sim N[(\Gamma (0)^{-1/2} \otimes \Gamma (0)^{-1/2} ) \delta(h), Id] $$

\nin It follows that the portmanteau statistic $\hat{\xi}_T(H)$ has asymptotically, under the sequence of local alternatives, a chi-square $\chi^2 (K^2 H, \lambda)$ distribution with the non-centrality parameter $\lambda$, where

\begin{equation}
    \lambda = \Sum_{h=1}^H \delta (h)' [\Gamma (0)^{-1/2} \otimes \Gamma (0)^{-1/2}] [\Gamma (0)^{-1/2} \otimes \Gamma (0)^{-1/2}] \delta (h)
    = \Sum_{h=1}^H \delta (h)' [\Gamma (0)^{-1} \otimes \Gamma (0)^{-1}]\delta (h),
\end{equation}
is the non-centrality parameter.

\bigskip
\color{black}
\nin {\bf B.7 Regularity Conditions for Bootstrapping the Test Statistic (finite number of autocovariance conditions)}

In the literature, an approach with a finite number of nonlinear moment conditions in $u_t$ has not yet been developed. Inoue and Shintani (2006) consider the block bootstrap with a finite number of moment conditions that are linear in $u$. Escanciano (2007), Section 3 considers wild bootstrap with an infinite number of moment conditions also linear in $u$, while Wan and Davis (2022) study \textcolor{black}{ a parametric bootstrap with infinite} number of nonlinear sine and cosine transformations\footnote{Cavaliere, Nielsen and Rahbek (2020) develop a bootstrap approach for the OLS estimator of autoregressive coefficient $\rho$ in a noncausal AR(1) model: $y_t = \rho y_{t+1} +u_t$, where $u_t$ has a stable distribution. In their framework the coefficient $\rho$ is not equal to $Cov(y_t, y_{t-1})/Var(y_{t-1})$, because the theoretical moments may not exist. Hence, the OLS estimator is not a moment (or covariance) estimator, and their theoretical results cannot be applied to our framework.}. The approach proposed in our paper is based on a statistic that is nonlinear in $u$ with a finite number of nonlinear transformations \textcolor{black}{and a finite lag $H$}, and differs from the literature in this respect.

For ease of exposition, let us consider $H=1$, $dim(u_t) =1$, $u_t(\theta_0)$ with a symmetric density $f_0$ satisfying:

\begin{equation}
\sup_u f_0(u)/\exp( b |u|) \textcolor{black}{<} \infty, \; \mbox{for}\; b >0, 
\end{equation}

\nin and a GCov estimator based on a finite number of transformations $a_k(u) = |u|^{p_k} \exp(-t_k |u|), k=1,...,K$, where $p_k, t_k$ are fixed. The tail condition (B.6) ensures the uniform integrability of all moments $Ea_k(u_t(\theta_0))$  by 
considering transformations that reduce the effect of the tail. We also assume $g(\tilde{y}_t; \theta) = g(y_t,..., y_{t-p};\theta)$, and we distinguish the following regularity conditions for bootstrap validity:

i) Regularity conditions for deriving the asymptotic distribution of the test statistic under the null hypotheses of i.i.d. errors $u_t=u_t(\theta_0)$.

ii) Conditions concerning the third term in the Edgeworth expansion for the refinement of the bootstrap procedure 
%if the test statistic is asymptotically pivotal under the null.

iii) Additional regularity conditions to ensure the consistency of the bootstrap approach.

\nin Let us now discuss these regularity conditions.

i) Regularity conditions for deriving the asymptotic distribution of the bootstrap-adjusted portmanteau test statistic under the null hypothesis of serial independence.

%\nin A set of sufficient conditions has been given in Gourieroux, Jasiak (2023), Assumptions A.1, A.2., which include the hypothesis of i.i.d. errros $u_t$, the asymptotic identifiability of parameter $\theta$, and the assumption of invertibility of $\lim_{T \rightarrow \infty} [\frac{1}{T}
%\frac{\partial^2 \xi_T(\theta_0)}{\partial \theta \partial \theta'}]$.

These conditions have already been discussed in Appendices B.1-B.4.
As a consequence of these regularity conditions, we have the asymptotic equivalence of the GCov estimator:

\begin{equation}
\sqrt{T} (\hat{\theta}_T - \theta_0) \equiv \frac{1}{\sqrt{T}} \sum_{t=1}^T m_{T}(\underline{u_t(\theta_0)}, \theta_0) + o_p(1),
\end{equation}

\nin where $\underline{u_t(\theta_0)}= (u_t, u_{t-1},...)$, and 
$m_T(\underline{u_t}, \theta_0)$ is a vector-valued function satisfying the martingale difference sequence condition $E(m_T(\underline{u_t}, \theta_0)|\underline{u_{t-1}})=0$, $E ||m_T(\underline{u_t}), \theta_0||^2 < \infty$. An expression of $m_T(u_t(\theta_0), \theta_0)$ can be easily deduced from (B.3). Then, we observe that the empirical process 
($\frac{1}{\sqrt{T}}
m_{T} (\underline{u_t(\theta_0)}, \theta_0)$ converges in distribution to a Gaussian process (as a triangular array), and we
deduce the asymptotic distribution of the test statistic from its asymptotic expansion.

\medskip

ii) Conditions for refinement in the third-order Edgeworth expansion

The function $m_T$ in (B.7) is not uniquely defined, since it can always be modified by a term $B(\theta_0)/T$ of order $1/T$. Additional conditions can be introduced, especially on the third-order derivatives of $g(\tilde{y}_t, \theta)$ with respect to $\theta$ [see, e.g., Leucht and Neuman (2003)]. Then, we get:

\begin{equation}
\sqrt{T} (\hat{\theta}_T - \theta_0) \equiv \frac{1}{\sqrt{T}} \sum_{t=1}^T m_{T}
(\underline{u_t(\theta_0)}, \theta_0) + o_p(1/T).
\end{equation}

\nin In this expansion, the function $m_{T}$ depends on the function $g$ and the transforms $a_k, k=1,...,K$. It has a closed-form expression, even though it is complicated. In fact, we only need the existence of such an expansion, and the convergence in distribution of the empirical process $ \frac{1}{\sqrt{T}} \sum_{t=1}^T m_{T}
(\underline{u_t(\theta_0)}, \theta_0)$ to a Gaussian process. This will imply an asymptotically pivotal test statistic at order 1 under (B.7), and at order $1/T$ under (B.8).

\medskip
iii) Regularity conditions to ensure the consistency of the bootstrap under the null hypothesis of serial independence.

These conditions have to be introduced for the validity of the bootstrap under the null hypothesis $H_{0,iid}$. Below, we describe sets of sufficient conditions introduced in  Wan and Davis (2022) that are specific to this bootstrap procedure\footnote{See also Escanciano (2007), Assumption A.5 for a sufficient high-level condition for the validity of the bootstrap, when the moment condition is linear in $u_t$.}  and concern functions $m_T$ in the expansion (B.7) (resp.(B.8)).  These high level regularity conditions correspond to the assumptions M3, M3', M1' in  Wan and Davis (2022), provided below and adapted to our framework of a finite set of nonlinear autocovariance conditions. They are valid for causal models when $u_t$ is a nonlinear innovation of process $(y_t)$ [Gourieroux and Jasiak (2005)]\footnote{They can also be used for pure noncausal processes, i.e., when $u_t$ is a nonlinear innovation in reversed time. In this case, the innovations as well as their bootstrapped values have to be defined in reverse time.}

Assumption M3: $\frac{1}{\sqrt{T}} \sum_{t=1}^T |a_k(\hat{u}_{T,t}) - a_k(\hat{u}_{\infty,t})|^{m} = o_p(1), k=1,...,K,\; m =1,2,$

\nin where $\hat{u}_{\infty,t}$ denotes the fitted residual based on an infinite sequence of observations.

Assumption M3': For any $\epsilon>0$,

$\textcolor{black}{P^*} [\frac{1}{\sqrt{T}} \sum_{t=1}^T |a_k(\hat{u}_{T,t}^s) - a_k(\hat{u}_{\infty,t}^s)|^{m} > \epsilon] \rightarrow 0, k=1,...,K,\; m =1,2,$

\nin when $T$ tends to infinity, where $\hat{u}_{\infty,t}^s$ denotes the bootstrapped residuals based on an infinite sequence of observations.

In comparison with the analogous assumption in Wan and Davis (2022), the condition $u_t(\theta_0) = g(y_t,..., y_{t-p}; \theta_0)$ is equivalent to  $\exp(-t_k u_t(\theta_0))$  $= \exp(-t_k g(y_t,...,y_{t-p};\theta_0))$ for example.

Assumption M1': For any $\epsilon>0$ and some $\tau>0$:

$\textcolor{black}{P^*} (|\frac{1}{T} \sum_{t=1}^T \textcolor{black}{E^*} [m_T'(\underline{\hat{u}^s_{T,t}}, \hat{\theta}_T) m_T(\underline{\hat{u}^s_{T,t}}, \hat{\theta}_T)| \underline{\hat{u}_{T,t-1}^s}] - \tau^2| > \epsilon) \stackrel{p}{\rightarrow} 0,$

\nin when $T \rightarrow \infty$ and

$\textcolor{black}{P^*} (\frac{1}{T} \sum_{t=1}^T \textcolor{black}{E^*}[m_T'(\underline{\hat{u}^s_{T,t}}, \hat{\theta}_T) m_T(\underline{\hat{u}^s_{T,t}}, \hat{\theta}_T) 1_{||m_T(\underline{\hat{u}^s_{T,t}},\hat{\theta}_T)||> \sqrt{T} \epsilon} | \underline{\hat{u}_{T,t-1}^s}] > \epsilon) \stackrel{p}{\rightarrow}0,$ 

\nin when $T \rightarrow \infty$,
where \textcolor{black}{$P^*, E^*$} are taken with respect to the bootstrap sampling conditional on \textcolor{black}{the observations $Y_1,...,Y_T$}, and $\stackrel{p}{\rightarrow}$ denotes the convergence in probability.

Assumption M1' ensures that the martingale difference sequence condition is asymptotically satisfied at order 1 by the bootstrapped residuals.
%\footnote{The rate $\sqrt{T}$ in the second condition in M1' can be modified for refined bootstrap.}.
Then, we can apply Theorem 4.2 in Wan and Davis (2022) to obtain the consistency of the bootstrap-adjusted GCov test statistic:

$$\sup_z |\textcolor{black}{P^*} [ \hat{\xi}_T^s < z ] - P[\tilde{\xi}_T < z]|\stackrel{p}{\rightarrow}0,$$

\nin where \textcolor{black}{$P^*$} stands for conditional on data, and $\tilde{\xi}_T$ is the value of $\xi_T$ adjusted for its first-order bias. 

\bigskip
The local power analysis in Section 4.2 is based on Escanciano (2007), Section 3, Theorem 6, where a two-step parametric bootstrap is considered. The validity of the second step bootstrap is established for the wild bootstrap and extended to parametric bootstrap with conventional drawing on page 126.

\color{black}

\subsection*{ B.8 
Additional Regularity Conditions for an Infinite Set of Nonlinear Autocovariance Conditions}

This Appendix provides the regularity conditions for the results on the asymptotic behavior of the GCov portmanteau test statistics. 
%including the lag $H_T$ and the selected set $\mathcal{A}_{n_T}$ when $T$ tends to infinity.
%\medskip
%\nin {\bf B.8.1 Asymptotic Behavior of the Test Statistic}

We consider below the GCov statistic $\tilde{\xi}_{n,T}$ \textcolor{black}{based on $\tilde{\theta}_{n,T}$.} The asymptotic behavior of this statistic is derived along the following steps:

\nin step 1: We consider the infeasible portmanteau test statistic:

$$\xi_{n,T}^0 = T \sum_{h=1}^H \sum_{j=1}^{K_n} \sum_{k=1}^{K_n}  \left( \frac{1}{T}
\sum_{t=1+h}^T a_{j,n}^* [g(\tilde{y}_t; \theta_0)] a_{k,n}^*[g(\tilde{y}_{t-h}; \theta_0)] \right)^2,$$

\nin where $a_{k,n}^*$ are the transformations obtained from orthonormalization performed given the true distribution $f_0$.
%and with the starting value $\tilde{\theta}_{n,T}$ replacing $\theta_0$. 
The analysis of the asymptotic properties of this theoretical infeasible statistic needs to take into account the increasing sequence $\mathcal{A}_n$ and the theoretical regularization used in the orthonormalization. 

\nin step 2: We take into account the uncertainties due to the estimation of $\theta_0$ and $f_0$, and the replacement of $\theta_0$ by the starting value  $\tilde{\theta}_{n,T}$.

\medskip
\nin Step 1 Regularity Conditions

\nin These regularity conditions concern the theoretical orthonormal bases of $\mathcal{A}_n$. i.e., $a^*_{j,n}, j=1,...,n$.
Let us assume that $u_t$ has non-negative components, and a continuous density $f_0(u)$ that satisfies a tail condition:

$$\sup_{u>0} f_0(u) /\exp(b'u) < \infty, \; \mbox{for some}\; b,$$

\nin with strictly positive components, ensuring that the transformations in $\mathcal{A}_n$ are uniformly integrable. Then, under the hypothesis $H_{0,iid}$ of i.i.d. $u_t = g(\tilde{y}_t, \theta_0)$, for any fixed $K_n$, the multidimensional Lindeberg-Feller condition for the convergence in distribution of:

$$\frac{1}{T}
\sum_{t=1+h}^T a_{j,n}^* [u_t] a_{k,n}^*[u_{t-h}], \; j,k=1,...,K_n, \; h=1,...,H.$$
 
\nin to a standard normal $N(0, Id_{K_n^2H})$ is satisfied. Then, under the assumption of an increasing sequence $\mathcal{A}_n$ with a dense $\bigcup_n \mathbf{A}_n$ and a tightness condition [see Koenker and Machado (1999)], we get the analogue of Proposition 5 for the theoretical infeasible portmanteau test statistic:

\nin {\bf Proposition 5'} [Koenker and Machado (1999)]

$$(\xi_{n,T}^0 - H K_n^2)/\sqrt{2H K_n^2} \stackrel{d}{\rightarrow} N(0,1),$$

\nin when $T$ and $K_n$ tend to infinity.

\medskip

\nin Step 2 Regularity Conditions

The two-step estimator-based statistic $\hat{\xi}_{n,T} = \xi_{n,T} (\tilde{\tilde{\theta}}_{n,T})$ has the standard normal asymptotic distribution
provided that:
\nin a) the first-step estimator $\tilde{\theta}_{n,T}$ is consistent, asymptotically normally distributed;  b) the regularization tuning parameter $\epsilon_{n,T}$ tends to zero at an adequate rate; c) $T$ and $K_n$ tend to infinity at a rate such that $K_n^2/T$ tends to zero.

\nin Under an additional tightness condition, the difference between $a^*_{j,n,T}$ and $a^*_{j,n}, j=1,...,K_n$ is asymptotically negligible when $K_n$ tends to infinity.

These conditions ensure that the uncertainty due to the approximation of $\theta_0$ by $\tilde{\theta}_{n,T}$ and of the true orthonormalization by the estimated one are negligible [see e.g., Dovonon and Gospodinov (2024) for a similar argument].

\newpage
\color{black}
\begin{center}
\textbf{Appendix C}
\end{center}
\textcolor{black}{
This Appendix provides additional results based on the simulations and empirical application.}

\subsection*{ C.1 Simulations for NLSD and GCov Test}

\textcolor{black}{
Table 5 presents the comprehensive results for the empirical size and power of NLSD, with some rows also provided in Table 1 of the main text. }

\begin{table}[ht!]
\footnotesize{
\centering
\caption{NLSD Test of the absence of (non)linear dependence against the fixed alternative of MAR(0,1) at 5\% significance level: size and power }
\begin{tabular}{|c|ccc|ccc|ccc|}
\hline
\multirow{2}{*}{$\gamma$} & \multicolumn{3}{c|}{T=100} & \multicolumn{3}{c|}{T=200} & \multicolumn{3}{c|}{T=500} \\
\cline{2-10} 
& Uniform       & Laplace      & t(5)      & Uniform       & Laplace      & t(5)      & Uniform       & Laplace      & t(5)     \\ \hline
    0.0 & 0.0414 & 0.0484 & 0.0512 & 0.0450 & 0.0502 & 0.0518 & 0.0496 & 0.0540 & 0.0480 \\
    \hline
    0.1 & 0.0878 & 0.0706 & 0.0684 & 0.1502 & 0.1360 & 0.1220 & 0.3760 & 0.3390 & 0.3572 \\
    \hline
    0.2 & 0.2682 & 0.2202 & 0.2240 & 0.5618 & 0.5390 & 0.5322 & 0.9586 & 0.9546 & 0.9570 \\
    \hline
    0.3 & 0.6014 & 0.5742 & 0.5674 & 0.9202 & 0.9266 & 0.9256 & 1 & 0.9998 & 1 \\
    \hline
    0.4 & 0.8754 & 0.8814 & 0.8754 & 0.9966 & 0.9988 & 0.9976 & 1 & 1 & 1 \\
    \hline
    0.5 & 0.9796 & 0.9822 & 0.9856 & 1 & 1 & 1 & 1 & 1 & 1 \\
    \hline
    0.6 & 0.998 & 0.9986 & 0.9998 & 1 & 1 & 1 & 1 & 1 & 1 \\
    \hline
    0.7 & 1 & 1 & 1 & 1 & 1 & 1 & 1 & 1 & 1 \\
    \hline
    0.8 & 1 & 1 & 1 & 1 & 1 & 1 & 1 & 1 & 1 \\
    \hline
    0.9 & 1 & 1 & 1 & 1 & 1 & 1 & 1 & 1 & 1 \\
    \hline
\end{tabular}
\label{table:dependenceMAR(0,1)app}
\caption*{The first row ($\gamma=0$) shows the size of the test and the remaining rows show the size-adjusted power against fixed alternatives.}
}
\end{table}

\textcolor{black}{
Table 6 illustrates the power of the GCov specification test for the values of $\gamma$ between 0.1 and 0.9 under the fixed alternative, completing the results given in Table 2. }

 \begin{table}[H]
 \centering
 \footnotesize{
\caption{Power of GCov specification test of MAR(0,1) against the fixed alternative of MAR(1,1) at 5\% significance level}
\begin{tabular}{|c|c|ccc|ccc|ccc|}
\hline
\multirow{2}{*}{$\psi$} & \multirow{2}{*}{$\gamma=\phi$} & \multicolumn{3}{c|}{T=100} & \multicolumn{3}{c|}{T=200} & \multicolumn{3}{c|}{T=500} \\ \cline{3-11} 
                        &                         & Uniform       & Laplace      & t(5)      & Uniform       & Laplace      & t(5)      & Uniform       & Laplace      & t(5)     \\ \hline
\multirow{9}{*}{0.3}    & 0.1                     & 0.0226  & 0.0528  & 0.0446  & 0.0394  & 0.0672  & 0.0616  & 0.0572  & 0.0758  & 0.0752  \\ \cline{2-11} 
                        & 0.2                     & 0.0276  & 0.0626  & 0.0568  & 0.0720  & 0.0956  & 0.0980  & 0.1986  & 0.1624  & 0.1880  \\ \cline{2-11} 
                        & 0.3                     & 0.0416  & 0.0850  & 0.0866  & 0.1364  & 0.1612  & 0.1748  & 0.5244  & 0.3898  & 0.4162  \\ \cline{2-11} 
                        & 0.4                     & 0.0568  & 0.1080  & 0.1242  & 0.1796  & 0.2286  & 0.2394  & 0.6136  & 0.5280  & 0.5590  \\ \cline{2-11} 
                        & 0.5                     & 0.0604  & 0.1238  & 0.1346  & 0.2174  & 0.2568  & 0.2696  & 0.8000  & 0.6812  & 0.7002  \\ \cline{2-11} 
                        & 0.6                     & 0.0642  & 0.1276  & 0.1332  & 0.2504  & 0.3070  & 0.3216  & 0.8286  & 0.7898  & 0.7938  \\ \cline{2-11} 
                        & 0.7                     & 0.0746  & 0.1460  & 0.1518  & 0.2796  & 0.3614  & 0.3796  & 0.8722  & 0.8644  & 0.8634  \\ \cline{2-11} 
                        & 0.8                     & 0.1016  & 0.1724  & 0.1788  & 0.3404  & 0.4468  & 0.4672  & 0.9092  & 0.9282  & 0.9300  \\ \cline{2-11} 
                        & 0.9                     & 0.2100  & 0.2478  & 0.2552  & 0.5486  & 0.6180  & 0.6174  & 0.9804  & 0.9846  & 0.9862  \\ \hline
\multirow{9}{*}{0.7}    & 0.1                     & 0.0182  & 0.0500  & 0.0378  & 0.0374  & 0.0702  & 0.0664  & 0.1174  & 0.1224  & 0.1140  \\ \cline{2-11} 
                        & 0.2                     & 0.0296  & 0.0718  & 0.0628  & 0.1012  & 0.1456  & 0.1454  & 0.5346  & 0.4464  & 0.4602  \\ \cline{2-11} 
                        & 0.3                     & 0.0656  & 0.1430  & 0.1526  & 0.2854  & 0.3712  & 0.3764  & 0.8908  & 0.8630  & 0.8654  \\ \cline{2-11} 
                        & 0.4                     & 0.1578  & 0.3120  & 0.3098  & 0.5718  & 0.7040  & 0.6852  & 0.9926  & 0.9950  & 0.9908  \\ \cline{2-11} 
                        & 0.5                     & 0.3392  & 0.5700  & 0.5490  & 0.8318  & 0.9316  & 0.9196  & 1       & 1       & 0.9996  \\ \cline{2-11} 
                        & 0.6                     & 0.6034  & 0.8144  & 0.7848  & 0.9690  & 0.9914  & 0.9898  & 1       & 1       & 1       \\ \cline{2-11} 
                        & 0.7                     & 0.8528  & 0.9424  & 0.9370  & 0.9990  & 0.9996  & 1       & 1       & 1       & 1       \\ \cline{2-11} 
                        & 0.8                     & 0.9680  & 0.9882  & 0.9896  & 1       & 1       & 1       & 1       & 1       & 1       \\ \cline{2-11} 
                        & 0.9                     & 0.9970  & 0.9980  & 0.9990  & 1       & 1       & 1       & 1       & 1       & 1       \\ \hline
\end{tabular}
\label{newwtable:power}
}
\end{table}

\bigskip
The following Tables 7 and 8 provide additional simulation results on the NLSD and  GCov specification tests, respectively.
Table \ref{table:locdependenceMAR(0,1)} provides the results on the size of the NLSD test for local alternatives and $\delta$ increasing from 0 to 0.9. 
Table \ref{newwtable:1} shows the size of the GCov specification test for fixed alternatives and different values of $\psi$.

\begin{table}[H]
\footnotesize{
\centering
\caption{Test of the absence of (non)linear dependence (MAR(0,0)) against local  MAR(0,1) alternatives at 5\% significance level: size and size-adjusted power}
\begin{tabular}{|c|ccc|ccc|ccc|}
\hline
\multirow{2}{*}{$\gamma_T =$} & \multicolumn{3}{c|}{T=100} & \multicolumn{3}{c|}{T=200} & \multicolumn{3}{c|}{T=500} \\
\cline{2-10} 
$\frac{\delta}{\sqrt{T}}$ & Uniform       & Laplace      & t(5)      & Uniform       & Laplace      & t(5)      & Uniform       & Laplace      & t(5)     \\ \hline
      $\frac{0}{\sqrt{T}}$ & 0.0414 & 0.0484 & 0.0512 & 0.0450 & 0.0502 & 0.0518 & 0.0496 & 0.0540 & 0.0480 \\
    \hline
    $\frac{0.1}{\sqrt{T}}$ & 0.0498 & 0.0494 & 0.0470 & 0.0496 & 0.0490 & 0.0504 & 0.0512 & 0.0494 & 0.0508 \\
    \hline
    $\frac{0.2}{\sqrt{T}}$ & 0.0498 & 0.0488 & 0.0460 & 0.0502 & 0.0496 & 0.0496 & 0.0528 & 0.0492 & 0.0514 \\
    \hline
    $\frac{0.3}{\sqrt{T}}$ & 0.0516 & 0.0474 & 0.0464 & 0.0506 & 0.0524 & 0.0486 & 0.0544 & 0.0504 & 0.0540 \\
    \hline
    $\frac{0.4}{\sqrt{T}}$ & 0.0526 & 0.0486 & 0.0474 & 0.0530 & 0.0536 & 0.0500 & 0.0564 & 0.0520 & 0.0564 \\
    \hline
    $\frac{0.5}{\sqrt{T}}$ & 0.0574 & 0.0502 & 0.0486 & 0.0556 & 0.0568 & 0.0524 & 0.0606 & 0.0550 & 0.0602 \\
    \hline
    $\frac{0.6}{\sqrt{T}}$ & 0.0596 & 0.0530 & 0.0492 & 0.0606 & 0.0610 & 0.0548 & 0.0648 & 0.0584 & 0.0642 \\
    \hline
    $\frac{0.7}{\sqrt{T}}$ & 0.0650 & 0.0552 & 0.0520 & 0.0664 & 0.0662 & 0.0606 & 0.0710 & 0.0620 & 0.0710 \\
    \hline
    $\frac{0.8}{\sqrt{T}}$ & 0.0714 & 0.0580 & 0.0564 & 0.0756 & 0.0742 & 0.0652 & 0.0772 & 0.0664 & 0.0774 \\
    \hline
    $\frac{0.9}{\sqrt{T}}$ & 0.0804 & 0.0640 & 0.0622 & 0.0838 & 0.0812 & 0.0714 & 0.0860 & 0.0758 & 0.0848 \\
    \hline
\end{tabular}
\label{table:locdependenceMAR(0,1)}
\caption*{The first row ($\psi=0$) shows the size of the test and the remaining rows show the power against local alternatives with $\gamma_T=\psi_T=\frac{\delta}{\sqrt{T}}$.}
}
\end{table}

\begin{table}[H]
\centering
\footnotesize{
\caption{Size of GCov specification test of MAR(0,1) at 5\% significance level}
\begin{tabular}{|c|ccc|ccc|ccc|}
\hline
\multirow{2}{*}{$\psi$}  & \multicolumn{3}{c|}{T=100} & \multicolumn{3}{c|}{T=200} & \multicolumn{3}{c|}{T=500} \\ \cline{2-10} 
         & Uniform       & Laplace      & t(5)      & Uniform       & Laplace      & t(5)      & Uniform       & Laplace      & t(5)     \\ \hline
    0.1 & 0.0212 & 0.0404 & 0.0378 & 0.0342 & 0.0538 & 0.0514 & 0.0408 & 0.0538 & 0.0544 \\ \hline
    0.2 & 0.0216 & 0.0414 & 0.0390 & 0.0344 & 0.0548 & 0.0534 & 0.0414 & 0.0558 & 0.0560 \\ \hline
    0.3 & 0.0224 & 0.0454 & 0.0386 & 0.0348 & 0.0566 & 0.0534 & 0.0406 & 0.0544 & 0.0560 \\ \hline
    0.4 & 0.0214 & 0.0460 & 0.0390 & 0.0344 & 0.0560 & 0.0522 & 0.0414 & 0.0548 & 0.0540 \\ \hline
    0.5 & 0.0194 & 0.0450 & 0.0382 & 0.0326 & 0.0550 & 0.0518 & 0.0414 & 0.0548 & 0.0540 \\ \hline
    0.6 & 0.0196 & 0.0428 & 0.0348 & 0.0318 & 0.0536 & 0.0492 & 0.0426 & 0.0560 & 0.0544 \\ \hline
    0.7 & 0.0200 & 0.0400 & 0.0338 & 0.0298 & 0.0528 & 0.0468 & 0.0408 & 0.0552 & 0.0528 \\ \hline
    0.8 & 0.0202 & 0.0392 & 0.0316 & 0.0296 & 0.0512 & 0.0454 & 0.0396 & 0.0532 & 0.0500 \\ \hline
    0.9 & 0.0178 & 0.0384 & 0.0310 & 0.0278 & 0.0508 & 0.0428 & 0.0350 & 0.0494 & 0.0498 \\ \hline
\end{tabular}
\label{newwtable:1}
}
\end{table}

\newpage
\subsection*{ C.2 GCov Specification Test for Processes with Cauchy Error Distribution}

% \begin{figure}[ht!]
%     \centering
%     \begin{subfigure}[b]{0.49\textwidth}
%         \centering
%         \includegraphics[width=\textwidth]{Csize2.png}
%         \caption{Size}
%     \end{subfigure}
%     \hfill
%         \begin{subfigure}[b]{0.49\textwidth}
%         \centering
%         \includegraphics[width=\textwidth]{Cpower2.png}
%         \caption{Power }
%     \end{subfigure}
%     \caption{Size and power of GCov specification test of the MAR(0,1) with Cauchy error distribution } 
%     \label{Csizenew}
% \end{figure}

\begin{table}[H]
\centering
\caption{Size and power of GCov specification test of the MAR(0,1) with Cauchy error distribution at 5\% significance level} 
\begin{tabular}{|c|c|c|ccccc|}
\hline
\multirow{2}{*}{S./P.} & \multirow{2}{*}{$\phi$} & \multirow{2}{*}{$\psi$} & \multicolumn{5}{c|}{T}                                                                                                         \\ \cline{4-8} 
                       &                         &                         & \multicolumn{1}{c|}{100}    & \multicolumn{1}{c|}{200}    & \multicolumn{1}{c|}{300}    & \multicolumn{1}{c|}{400}    & 500    \\ \hline
\multirow{2}{*}{S.}    &                         & 0.3                     & \multicolumn{1}{c|}{0.0306} & \multicolumn{1}{c|}{0.0468} & \multicolumn{1}{c|}{0.0472} & \multicolumn{1}{c|}{0.0552} & 0.0538 \\ \cline{2-8} 
                       &                         & 0.7                     & \multicolumn{1}{c|}{0.0342} & \multicolumn{1}{c|}{0.0440} & \multicolumn{1}{c|}{0.0470} & \multicolumn{1}{c|}{0.0558} & 0.0518 \\ \hline
\multirow{2}{*}{P.}    & 0.8                     & 0.3                     & \multicolumn{1}{c|}{0.7526} & \multicolumn{1}{c|}{0.9724} & \multicolumn{1}{c|}{0.9990} & \multicolumn{1}{c|}{1}      & 1      \\ \cline{2-8} 
                       & 0.8                     & 0.7                     & \multicolumn{1}{c|}{0.9994} & \multicolumn{1}{c|}{1}      & \multicolumn{1}{c|}{1}      & \multicolumn{1}{c|}{1}      & 1      \\ \hline
\end{tabular}

\label{Csizenew}
\end{table}

\begin{table}[H]
\centering
\caption{Size and power of bootstrap test with and without replacement, MAR(0,1) with Cauchy error distribution at 5\% significance level with estimation by OLS} 
\begin{tabular}{|c|c|c|cc|cc|cc|}
\hline
\multirow{2}{*}{S./P.} & \multirow{2}{*}{$\phi$} & \multirow{2}{*}{$\psi$} 
& \multicolumn{2}{c|}{T=100} & \multicolumn{2}{c|}{T=200} & \multicolumn{2}{c|}{T=500} \\ \cline{4-9} 
& & & \multicolumn{1}{c|}{with} & without & \multicolumn{1}{c|}{with} & without & \multicolumn{1}{c|}{with} & without \\ \hline

\multirow{2}{*}{S.} & & 0.3 & \multicolumn{1}{c|}{0.077} & 0.080 & \multicolumn{1}{c|}{0.056} & 0.078 & \multicolumn{1}{c|}{0.050} & 0.073 \\ \cline{3-9} 
& & 0.7 & \multicolumn{1}{c|}{0.075} & 0.091 & \multicolumn{1}{c|}{0.054} & 0.087 & \multicolumn{1}{c|}{0.048} & 0.077 \\ \hline

\multirow{2}{*}{P.} & 0.8 & 0.3 & \multicolumn{1}{c|}{0.808} & 0.802 & \multicolumn{1}{c|}{0.880} & 0.915 & \multicolumn{1}{c|}{0.986} & 0.988 \\ \cline{2-9} 
& 0.8 & 0.7 & \multicolumn{1}{c|}{0.999} & 1.000 & \multicolumn{1}{c|}{1.000} & 1 & \multicolumn{1}{c|}{1} & 1 \\ \hline
\end{tabular}
\label{bootCSizeOLS}
\end{table}

\subsection*{ C.3 Distribution of GCov bootstrap test statistic}

\begin{figure}[H]
    \centering
    
    \begin{subfigure}[b]{0.49\textwidth}
        \centering
        \includegraphics[width=\textwidth]{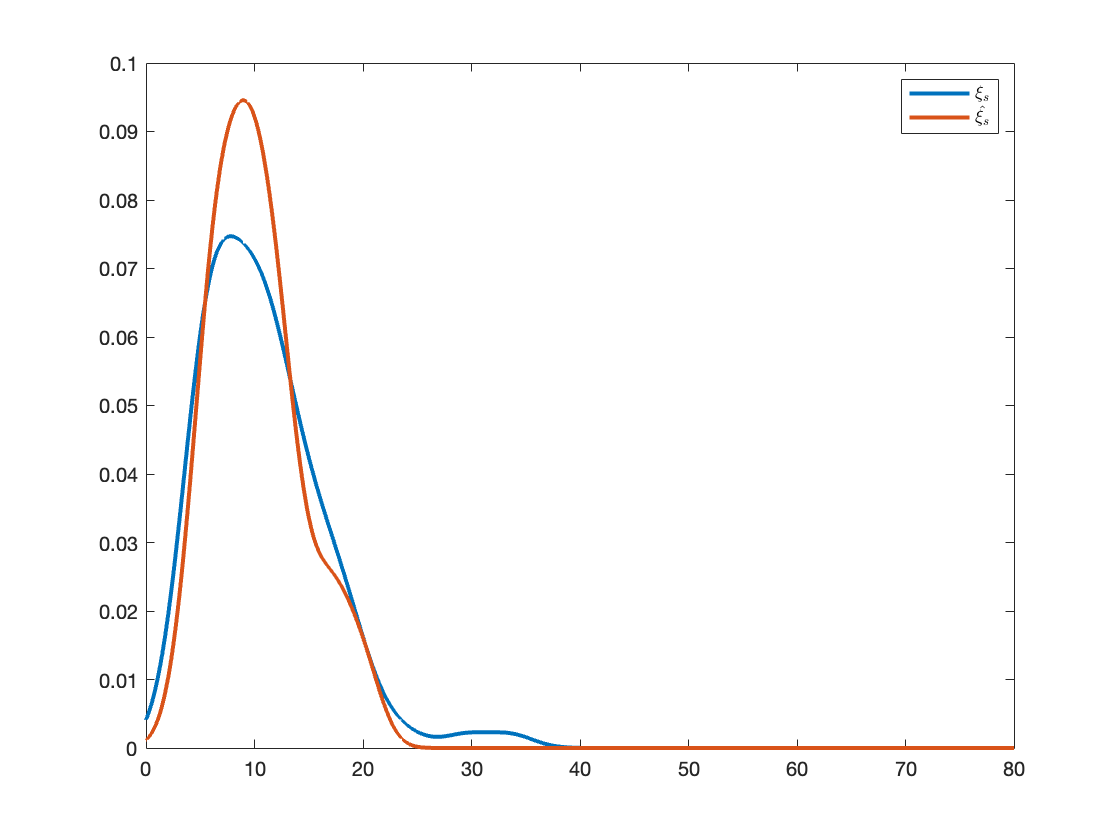}
        \caption{$T=100$  }
        \label{local03}
    \end{subfigure}
    \hfill
        \begin{subfigure}[b]{0.49\textwidth}
        \centering
        \includegraphics[width=\textwidth]{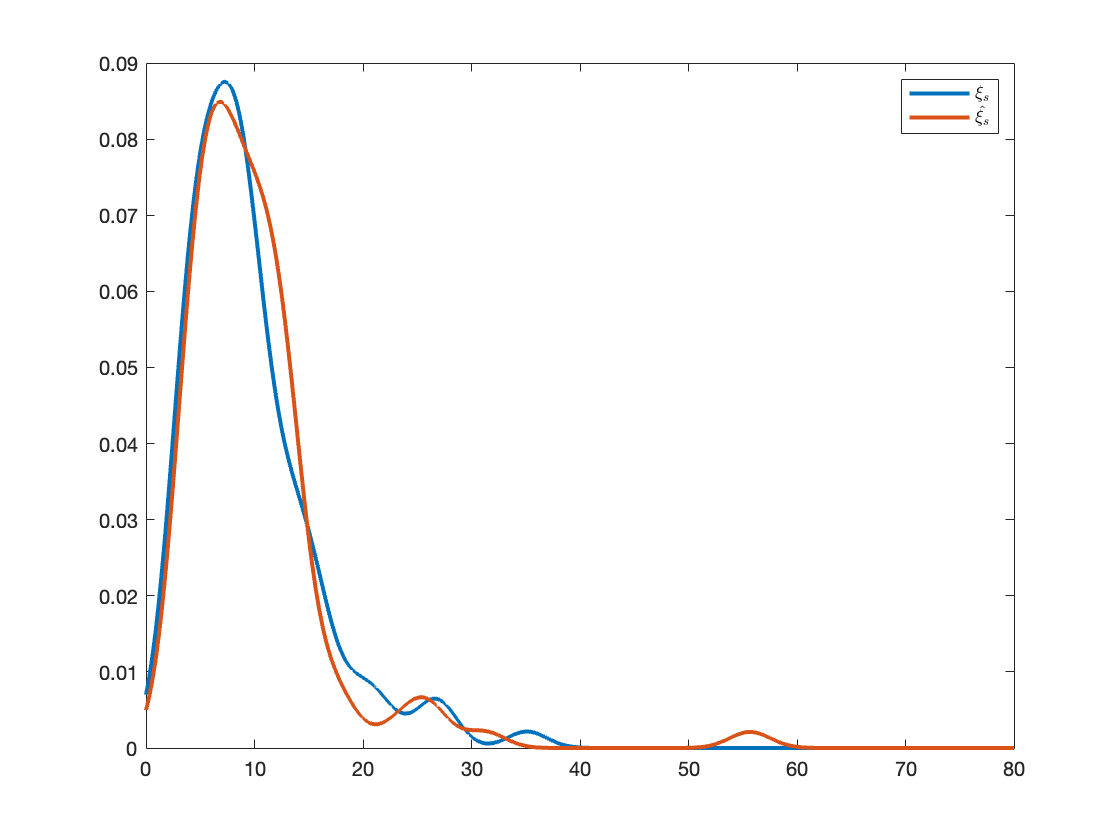}
        \caption{$T=500$ }
        \label{local07}
    \end{subfigure}
    \caption{Comparison of the distributions of $\xi_s$ and $\hat{\xi_s}$ for MAR(0,1) with t(5) error distribution  and $\psi=0.7$ } 
    \label{boot MAR(0,1)}
\end{figure}

\vspace{-0.3cm}

\begin{table}[H]
\footnotesize{
\centering
\caption{Bootstrap approximation  size and power of GCov specification based on GCov estimation of MAR(0,1), at 5\% significance level \textbf{A:} without replacement and \textbf{B:} with replacement}

\begin{tabular}{|c|c|cc|ccc|ccc|ccc|}
\hline
\multirow{2}{*}{Panel} & \multirow{2}{*}{S./P.} & \multirow{2}{*}{$\phi$} & \multirow{2}{*}{$\psi$} & \multicolumn{3}{c|}{T=100}                                  & \multicolumn{3}{c|}{T=200}                              & \multicolumn{3}{c|}{T=500}                              \\ \cline{5-13} 
                     &  &                      &                      & \multicolumn{1}{c|}{Uniform} & \multicolumn{1}{c|}{Laplace}     & t(5)  & \multicolumn{1}{c|}{Uniform} & \multicolumn{1}{c|}{Laplace} & t(5)  & \multicolumn{1}{c|}{Uniform} & \multicolumn{1}{c|}{Laplace} & t(5)  \\ \hline
\multirow{4}{*}{A} &\multirow{2}{*}{S.}    &            0          & 0.3                  & 0.051 & 0.067 & 0.064 & 0.060 & 0.056 & 0.053 & 0.063 & 0.061 & 0.046  \\ \cline{3-13} 
                     &  &          0            & 0.7                  & 0.052 & 0.066 & 0.066 & 0.062 & 0.054 & 0.053 & 0.064 & 0.064 & 0.046  \\ \cline{2-13} 
& \multirow{2}{*}{P.}    & 0.8                  & 0.3                  & 0.227  & 0.361  & 0.365  & 0.454  & 0.529  & 0.566  & 0.946  & 0.936  & 0.941 \\ \cline{3-13} 
                  &     & 0.8                  & 0.7                  & 0.982  & 0.990  & 0.998  & 1       & 1       & 1       & 1       & 1       & 1      \\ \hline

\multirow{4}{*}{B} & \multirow{2}{*}{S.}    
& 0   & 0.3 & 0.052 & 0.068 & 0.060 & 0.062 & 0.060 & 0.047 & 0.056 & 0.065 & 0.046 \\ \cline{3-13} 
& & 0   & 0.7 & 0.058 & 0.066 & 0.062 & 0.054 & 0.064 & 0.050 & 0.051 & 0.066 & 0.042 \\ \cline{2-13} 
& \multirow{2}{*}{P.}    
& 0.8 & 0.3 & 0.219 & 0.349 & 0.375 & 0.456 & 0.540 & 0.589 & 0.936 & 0.936 & 0.950 \\ \cline{3-13} 
& & 0.8 & 0.7 & 0.985 & 0.989 & 0.996 & 1 & 1 & 1 & 1 & 1 & 1 \\ \hline
\end{tabular}
\label{boostrapapp}
\caption*{ S.: size, P.: power }
}
\end{table}

\begin{table}[H]
\footnotesize{
\centering
\caption{Size and power of GCov bootstrap test without replacement based on AML estimation of MAR(0,1), at 5\% significance level}

\begin{tabular}{|c|cc|ccc|ccc|ccc|}
\hline
\multirow{2}{*}{S./P.} & \multirow{2}{*}{$\phi$} & \multirow{2}{*}{$\psi$} 
& \multicolumn{3}{c|}{T=100} 
& \multicolumn{3}{c|}{T=200} 
& \multicolumn{3}{c|}{T=500} \\ \cline{4-12} 
& & & \multicolumn{1}{c|}{t(4)} & \multicolumn{1}{c|}{t(5)} & \multicolumn{1}{c|}{t(6)} 
& \multicolumn{1}{c|}{t(4)} & \multicolumn{1}{c|}{t(5)} & \multicolumn{1}{c|}{t(6)} 
& \multicolumn{1}{c|}{t(4)} & \multicolumn{1}{c|}{t(5)} & \multicolumn{1}{c|}{t(6)} \\ \hline
% \multirow{2}{*}{S.}    
% & 0   & 0.3 & 0.060  & 0.061 & 0.067  &0.064  & 0.048  & 0.051  & 0.058  & 0.045 &  0.056 \\ \cline{2-12} 
% & 0   & 0.7 & 0.061  & 0.064  & 0.063 &   0.066  & 0.050  & 0.049  & 0.053  & 0.044 & 0.055 \\ \hline
% \multirow{2}{*}{P.}    
% & 0.8 & 0.3 & 0.430 & 0.410 & 0.388 & 0.704& 0.664 & 0.633 & 0.982  & 0.980 & 0.975 \\ \cline{2-12} 
% & 0.8 & 0.7 & 0.991  & 0.993  & 0.911 & 1 & 1 & 1 & 1 & 1 & 1 \\ \hline
 \multirow{2}{*}{S.}    &            0          & 0.3                  & 0.061                  & 0.062 & 0.063 & 0.059                  & 0.046                  & 0.049 & 0.062                  & 0.046                  & 0.058 \\ \cline{2-12} 
                       &          0            & 0.7                  & 0.063                  & 0.063                      & 0.066 & 0.060                  & 0.047                  & 0.056 & 0.062                  & 0.049                  & 0.060  \\ \hline
 \multirow{2}{*}{P.}    & 0.8                  & 0.3                  & 0.438                  & 0.412                     & 0.399 & 0.706                  & 0.671                  & 0.633 & 0.982                  & 0.986                  & 0.980 \\ \cline{2-12} 
                       & 0.8                  & 0.7                  & 0.991                  & 0.992                      & 0.993  & 1                      & 1                      & 1     & 1                      & 1                      & 1     \\ \hline
\end{tabular}

\label{comparetable3}
\caption*{ S.: size, P.: power  }

}

\end{table}

\begin{table}[H]
\footnotesize{
\centering
\caption{ Empirical size and power of GCov test with many transformations at  5\% significance level. For power, we generate MAR(1,1) with $\phi=0.8$.}
\begin{tabular}{|c|c|c|c|c|c|c|c|c|}
\hline
\multirow{2}{*}{S./P.} & \multirow{2}{*}{$H$} & \multirow{2}{*}{$T$} 
& \multicolumn{2}{c|}{$K=7$} & \multicolumn{2}{c|}{$K=8$} & \multicolumn{2}{c|}{$K=9$} \\ \cline{4-9}
& & & $\psi=0.3$ & $\psi=0.7$ & $\psi=0.3$ & $\psi=0.7$ & $\psi=0.3$ & $\psi=0.7$ \\ \hline

\multirow{9}{*}{S.} 
& \multirow{3}{*}{$H=3$} & $T=100$ & 0.007 & 0.012 & 0.028 & 0.031 & 0.001 & 0.002 \\ \cline{3-9}
& & $T=200$ & 0.012 & 0.022 & 0.030 & 0.042 & 0.029 & 0.024 \\ \cline{3-9}
& & $T=500$ & 0.033 & 0.051 & 0.035 & 0.051 & 0.054 & 0.050 \\ \cline{2-9}

& \multirow{3}{*}{$H=4$} & $T=100$ & 0.013 & 0.006 & 0.027 & 0.029 & 0.002 & 0.006 \\ \cline{3-9}
& & $T=200$ & 0.010 & 0.016 & 0.030 & 0.028 & 0.021 & 0.019 \\ \cline{3-9}
& & $T=500$ & 0.031 & 0.049 & 0.034 & 0.059 & 0.050 & 0.056 \\ \cline{2-9}

& \multirow{3}{*}{$H=5$} & $T=100$ & 0.009 & 0.006 & 0.020 & 0.023 & 0.002 & 0.002 \\ \cline{3-9}
& & $T=200$ & 0.010 & 0.019 & 0.038 & 0.025 & 0.026 & 0.025 \\ \cline{3-9}
& & $T=500$ & 0.031 & 0.053 & 0.029 & 0.052 & 0.053 & 0.060 \\ \hline

\multirow{9}{*}{P.} 
& \multirow{3}{*}{$H=3$} & $T=100$ & 0.549 & 0.503 & 0.201 & 0.356 & 0.139 & 0.213 \\ \cline{3-9}
& & $T=200$ & 0.703 & 0.771 & 0.231 & 0.318 & 0.454 & 0.665 \\ \cline{3-9}
& & $T=500$ & 0.703 & 0.897 & 0.535 & 0.529 & 0.567 & 0.877 \\ \cline{2-9}

& \multirow{3}{*}{$H=4$} & $T=100$ & 0.498 & 0.467 & 0.198 & 0.327 & 0.141 & 0.193 \\ \cline{3-9}
& & $T=200$ & 0.695 & 0.752 & 0.215 & 0.307 & 0.437 & 0.626 \\ \cline{3-9}
& & $T=500$ & 0.664 & 0.889 & 0.528 & 0.514 & 0.571 & 0.872 \\ \cline{2-9}

& \multirow{3}{*}{$H=5$} & $T=100$ & 0.440 & 0.406 & 0.179 & 0.320 & 0.120 & 0.168 \\ \cline{3-9}
& & $T=200$ & 0.680 & 0.723 & 0.194 & 0.298 & 0.428 & 0.594 \\ \cline{3-9}
& & $T=500$ & 0.673 & 0.891 & 0.493 & 0.486 & 0.552 & 0.863 \\ \hline
\end{tabular}
\label{Manytransformationsapp}

}
\end{table}

\subsection*{ C.4 Empirical Power of Specification Test Based on Many Transformations}

\begin{figure}[H]
    \centering
    \begin{subfigure}[b]{0.49\textwidth}
        \centering
        \includegraphics[width=\textwidth]{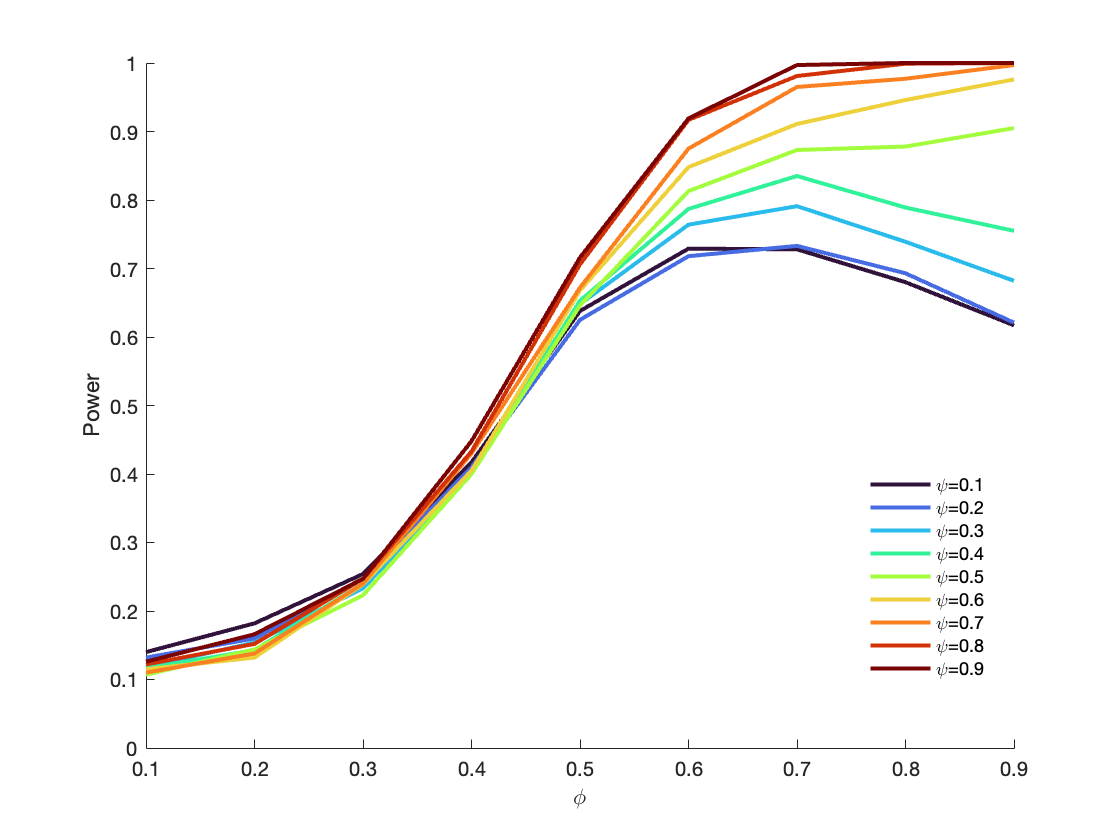}
        \caption{$T=500$  }
        \label{powermany500}
    \end{subfigure}
    \hfill
        \begin{subfigure}[b]{0.49\textwidth}
        \centering
        \includegraphics[width=\textwidth]{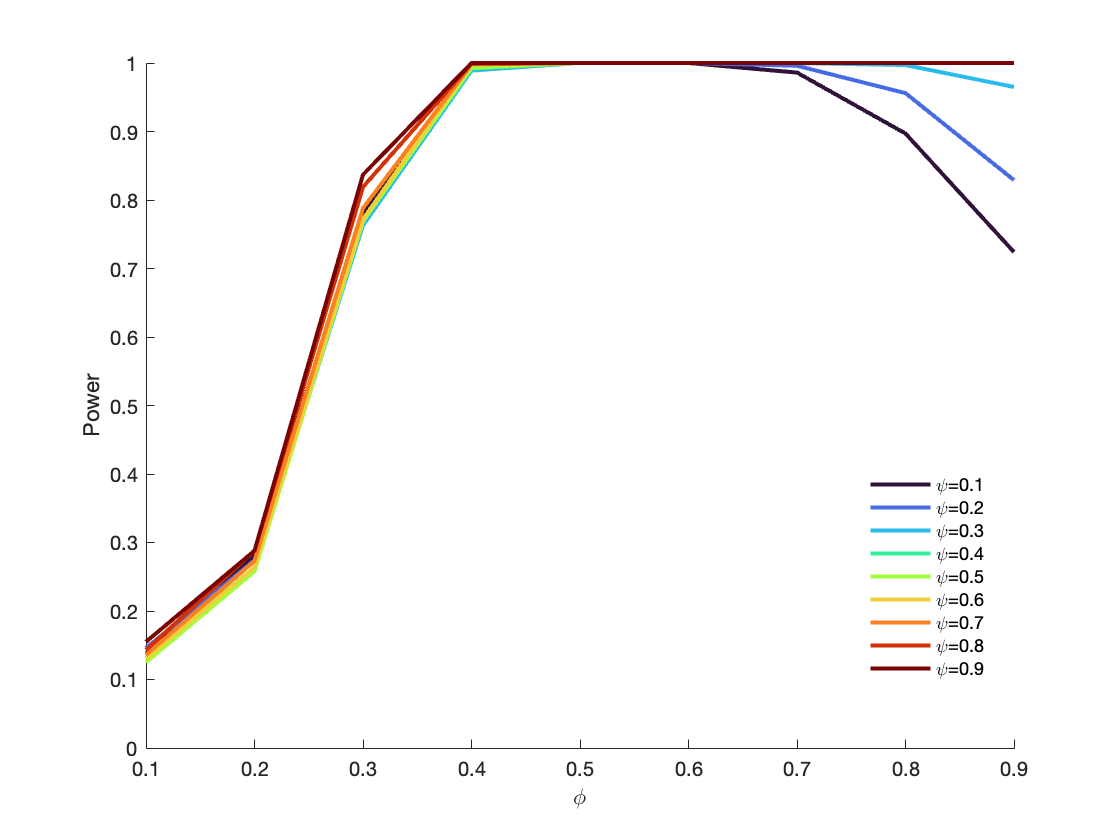}
        \caption{$T=5000$ }
        \label{powermany5000}
    \end{subfigure}
    \caption{Empirical power of specification test based on $H=4$ and $K=7$ as a function of coefficient $\phi$ under the alternative with $t(5)$ error distribution. } 
    \label{Manypowerbyphi}
\end{figure}

\begin{figure}[H]
    \centering
         \includegraphics[width=0.5\textwidth]{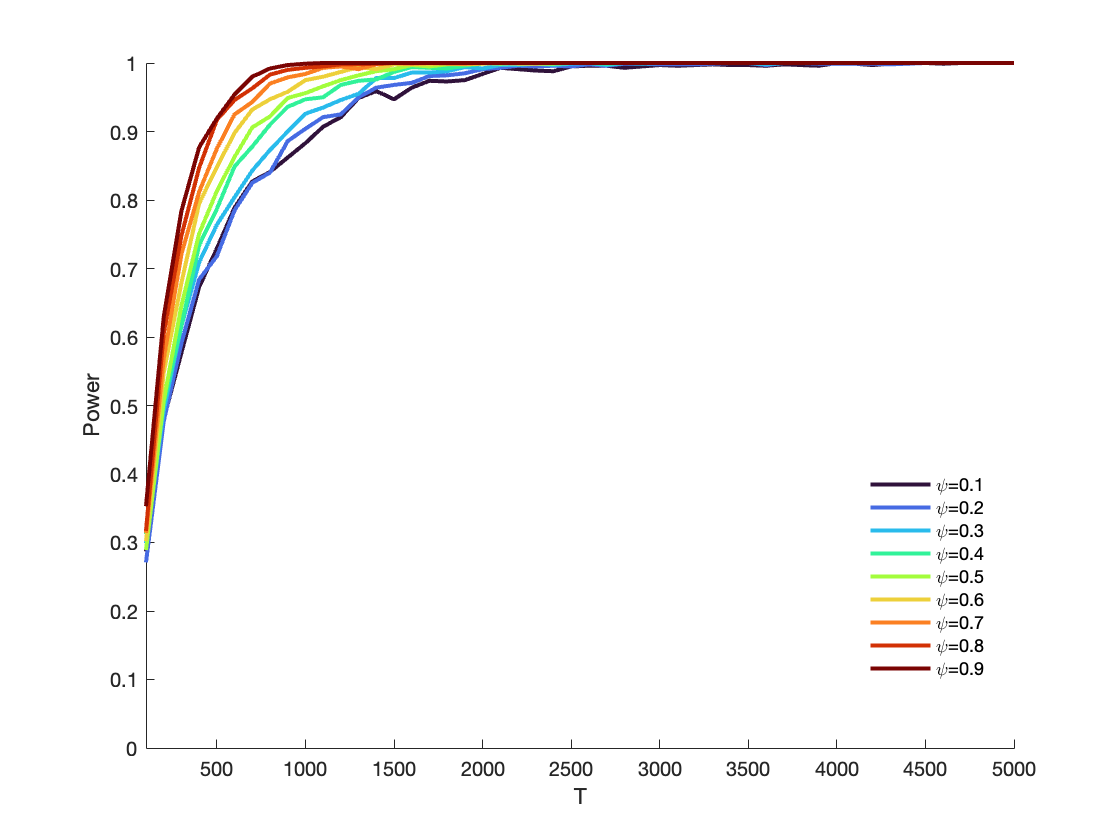}
    \caption{Empirical power of specification test based on $H=4$ and $K=7$ with fixed $\phi=0.6$ as a function of the number of observation T with $t(5)$ error distribution. }
    \label{ManypowerbyT}
\end{figure}

Figure \ref{Manypowerbyphi} provides empirical power of the specification test with many transformations. We generate MAR(1,1) and fit MAR(0,1) to the process; therefore, $\phi$ is our distance to the null hypothesis. For high persistence of noncausality, the power converges to one with both $T=500$ and $T=5000$. In both sample sizes, power decreases when $\psi$ is low and $\phi$ gets closer to the unit root. It is due to the identification problem that when we get close to a unit root, the roots flip. Moreover, Figure \ref{ManypowerbyT} demonstrates the convergence of empirical power to one when we increase the sample size. The more persistence in noncausality, the faster we converge to one.

\subsection*{C.5 Empirical Application}
\begin{figure}[H]
    \centering
    
    \begin{subfigure}[b]{0.49\textwidth}
        \centering
        \includegraphics[width=\textwidth]{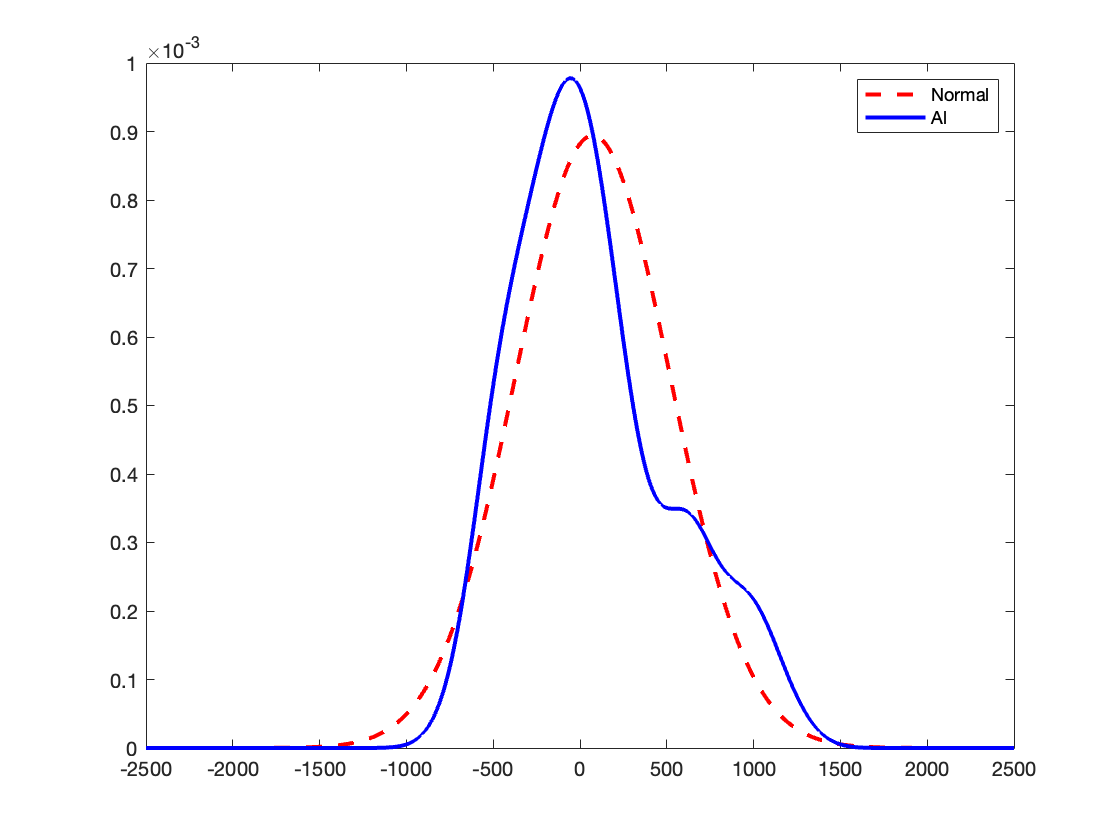}
        \caption{Demeaned-Detrended aluminum price }
        \label{4-a}
    \end{subfigure}
    \hfill
        \begin{subfigure}[b]{0.49\textwidth}
        \centering
        \includegraphics[width=\textwidth]{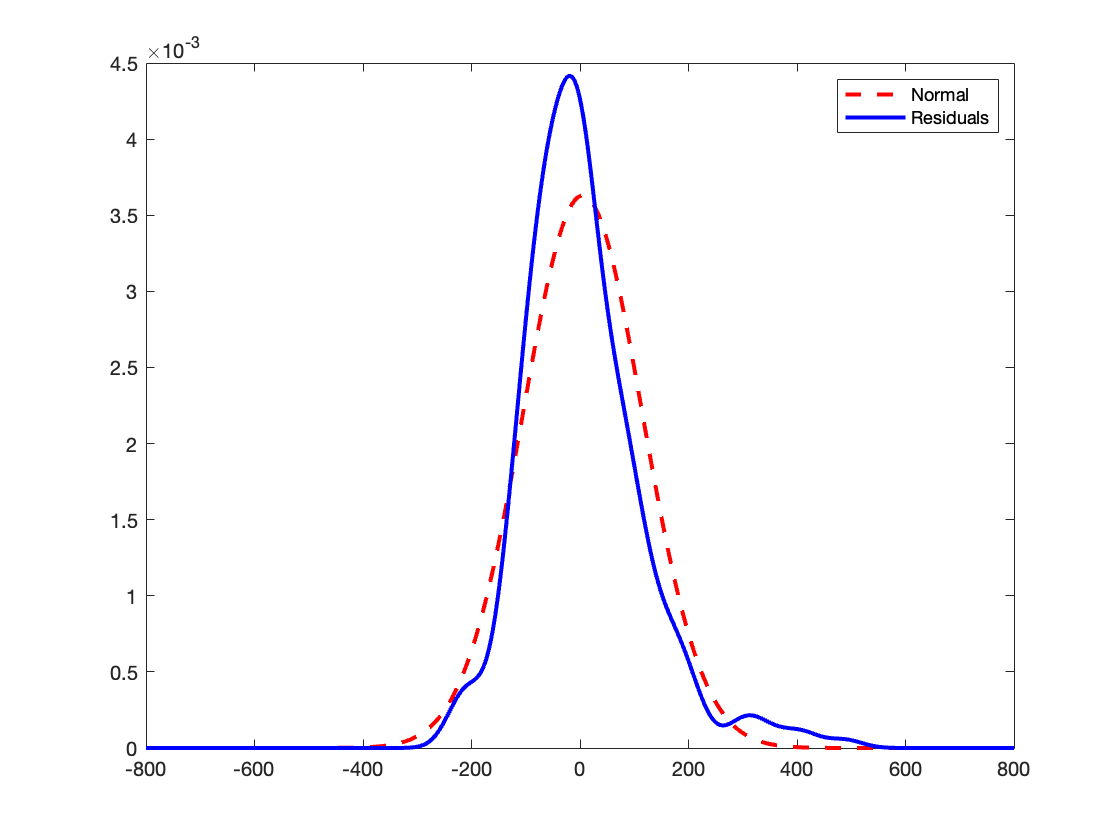}
        \caption{MAR(1,1) residuals}
        \label{4-b}
    \end{subfigure}

    \caption{Densities of demeaned aluminum price and MAR(1,1) residuals, compared with the normal density}
    \label{figure 5}
\end{figure}

 \begin{figure}[H]
    \centering
    
    \begin{subfigure}[b]{0.49\textwidth}
        \centering
        \includegraphics[width=\textwidth]{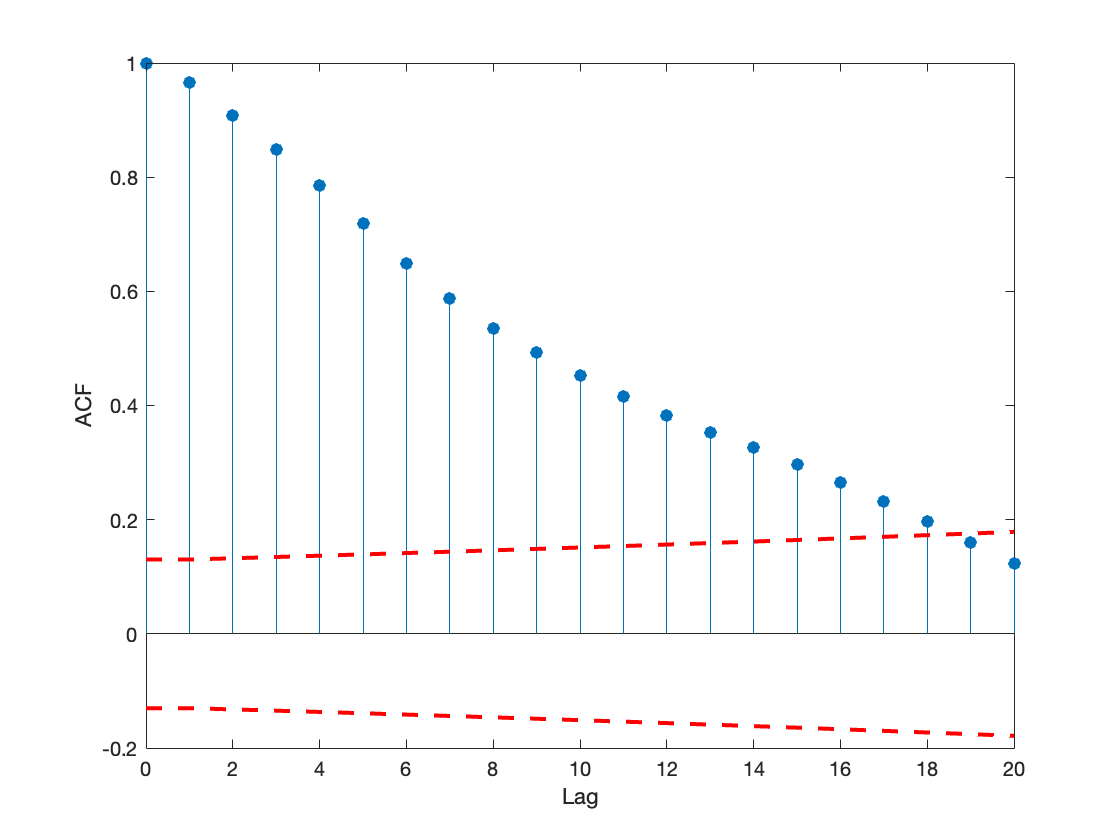}
        \caption{ACF of the series }
        \label{3-a}
    \end{subfigure}
    \hfill
        \begin{subfigure}[b]{0.49\textwidth}
        \centering
        \includegraphics[width=\textwidth]{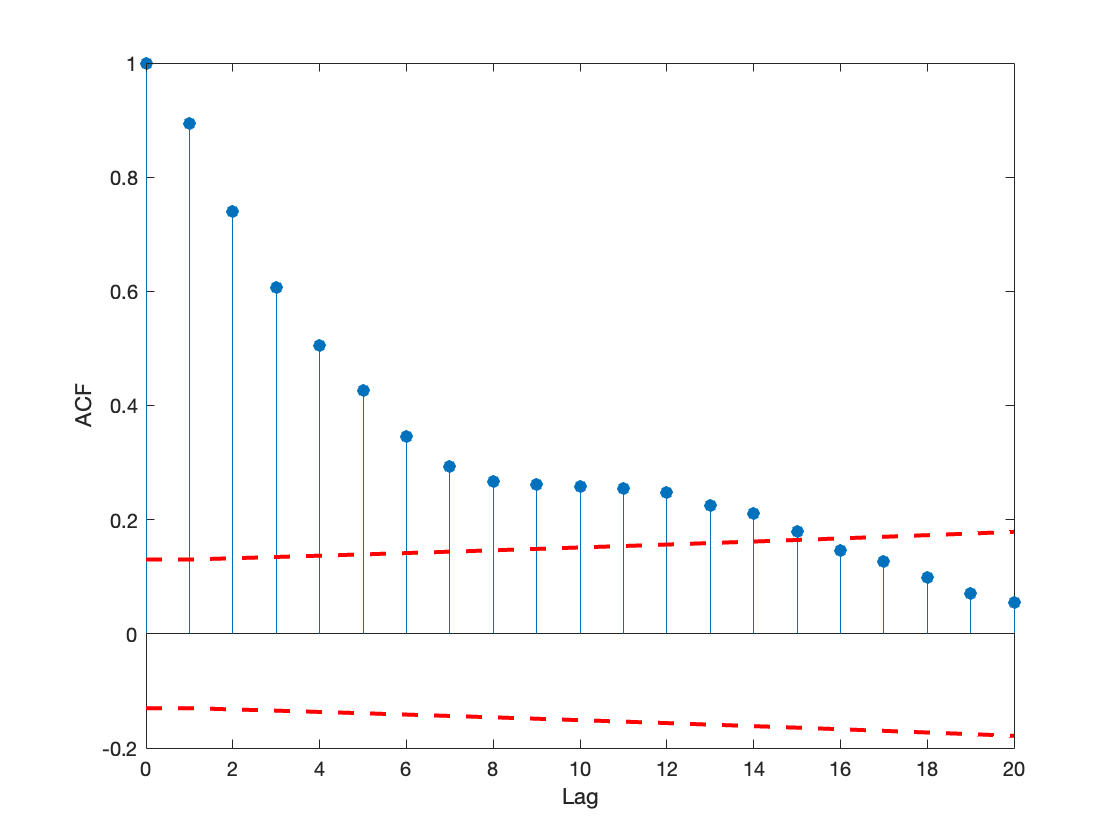}
        \caption{ACF of the series square}
        \label{3-b}
    \end{subfigure}

    \begin{subfigure}[b]{0.49\textwidth}
        \centering
        \includegraphics[width=\textwidth]{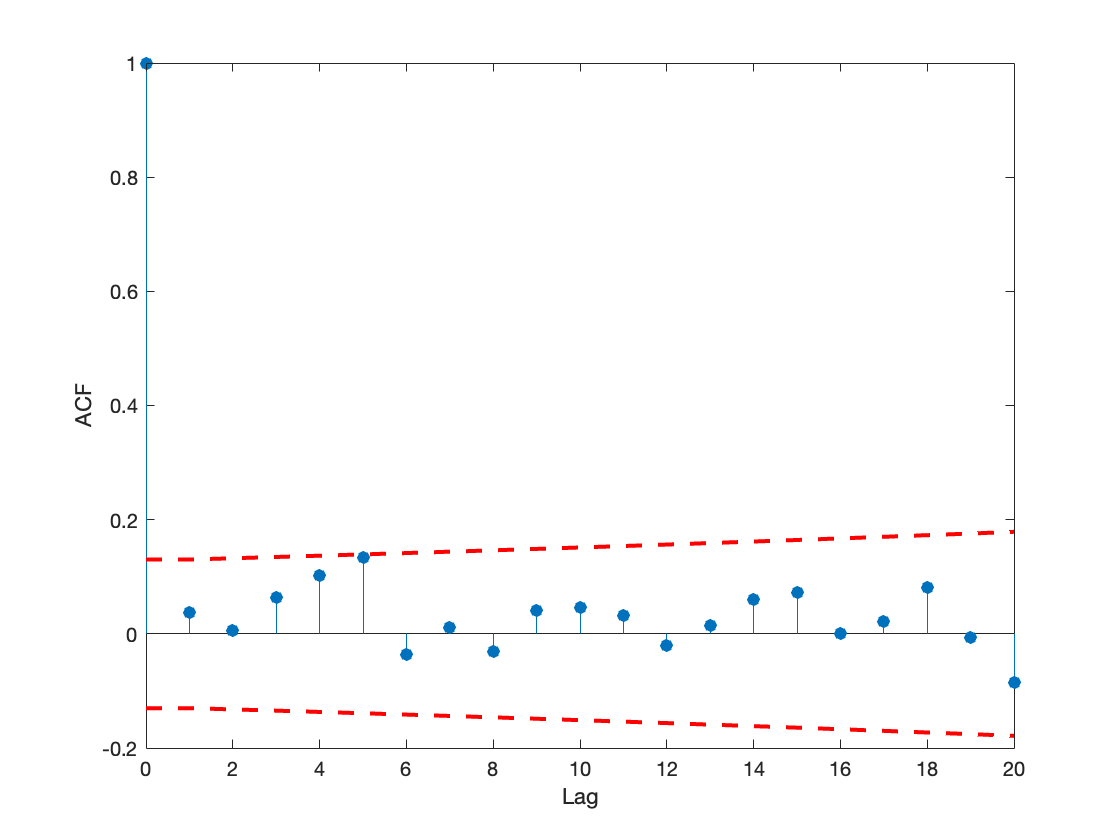}
        \caption{ACF of the residuals}
        \label{3-c}
    \end{subfigure}
    \hfill
    \begin{subfigure}[b]{0.49\textwidth}
        \centering
        \includegraphics[width=\textwidth]{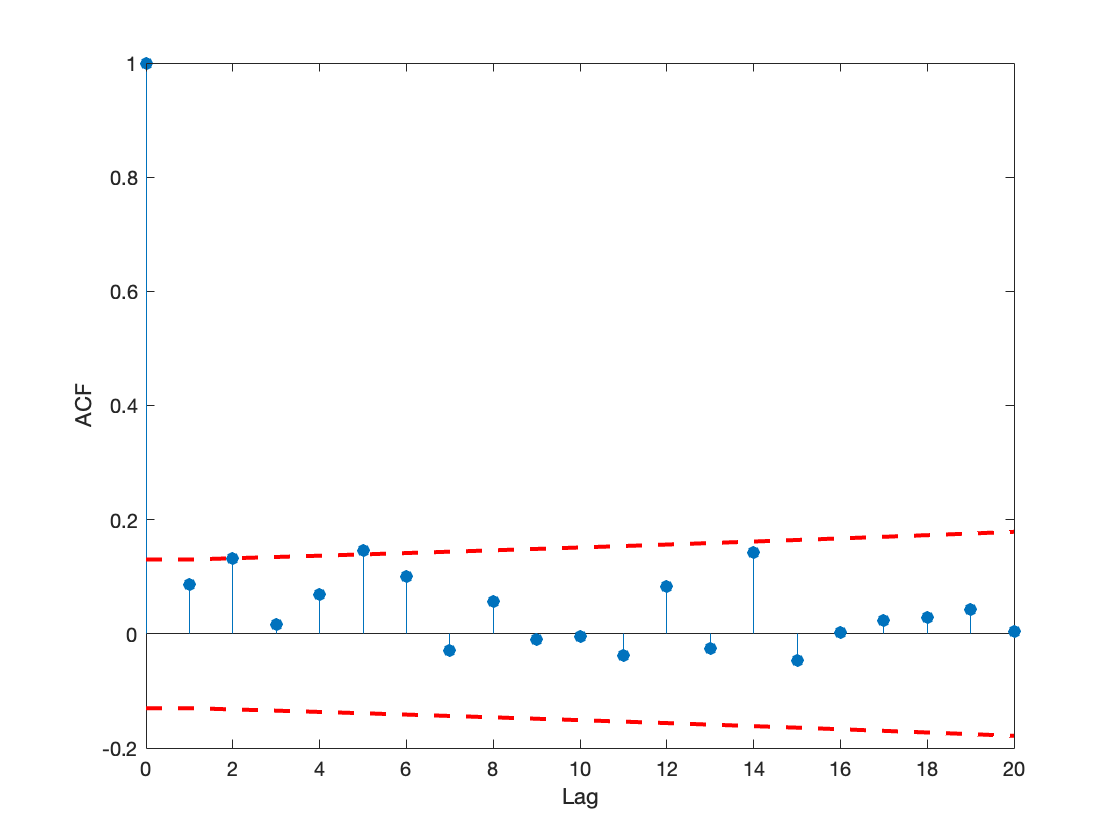}
        \caption{ACF of the square of the residuals}
        \label{3-d}
    \end{subfigure}

    \caption{ACF of aluminum prices and squared prices (panels a and b) and  MAR(1,1) residuals and squared residuals (panels c and d)}
    \label{figure 3}
\end{figure}

 \begin{figure}[H]
    \centering

     % \hfill
     % \begin{subfigure}[b]{0.49\textwidth}
     %     \centering
     %     \includegraphics[width=\textwidth]{Feb06-residuals.png}
     %     \caption{MAR(1,1) residuals}
     %     \label{2-c}
     % \end{subfigure}
     % \hfill
     % \begin{subfigure}[b]{0.49\textwidth}
     %     \centering
         \includegraphics[width=0.5\textwidth]{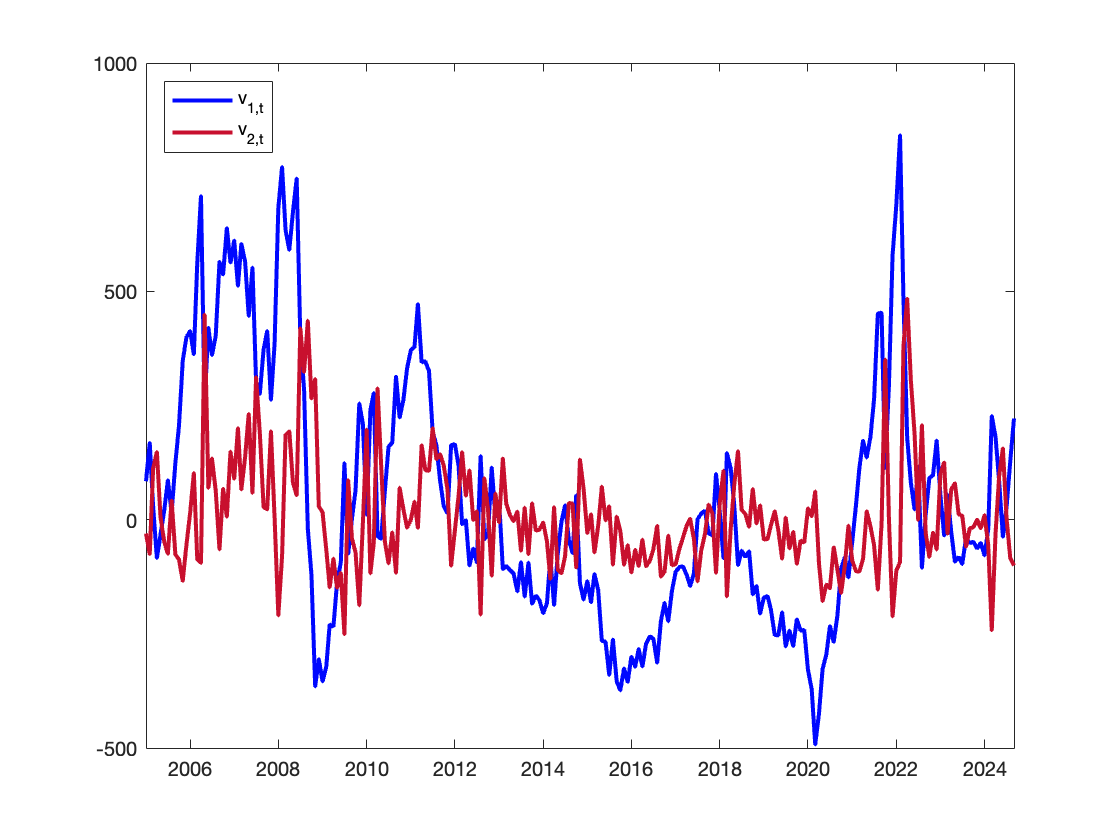}
     %     \caption{MAR(1,1) causal-noncausal components}
     %     \label{2-d}
     % \end{subfigure}
    
    \caption{ MAR(1,1)  causal-noncausal components}
    \label{2-d}
\end{figure}

\newpage

\clearpage
\newpage

\begin{center}
REFERENCES TO APPENDICES A-C
\end{center}

\color{black}

% \nin Anderson, T. (2002): "Canonical Correlation Analysis and Reduced Rank Regression in Autoregressive Models", Annals of Statistics, 30, 1134-1154.

\medskip
\nin Andrews, D. (1988): "Laws of Large Numbers for Dependent Non-Identically Distributed Random Variables", Econometric Theory, 4, 458-467.

\medskip
\nin Andrews, D. (1992): "Generic Uniform Convergence", Econometric Theory, 8, 241-257.

\medskip
\nin Brown, B. (1971): "Martingale Central Limit Theorem", Annals of Mathematical Statistics, 42, 59-66

% \color{black}
\medskip
\nin Cavaliere, G., Nielsen, H, and A. Rahbek (2020): "Bootstrapping Noncausal Autoregressions with Applications to Explosive Bubble Modelling", Journal of Business and Economic Statistics, 38, 55-67.
\color{black}

\medskip
\nin Chitturi, R. (1974): "Distribution of Residual Autocorrelations in Multiple Autoregressive Schemes", JASA, 69, 928-934.

\medskip
\nin Chitturi, R. (1976): "Distribution of Multivariate White Noise Autocorrelations", JASA, 71, 223-226.

\color{black}

% \medskip
% \nin De Jong, R. (1995): "Laws of Large Numbers for Dependent Heterogeneous Processes", Econometric Theory, 11, 347-358.

\medskip
\nin De Jong, R. (1998): "Weak Laws of Large Numbers for Dependent Random Variables", Annales d'Economie et de Statistique, 51, 210-225.

\medskip
\nin \textcolor{black}{ Dovonon, P. and N. Gospodinov (2024): "Specification Testing for Conditional Moment Restrictions Under Local Identification Failure",  Quantitative Economics, forthcoming}

\medskip
\nin Dvoretski, A. (1970): "Asymptotic Normality for Sums of Dependent Random Variables", Proc. of the Sixth Berkeley Symposium on Math. Stat. and Prob., Vol II, 513-535.

\color{black}
\medskip
\nin Escanciano, M. (2007): "Model Checks Using Residual Marked Empirical Processes", Statistica Sinica, 17, 115-138
\color{black}

\medskip
\nin  Gourieroux, C. and J. Jasiak (2005): "Nonlinear Innovations and Impulse Responses", Annals of Economics and Statistics, 78, 1-30

\medskip
\nin Gourieroux, C. and J. Jasiak (2023): "Generalized Covariance Estimator", Journal of Business and Economic Statistics, 41, 1315-1327.

% \color{black}
% \medskip
% \nin Hahn, J. (1996): "A Note on Bootstrapping Generalized Method of Moments Estimators", Econometric Theory, 12, 187-197.
% \color{black}

% \medskip
% \nin Hannan, J. (1967): "Canonical Correlation and Multiple Equation Systems in Economics", Econometrica, 35, 123-138.

\medskip
\nin Hannan, J. (1976): "The Asymptotic Distribution of Serial Covariances", Annals of Statistics, 4, 396-399.

\color{black}
\medskip
\nin Inoue, A. and M. Shintani (2006): Bootstrapping GMM Estimators for Time Series", Journal of Econometrics, 133, 531-555.

% \medskip
% \nin Jennrich, R. (1969): "Asymptotic Properties of Nonlinear Least Squares", Annals of Mathematical Statistics, 40, 633-643.

%\medskip
%\nin  Keweloh, S.A. (2020): "A Generalized Method of Moments Estimator for Structural Autoregressions Based on Higher Moments", Journal of Business \& Economic Statistics, 39(3), 1-29.

\medskip
\nin \textcolor{black}{Koenker, R. and J. Machado (1999) "GMM Inference when the Number of Moment Conditions is Large", Journal of Econometrics, 93, 327-344}

\medskip
\nin Kundu, S., Majumdar, S., and K. Mukherjee (2000): "Central Limit Theorems Revisited", Statistics and Probability Letters, 47, 265-275.

\color{black}
\medskip
\nin Leucht, A., and M. Neumann (2003): " Consistency of General Bootstrap Methods for degenerate U-Type and V-Type Statistics", Journal of Multivariate Analysis, 100, 1622-1633.

\medskip
\nin Wan, P., and R. Davis (2022): "Goodness-of-Fit Testing for Time Series Models via Distance Covariance", Journal of Econometrics, 227(1), 4-24.

\end{document}